\documentclass[amssymb,aps,prl,superscriptaddress,noeprint,reprint,twocolumn]{revtex4-1}

\usepackage{graphicx}
\usepackage{mathtools}
\usepackage{tikz}
\usepackage{siunitx}
\usepackage{xcolor}
\usepackage[normalem]{ulem}
\usepackage{hyperref}

\usepackage{xcolor}
\hypersetup{
    colorlinks,
    linkcolor={red!50!black},
    citecolor={blue!50!black},
    urlcolor={blue!80!black}
}

\usepackage{color}
\definecolor{blue}{rgb}{0.00,0.00,0.95}

\usepackage{pgffor}
\usepackage{pdfpages} 
\makeatletter
\AtBeginDocument{\let\LS@rot\@undefined}
\makeatother

\renewcommand{\d}{{\rm d}}

\begin{document}

\title{Learning the dynamics of cell-cell interactions in confined cell migration}

\author{David B. Br\"uckner}
\affiliation{Arnold Sommerfeld Center for Theoretical Physics and Center for NanoScience, Department of Physics, Ludwig-Maximilian-University Munich, Theresienstr. 37, D-80333 Munich, Germany}
\author{Nicolas Arlt}
\affiliation{Arnold Sommerfeld Center for Theoretical Physics and Center for NanoScience, Department of Physics, Ludwig-Maximilian-University Munich, Theresienstr. 37, D-80333 Munich, Germany}
\author{Alexandra Fink}
\affiliation{Faculty of Physics and Center for NanoScience, Ludwig-Maximilian-University, Geschwister-Scholl-Platz 1, D-80539 Munich, Germany}
\author{Pierre Ronceray}
\affiliation{Center for the Physics of Biological Function, Princeton University, Princeton, NJ 08544, USA}
\author{Joachim O. R\"adler}
\affiliation{Faculty of Physics and Center for NanoScience, Ludwig-Maximilian-University, Geschwister-Scholl-Platz 1, D-80539 Munich, Germany}
\author{Chase P. Broedersz}
\email[]{c.broedersz@lmu.de}
\affiliation{Arnold Sommerfeld Center for Theoretical Physics and Center for NanoScience, Department of Physics, Ludwig-Maximilian-University Munich, Theresienstr. 37, D-80333 Munich, Germany}
\affiliation{Department of Physics and Astronomy, Vrije Universiteit Amsterdam, 1081 HV Amsterdam, The Netherlands}

\begin{abstract}
  The migratory dynamics of cells in physiological processes, ranging from wound healing to cancer metastasis, rely on contact-mediated cell-cell interactions. These interactions play a key role in shaping the stochastic trajectories of migrating cells. While data-driven physical formalisms for the stochastic migration dynamics of single cells have been developed, such a framework for the behavioral dynamics of interacting cells still remains elusive. Here, we monitor stochastic cell trajectories in a minimal experimental cell collider: a dumbbell-shaped micropattern on which pairs of cells perform repeated cellular collisions. We observe different characteristic behaviors, including cells reversing, following and sliding past each other upon collision. Capitalizing on this large experimental data set of coupled cell trajectories, we infer an interacting stochastic equation of motion that accurately predicts the observed interaction behaviors. Our approach reveals that interacting non-cancerous MCF10A cells can be described by repulsion and friction interactions. In contrast, cancerous MDA-MB-231 cells exhibit attraction and anti-friction interactions, promoting the predominant relative sliding behavior observed for these cells. Based on these experimentally inferred interactions, we show how this framework may generalize to provide a unifying theoretical description of the diverse cellular interaction behaviors of distinct cell types.
\end{abstract}

\maketitle

Collective cellular processes such as morphogenesis, wound healing, and cancer invasion, rely on cells moving and rearranging in a coordinated manner. For example, in epithelial wound healing, cells collectively migrate towards the injury and assemble to close the wound~\cite{Poujade2007a,Stramer2005,Weavers2016}. In contrast, in metastasizing tumors, cancer cells migrate outwards in a directed fashion and invade surrounding tissue~\cite{Friedl2003}. At the heart of these emergent collective behaviors lie contact-mediated cell-cell interactions ~\cite{Weavers2016,Carmona-Fontaine2008,Villar-Cervino2013,Theveneau2010,Davis2012,Smeets2016,Stramer2017}, which are apparent in two-body collisions of cells~\cite{Stramer2017,Astin2010,Teddy2004,Abercrombie1954a}. These cellular interactions depend on complex molecular mechanisms, including cadherin-dependent pathways and receptor-mediated cell-cell recognition~\cite{Carmona-Fontaine2008,Stramer2017,Astin2010,Davis2015,Moore2013,Matthews2008,Kadir2011}. At the cellular scale, this molecular machinery leads to coordinated, functional behaviors of interacting cells~\cite{Weavers2016,Carmona-Fontaine2008,Villar-Cervino2013,Theveneau2010,Davis2012,Smeets2016,Stramer2017}, which are highly variable and may take distinct forms in different biological contexts~\cite{Stramer2017,Abercrombie1979,Milano2016,Li2018,Hayakawa2020}.

Achieving a quantitative understanding of the stochastic migratory dynamics of cells at the behavioral level could yield key insights into both the underlying molecular mechanisms~\cite{Maiuri2015,Lavi2016} and the biological functions~\cite{Stramer2017} associated to these behaviors. For non-interacting, single migrating cells, data-driven approaches have revealed quantitative frameworks to describe the behavior of free unconstrained migration~\cite{Selmeczi2005,Li2011,Pedersen2016} and confined migration in structured environments~\cite{Brueckner2019,Brueckner2020,Fink2019}. However, it is still poorly understood how the migratory dynamics of cells are affected by cell-cell interactions and a quantitative formalism for the emergent behavioral dynamics of interacting cells is still lacking~\cite{Alert2020}. Indeed, it is unclear whether cellular collision behaviors follow a simple set of interaction rules, and if so, how these rules vary for different types of cells.


\begin{figure*}[ht]
\centering
	\includegraphics[width=0.8\textwidth]{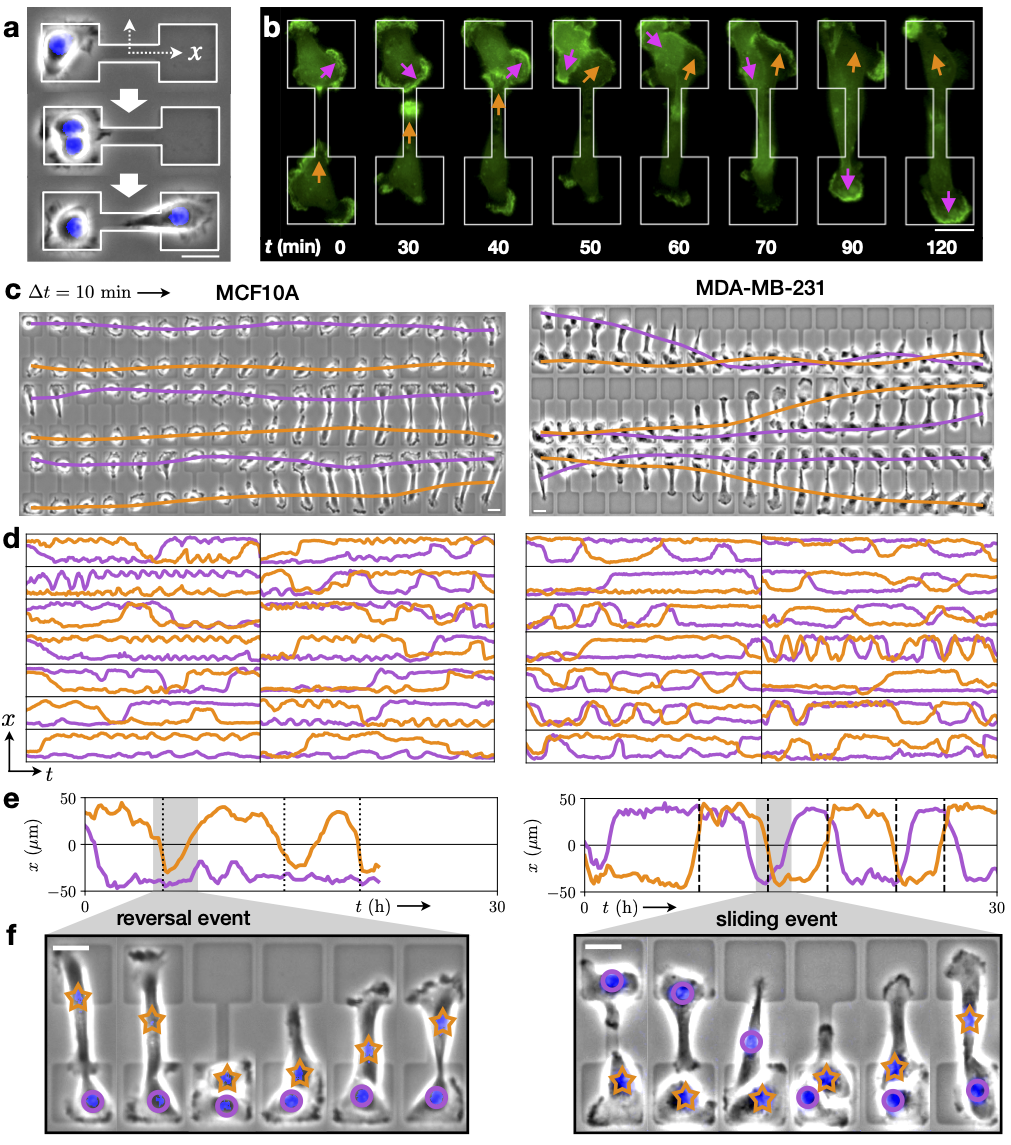}
	\caption{
		\textbf{Stochastic switching dynamics of confined cell pairs.} 
		\textbf{a.} Experimental design: single cells are confined to two-state micropatterns (white outline). We track cell pairs resulting from cell divisions. The stained nucleus is colored in blue.
		\textbf{b.} Time-series of two interacting MDA-MB-231 cells transfected with LifeAct-GFP. Arrows highlight regions of pronounced actin activity, and the arrow color indicates the cell identity.
		\textbf{c.} Brightfield image series with overlaid nuclear trajectories (orange, violet). Images are taken at a time interval $\Delta t$ = 10min.
		\textbf{d.} Sample set of nuclear trajectories $x_{1,2}(t)$ as a function of time, shown for 14 cell pairs. Axes limits are $0 < t < 30$ h and $-60 \ \si{\micro\meter} < x < 60 \ \si{\micro\meter}$, with $x = 0$ at the centre of the bridge. In total, we tracked 155 MCF10A cell pairs (corresponding to a total trajectory length of 3200 h) and 90 MDA-MB-231 cell pairs (2700 h).  
		\textbf{e.} Single cell-pair trajectory, with highlighted reversal (dotted lines) and sliding events (dashed lines).
		\textbf{f.} Key stages of the reversal and sliding events, corresponding to the sections highlighted in grey in \textbf{e}. Images are shown at 40 min time intervals for MCF10A, and 30 min intervals for MDA-MB-231. Orange stars and violet circles indicate the identities of the cells. 
		In panels \textbf{c}-\textbf{f}, the left column corresponds to MCF10A cells, and the right column to MDA-MB-231 cells. All scale bars correspond to $25 \si{\micro\meter}$.
		}
	\label{fig1}
\end{figure*}


The study of interacting cell dynamics is complicated by the complex settings in which they take place, confounding contributions of single-cell behavior, interaction with the local micro-environment, and cell-cell interactions. Thus, simplified assays have been developed where cells are confined by one-dimensional micro-patterned patches~\cite{Huang2005,Segerer2015} or tracks~\cite{Milano2016,Li2018,Desai2013,Scarpa2013}, microfluidics~\cite{Lin2015}, and suspended fibers~\cite{Singh2020}. In these systems, cells exhibit characteristic behaviors upon pair-wise collisions, including reversal, sliding and following events. Upon contact, many cell types exhibit a tendency to retract, repolarize and migrate apart - termed Contact Inhibition of Locomotion (CIL)~\cite{Stramer2017,Abercrombie1954a,Mayor2010}. Indeed, diverse cell types, including epithelial and neural crest cells, predominantly reverse upon collision~\cite{Milano2016,Desai2013,Scarpa2013}. In contrast, the breakdown of CIL is commonly associated with cancer progression~\cite{Astin2010,Abercrombie1979,Milano2016,Abercrombie1954,Milano2016}, and cancerous cells have been observed to move past each other more readily than non-cancerous cells~\cite{Milano2016}. However, it is unclear how to describe these distinct collision behaviors in terms of physical interactions.

Models for collective cell migration often assume repulsive potentials or alignment terms~\cite{Smeets2016,Alert2020,Sepulveda2013,Basan2013,Copenhagen2018,Garcia2015}, but the form of these interactions is not derived directly from experimental data. Such data-driven approaches have been developed for single cell migration~\cite{Selmeczi2005,Li2011,Pedersen2016,Brueckner2019,Brueckner2020,Fink2019}, but have not yet been extended to interacting systems. The search for unifying quantitative descriptions of the dynamics of interacting cell trajectories is further complicated by their intrinsic stochasticity, resulting in highly variable migration and collision behavior~\cite{Milano2016,Desai2013,Scarpa2013,Singh2020}. Thus, developing a system-level understanding of cell-cell interactions requires a quantitative data-driven approach to learn the full stochastic dynamics of interacting migrating cells.

Here, we develop a theoretical framework for the dynamics of interacting cells migrating in confining environments, inferred directly from experiments. Specifically, we confine pairs of migrating cells into a minimal 'cell collider': a two-state micropattern consisting of two square adhesive sites connected by a thin bridge. Both non-cancerous (MCF10A) and cancerous (MDA-MB-231) human breast tissue cells frequently migrate across the bridge, giving rise to repeated cellular collisions. In line with prior observations~\cite{Milano2016}, we find that while MCF10A cells predominantly reverse upon collision, MDA-MB-231 cells tend to interchange positions by sliding past each other. To provide a quantitative dynamical framework for these distinct interacting behaviors, we focus on a simplified, low-dimensional representation of these collision dynamics by measuring the trajectories of the cell nuclei. The cell collider experiments yield large data sets of such interacting trajectories, allowing us to infer the stochastic equation of motion governing the two-body dynamics of interacting cells. Our data-driven approach reveals the full structure of the cellular interactions in terms of the relative position and velocity of the cells. Specifically, the dynamics of MCF10A cells are captured by repulsion and friction interactions. In contrast, MDA-MB-231 cells exhibit novel and surprising dynamics, combining attractive and 'anti-friction' interactions, which have no equivalent in equilibrium systems. This inferred model quantitatively captures the key experimental observations, including the distinct collision phenotypes of both cell lines. Our framework can be generalized to provide a conceptual classification scheme for the system-level dynamics of cell-cell interactions, and is able to capture various previously observed types of cell-cell collision behaviors.


\begin{figure}[]
	\includegraphics[width=0.5\textwidth]{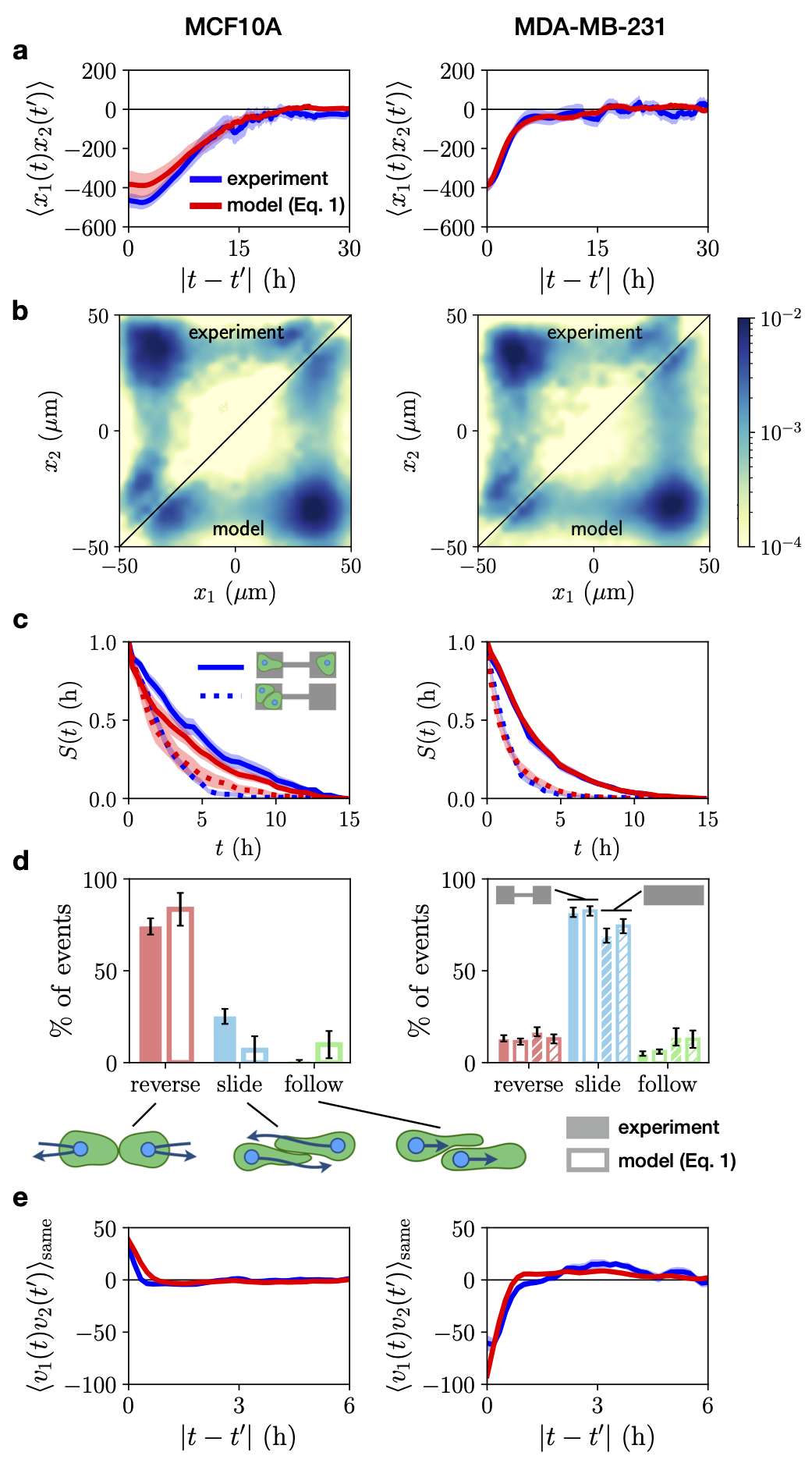}
	\caption{
	\textbf{Statistics of the stochastic interaction dynamics.} 
		\textbf{a.} Cross-correlation function of cell positions $\langle x_1 (t)x_2 (t')\rangle$.
		\textbf{b.} Joint probability distributions $p(x_1,x_2)$ of cell positions, plotted logarithmically. The top triangle of the symmetric distribution shows the experimental result, the bottom triangle shows the model prediction (for full distributions and linear plots, see Supplementary Fig. S12,13). 
		\textbf{c.} Probability distribution $S(t)$ giving the probability that a configuration switch has \textit{not} occurred after time $t$, for the opposite-side configuration (solid) and the same-side configuration (dotted). 
		\textbf{d.} Percentages of each of the three types of collision events observed, which are sketched below. For MDA-MB-231 cells, dashed bars correspond to data from cells on micropatterned tracks, with the corresponding model prediction obtained using a single-cell term inferred from single cells on a track, and interaction terms inferred from cell pairs on two-state patterns.
		\textbf{e.} Velocity cross-correlation function $\langle v_1 (t)v_2 (t')\rangle_\mathrm{same}$, calculated for times where the cells occupy the same island. 
		In panels \textbf{a} and \textbf{c}, experimental data are shown in blue, and model predictions (corresponding to Eqn.\eqref{eq1}) in red. Shaded regions and errorbars denote bootstrap errors (Supplementary Section S3).
		}
	\label{fig2}
\end{figure}


\subsection{Two-state micropatterns provide minimal cell collider}
To investigate the two-body interaction dynamics of migrating cells, we designed a micropatterned system in which two cells repeatedly collide. The micropattern confines the cells to a fibronectin-coated adhesive region, consisting of a narrow bridge separating two square islands. Outside this dumbbell-shaped region the substrate is passivated with PLL-PEG, to which the cells do not adhere. We first confine single cells to these patterns, as described in previous work~\cite{Brueckner2019}. Here, we identify cells which undergo division from which we obtain confined, isolated pairs of daughter cells (Fig.~\ref{fig1}a). We employ phase-contrast time-lapse microscopy to study the homotypic interactions of pairs of non-cancerous (MCF10A) and cancerous (MDA-MB-231) human mammary epithelial cells. The confining bridge between the two islands leads to two well-defined configurations, with either both cells on the same island, or on opposite sides of the pattern, between which the system repeatedly switches (Fig.~\ref{fig1}c,d and Supplementary Videos S1-4). During these switching events, the cells interact with each other. Therefore, our experimental setup offers a simple platform to study the interactions of confined migrating cells in a standardized manner: a minimal 'cell collider'.

Within this cell collider, cells are highly motile and exhibit actin-rich lamellipodia-like protrusions forming at the cell periphery (Fig.~\ref{fig1}b, Supplementary Video S5). As a simplified, low-dimensional representation of the interaction dynamics, we use the trajectories of the cell nuclei, which reflect the long time-scale interacting behavior of the cells (Fig.~\ref{fig1}c). These coupled cell trajectories are highly stochastic. Using this assay, we monitor the stochastic two-body dynamics of hundreds of cells over long periods of time (up to 40h per cell pair) in standardized micro-environments, yielding an unprecedented amount of statistics on cell-cell interactions (Fig.~\ref{fig1}d). Importantly, we find that most of the interactive behavior is captured by the $x$ position along the long axis of the pattern (Supplementary Section S3). Thus, our cell-collider experiments provide a large data set of low-dimensional trajectories of interacting cells, allowing in-depth statistical analysis of the cellular dynamics.

\subsection{Cell pairs exhibit mutual exclusion}
A key feature of the trajectories for both cell lines is the apparent preference for the configuration in which the cells are on opposite islands (Fig.~\ref{fig1}d). Indeed, the positions of the two cells are strongly correlated: the cross-correlation function $\langle x_1 (t)x_2 (t')\rangle$ exhibits a pronounced negative long-time scale correlation for both cell lines (Fig.~\ref{fig2}a). Correspondingly, the joint probability distribution of positions $p(x_1,x_2)$ exhibits prominent peaks where cells occupy opposite sides, and only faint peaks where they are on the same side (Fig.~\ref{fig2}b), suggesting two distinct configurations. These configurations are connected by 'paths' in the probability density, along which transitions occur. We find that the cumulative probability $S(t)$ that a configuration switch has not occurred after time t decays more slowly for opposite-side than same-side configurations (Fig.~\ref{fig2}c). Taken together, these results indicate that both MCF10A and MDA-MB-231 cells exhibit a mutual exclusion behavior.

\subsection{MCF10A and MDA-MB-231 cells exhibit distinct collision behavior}
While the cells mutually exclude each other, they are also highly migratory and thus frequently transit the constricting bridge. This results in repeated stochastic collision events, providing statistics for how these cells interact during a collision. Following a collision, we observe three distinct types of behaviors: reversal events, where the cells turn around upon collision; sliding events, where the cells interchange positions by sliding past each other; and following events where the cells remain in contact and perform a joint transition (Fig.~\ref{fig1}e,f, Supplementary Section S3). These three behaviors have been previously used as observables of cell-cell interactions in one-dimensional and fibrillar environments~\cite{Milano2016,Desai2013,Scarpa2013,Singh2020,Kulawiak2016}.

To quantify the interaction behavior of MCF10A and MDA-MB-231 cells, we identify collision events and measure the percentage that result in reversal, sliding or following events (Fig.~\ref{fig2}d). Both cell lines exhibit only a small fraction of following events. Remarkably however, we find that collisions of MCF10A cells predominantly result in reversals, while MDA-MB-231 cells typically slide past each other upon collision, in line with observations in other confining geometries~\cite{Milano2016}. To further explore the generality of this result, we perform additional experiments with  MDA-MB-231 cells on micropatterned tracks without constrictions, but the same overall dimensions of the two-state micropatterns. We find that sliding events similarly dominate for MDA-MB-231 cells on this pattern, with similar overall event ratios. The different responses to cell-cell contacts are also reflected by the velocity cross-correlation of the two cells when occupying the same side of the two-state micropatterns: $\langle v_1 (t)v_2 (t')\rangle_\mathrm{same}$: MCF10A cells exhibit a positive velocity correlation while MDA-MB-231 cells exhibit a negative velocity correlation (Fig.~\ref{fig2}e). Taken together, these findings show that while both cell lines exhibit similar mutual exclusion behavior, there are clear differences in their collision dynamics. This raises a key question: is there an overarching dynamical description which captures both the similarities and differences of these interaction behaviors?

\begin{figure}[]
	\includegraphics[width=0.5\textwidth]{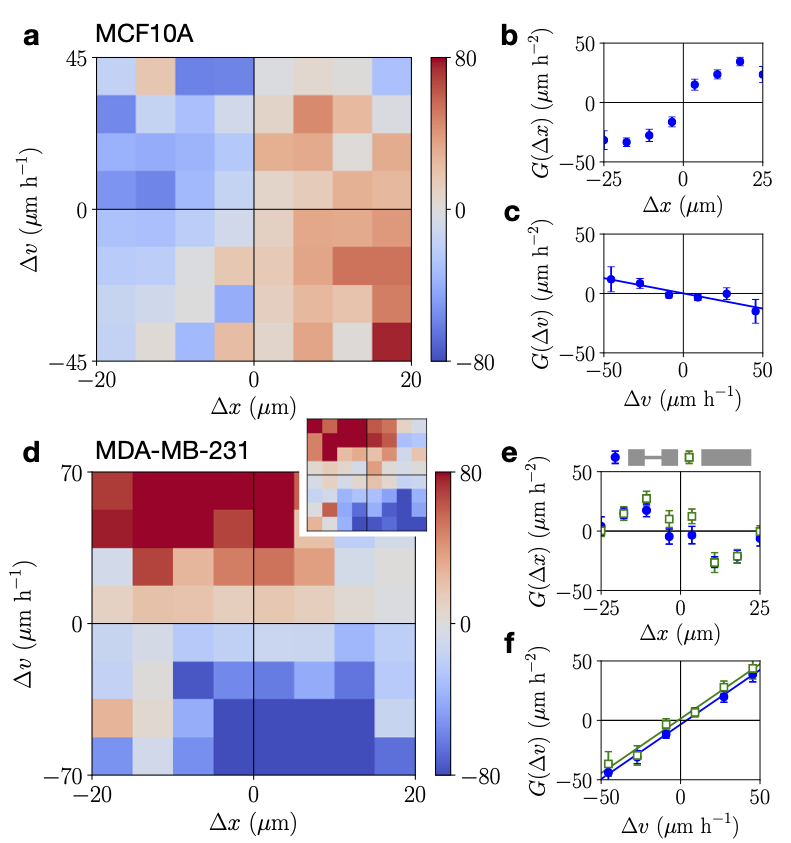}
	\caption{
	\textbf{Contact acceleration maps.} 
		\textbf{a,d.} Contact acceleration maps $G(\Delta x,\Delta v)$, measured in units of $\si{\micro\meter}/\mathrm{h}^{2}$. Inset in \textbf{d}: Map for MDA-MB-231 cells on micropatterned tracks.
		\textbf{b,e.} Contact accelerations as a function of the cell separation $\Delta x$: $G(\Delta x)=\langle \dot{v}-F(x,v)|\Delta x\rangle$.
		\textbf{c,f.} Contact accelerations as a function of the relative velocity of the cells $\Delta v$: $G(\Delta v)=\langle \dot{v}-F(x,v)|\Delta v\rangle$.
		Lines indicate linear fits. Error bars show bootstrap errors. Panels \textbf{a}-\textbf{c} show data for MCF10A cells, and panels \textbf{d}-\textbf{f} for MDA-MB-231 cells. In panels \textbf{e},\textbf{f}, open green symbols correspond to data from experiments on micropatterned tracks.
		}
	\label{fig3}
\end{figure}


\subsection{Contact acceleration maps reveal dynamics of cell-cell interactions}
Here, we aim to describe the underlying interaction dynamics that capture the full stochastic long time-scale behavior of repeatedly colliding cell pairs. The dynamics of single migrating cells is well described by an equation of motion that is second order in time~\cite{Selmeczi2005,Li2011,Pedersen2016,Brueckner2019,Brueckner2020,Fink2019}, making accelerations the natural quantity to describe cell motility. Specifically, we previously showed that the migration dynamics of single cells in confinement can be described by the average acceleration as a function of cell position $x$ and velocity $v=\d x/\d t$, given by the conditional average $F(x,v)=\langle \dot{v}|x,v\rangle$, where $\dot{v}=\d v/\d t$~\cite{Brueckner2019,Brueckner2020,Fink2019}. To uncover the general structure of the cell-cell interactions in our experiments, we therefore first focus on the observed cellular accelerations upon contact as a function of the distance and relative velocity of the cells. We anticipate contributions from cell-cell interactions to depend on the relative position $\Delta x$ and relative velocity $\Delta v$ of the cell pair. Under certain assumptions, which we test in the next section, we can estimate the interactive contribution to cellular accelerations by first subtracting the single-cell contribution $F(x,v)$, and then determining the remaining acceleration as a function of $\Delta x$ and $\Delta v$: $G(\Delta x,\Delta v)=\langle \dot{v}-F(x,v)|\Delta x,\Delta v\rangle$ (see Methods and Supplementary Section S3). To further illustrate this approach, we verify that it accurately recovers the functional dependence of simple interactions from simulated trajectories (Supplementary Section S3). Thus, we interpret this 'contact acceleration map' as the average acceleration due to the interactions of a cell pair. 

Strikingly, we find that MCF10A and MDA-MB-231 cells exhibit qualitatively different contact acceleration maps (Fig.~\ref{fig3}a,d). Indeed, for MCF10A cells, the contact acceleration exhibits a clear dependence on the relative position, while MDA-MB-231 cells exhibit accelerations that mainly depend on the relative velocity. We investigate these differences by measuring the 1D-dependence of the contact accelerations as a function of just $\Delta x$ or $\Delta v$. These plots reveal that MCF10A cells exhibit a combination of repulsive accelerations (Fig.~\ref{fig3}b) and a weak friction-like component (Fig.~\ref{fig3}c). By contrast, MDA-MB-231 cells exhibit contact accelerations with opposite sign, suggesting an attractive component (Fig.~\ref{fig3}e) and an effective linear 'anti-friction' (Fig.~\ref{fig3}f). Interestingly, we find that the contact accelerations on micropatterned tracks are qualitatively and quantitatively similar, suggesting that these findings are not very sensitive to the confinement geometry (Inset Fig.~\ref{fig3}d). These findings suggest that the contact accelerations of these cells exhibit features that could be described as combinations of cohesive (repulsion/attraction) and frictional terms. This raises the question: are the simple physical interactions suggested by these maps sufficient to describe the complex interaction dynamics of these cell pairs?


\begin{figure*}[]
\centering
	\includegraphics[width=0.7\textwidth]{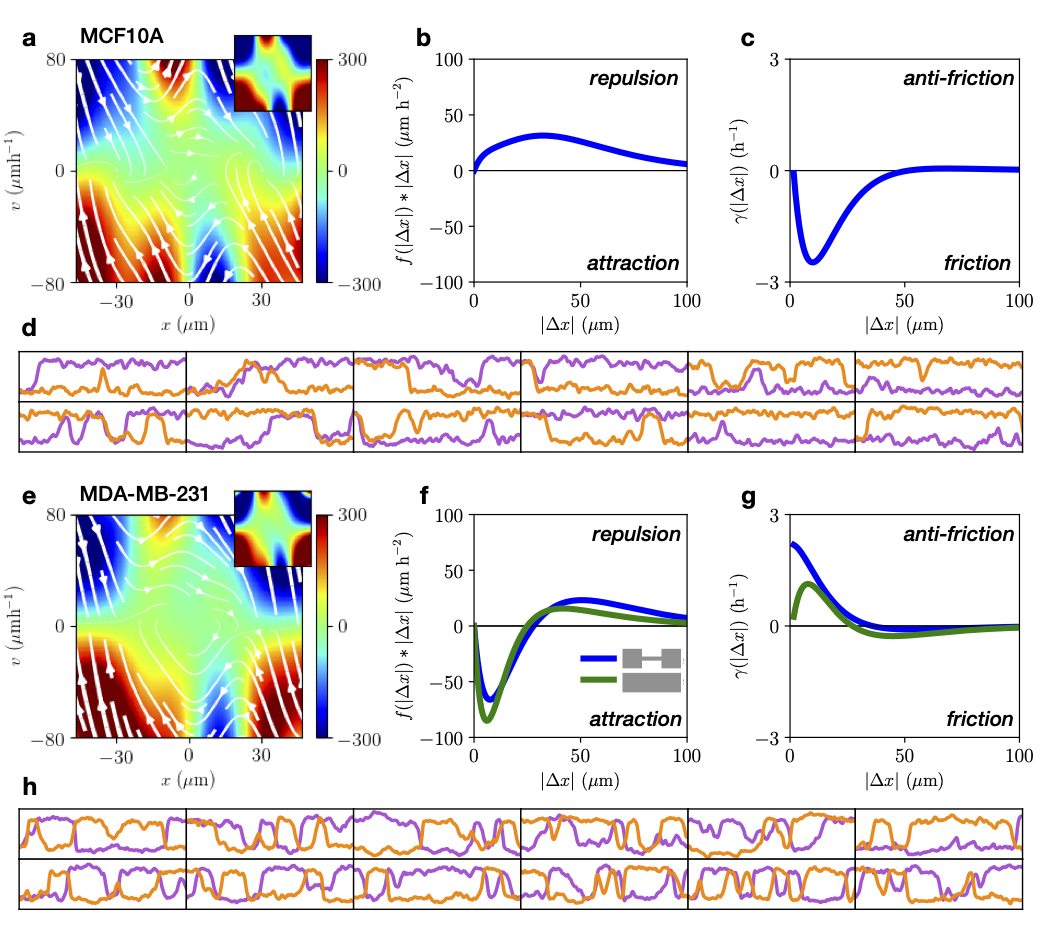}
	\caption{
		\textbf{Equation of motion for interacting cells.} 
		\textbf{a,e.} Single-cell contribution $F(x,v)$ to the interacting dynamics, measured in units of $\si{\micro\meter}/\mathrm{h}^{2}$. White lines indicate the flow field given by $(\dot{x},\dot{v})=(v,F(x,v))$. Inset: corresponding term inferred from experiments with single cells~\cite{Brueckner2019}.
		\textbf{b,f.} Cohesive interaction term $f(|\Delta x|)\Delta x$. Positive values indicate repulsive interactions, while negative values correspond to attraction.
		\textbf{c,g.} Effective frictional interaction term $\gamma(|\Delta x|)$. Here, positive values indicate an effective anti-friction, and negative values an effective frictional interaction.
		\textbf{d,h.} Trajectories obtained from model simulations. Axes limits are 0 < $t$ < 30 h and -60 \si{\micro\meter} < $x$ < 60 \si{\micro\meter}.
		Panels \textbf{a}-\textbf{d} show data for MCF10A cells, and panels \textbf{e}-\textbf{h} for MDA-MB-231 cells. For MDA-MB-231 cells, green lines show the interactions inferred from cell pairs interacting on micropatterned tracks.
		}
	\label{fig4}
\end{figure*}


\subsection{Interacting equation of motion captures experimental statistics}
To investigate whether the interacting dynamics of MDA-MB-231 and MCF10A cells can be described by the physical interactions implied by the contact acceleration maps, we consider a simple model for cell-cell interactions in confining environments. Motivated by the structure of the contact accelerations, we postulate that the dynamics of the cells can be described by a stochastic equation of motion of the form
\begin{equation}
\frac{\d v}{\d t}  = F(x,v)+ f(|\Delta x|)\Delta x + \gamma(|\Delta x|)\Delta v + \sigma \eta(t)
\label{eq1}
\end{equation}
Here, we assume that the interactions between each cell and the confinement can be described by a term $F(x,v)$, similar to single cell experiments~\cite{Brueckner2019}. Furthermore, we assume that the interactions between the two cells can be separately described by two interaction terms: a cohesive term $f(|\Delta x|)\Delta x$, which captures repulsion and attraction; and an effective friction term $\gamma(|\Delta x|)\Delta v$ that may depend on the distance between the cells. The intrinsic stochasticity of the migration dynamics is accounted for by a Gaussian white noise $\eta(t)$, with $\langle \eta(t)\rangle=0$ and $\langle \eta(t)\eta(t') \rangle=\delta(t-t')$. Note that this equation of motion captures the effective dynamics that describe the cellular accelerations, rather than mechanical forces acting on the cell.

To investigate this model, we first require a systematic approach to infer the systems' stochastic dynamics and delineate single-cell (one-body) and interactive (two-body) contributions to the dynamics. Thus, we employ a rigorous inference method, Underdamped Langevin Inference (ULI)~\cite{Brueckner2020a}, to infer the terms of this equation of motion from the experimentally measured trajectories. In this approach, the inferred terms are completely constrained by the short time-scale information in the measured trajectory, i.e. the velocities and accelerations of the cells (see Methods and Supplementary Section S4). 

Importantly, there is no a priori reason why \eqref{eq1} should provide a reasonable ansatz to correctly capture cell-cell interactions, which could require a more complex description. Thus, we investigate the predictive power of our model by testing whether it correctly captures experimental statistics that were not used to constrain the terms in \eqref{eq1}. Specifically, while the model is learnt on the experimental short time-scale dynamics, we aim to make predictions for long time-scale statistics such as correlation functions. To this end, we simulate stochastic trajectories of interacting cell pairs based on our model (Fig.~\ref{fig4}d,h) to make a side-by-side comparison with the experiments. Remarkably, we find that the model performs well in predicting key experimental statistics for both cell lines, including the joint probability distributions (Fig.~\ref{fig2}b), the distributions of switching times (Fig.~\ref{fig2}c), the cross-correlations of positions and velocity (Fig.~\ref{fig2}a,e), as well as the relative fractions of reversal, sliding and following events (Fig.~\ref{fig2}d). In contrast, performing the same inference procedure with simpler models than \eqref{eq1}, e.g. with only cohesive or friction interactions, shows that simulated trajectories of these models do not capture the observed statistics (Supplementary Section S4). To further challenge our approach, we test whether we can use the interactions learnt from experiments on two-state micropatterns to predict the collision behavior in a different confinement geometry. Specifically, we use the single-cell term $F(x,v)$ inferred from single cell data of MDA-MB-231 cells migrating on micropatterned tracks, together with the interactions inferred from cell pair experiments on two-state micropatterns, to predict the collision ratios of cell pairs on tracks. We find that this model accurately predicts the observed event ratios (Fig.~\ref{fig2}d), showing that the inferred interactions have predictive power also beyond the data set on which they are learnt.

Remarkably, our inference approach reveals that the inferred single-cell contributions $F(x,v)$ on two-state micropatterns are qualitatively and quantitatively similar to the equivalent term inferred from experiments with single cells for both cell lines~\cite{Brueckner2019} (Fig.~\ref{fig4}a,e, Supplementary Section S4). Also, the inferred noise amplitudes are similar to those inferred from single cell experiments for both cell lines, $\sigma \approx 50 \ \si{\micro\meter}/\mathrm{h}^{3/2}$. This suggests that the presence of another cell does not significantly alter the confinement dynamics experienced by one of the cells, and instead manifests in the interaction terms of the equation of motion. Our inference yields the spatial dependence of the cohesion term (Fig.~\ref{fig4}b,f) and the effective friction term (Fig.~\ref{fig4}c,g). Importantly, the functional dependence of the inferred terms is in accord with our interpretation of the contact acceleration maps (Fig.~\ref{fig3}): MCF10A cells exhibit a repulsive cohesive interaction, and a regular effective friction, which reflects that cells slow down as they move past each other. In contrast, MDA-MB-231 cells interact through a predominantly attractive cohesion term, becoming weakly repulsive at long distances, and exhibit effective 'anti-friction'. We infer a similar 'anti-friction' interaction from MDA-MB-231 cell pairs migrating on micropatterned tracks, suggesting that this result is not sensitive to the presence of the constriction (Fig.~\ref{fig4}f,g). This anti-friction generates sliding behavior, where cells on average accelerate as they move past each other with increasing relative speed. These results are robust with respect to the details of the inference procedure (Supplementary Section S4). Taken together, these findings demonstrate that the dynamics of interacting MCF10A and MDA-MB-231 cells on confining micropatterns are well described by our model (\eqref{eq1}) with distinct types of interactions for the two cell lines.

\subsection{Interaction behavior space: a theoretical framework for cell-cell interactions}
\begin{figure*}[]
\centering
	\includegraphics[width=\textwidth]{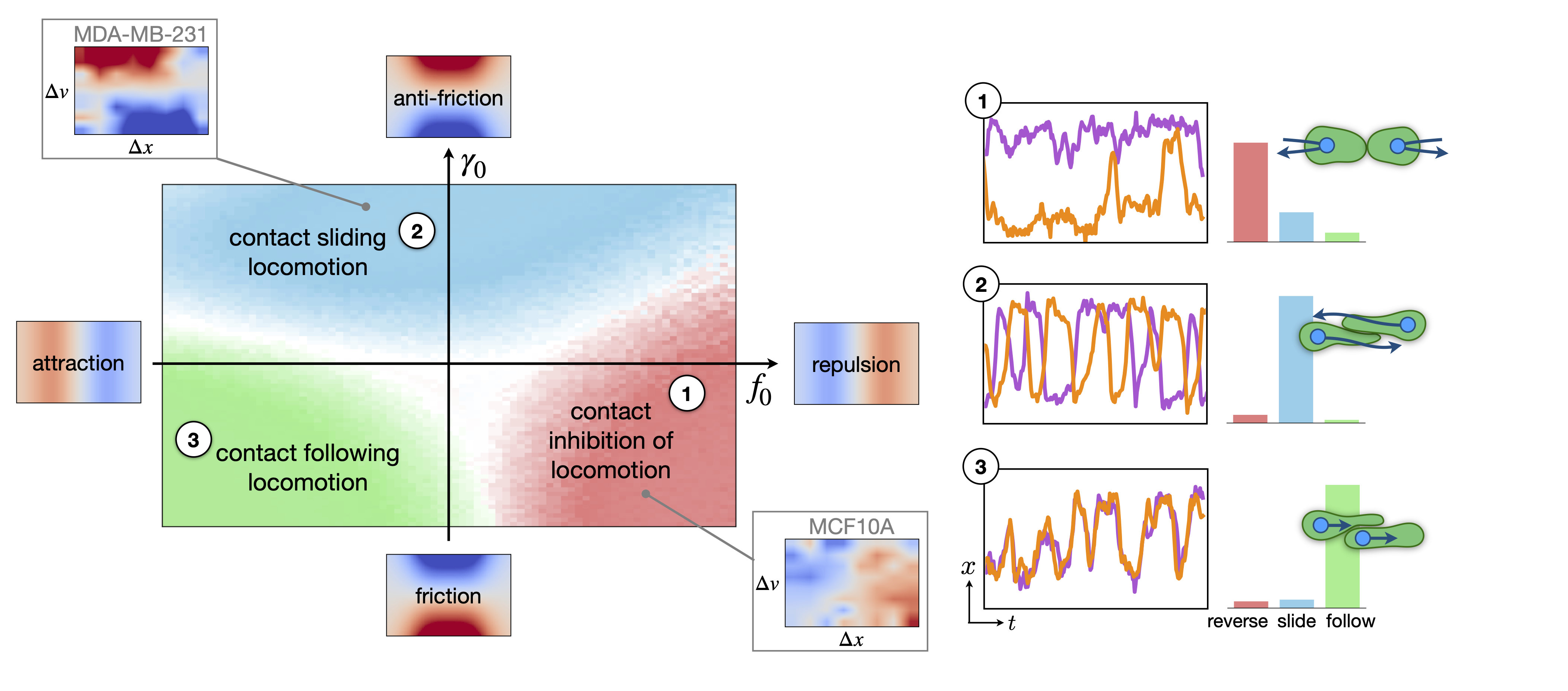}
	\caption{
		\textbf{Interaction behavior space.} 
We construct an interaction space by varying the amplitude of the cohesive and friction interactions, $f_0$ and $\gamma_0$, respectively. Contact acceleration maps for purely attractive, repulsive, frictional and anti-frictional interactions are indicated on the axes. Based on the inferred short-range interactions, we place MDA-MB-231 and MCF10A cells into diagonally opposed quadrants. Predicted behaviors in the interaction space are obtained by varying the cohesion and friction interactions in our model. Specifically, we simulate a model including the inferred MDA-MB-231 single-cell term $F(x,v)$ together with a cohesive term $f(|\Delta x|)=f_0 g_\mathrm{c}(|\Delta x|)$ and an effective friction term $\gamma(|\Delta x|)=\gamma_0 g_\mathrm{f}(|\Delta x|)$, for varying $f_0$ and $\gamma_0$. The distance-dependent functions $g_\mathrm{c,f}$ are positive and monotonically decreasing. These results do not sensitively depend on the specific choice of $F(x,v)$ or $g_\mathrm{c,f}$ ($g_\mathrm{c,f}=\exp[(-|\Delta x|/R_0)]$ is used here) (Supplementary Section S5). For each parameter combination, reversal, sliding and following events where identified. At each point, the dominant behavior is indicated by the color scheme, and white regions correspond to states where no single behavior contributes more than 50\% of events. Numbered insets show sample trajectories from different parts of the interaction map, and the corresponding percentages of reversal (red), sliding (blue), and following events (green).
		}
	\label{fig5}
\end{figure*}

To conceptualize the distinct interactions of MCF10A and MDA-MB-231 cells, we propose an \textit{interaction behavior space}, spanned by the amplitudes of the cohesive and frictional contributions (Fig.~\ref{fig5}). Based on our inference, the two cell lines occupy diagonally opposed quadrants in this space. To investigate whether our model (\eqref{eq1}) is able to capture cellular interaction behaviors more broadly, we predict trajectories for various locations within this interaction map. For interactions consisting of repulsion and friction, we find that collisions predominantly result in reversal events, as we have observed for MCF10A cells. In contrast, for positive friction coefficients, corresponding to effective 'anti-friction', we find that sliding events dominate for all parameter values. This regime thus corresponds to the dynamics we have observed for MDA-MB-231 cells. Finally, attractive interactions with regular friction result in a dominance of following events. The interaction behavior space thus provides an insightful connection between the inferred interaction terms governing the instantaneous dynamics of the system, and the emergent macroscopic, long time-scale collision behavior.


\subsection{Discussion}
In this study, we introduced a conceptual framework for the stochastic behavioral dynamics of interacting cells. To this end, we designed a micropatterned 'cell collider' in which pairs of cells repeatedly collide with each other, providing large amounts of statistics on the long time-scale interactions of migrating cell pairs. A key advantage of this setup is that it yields a large number of collisions under controllable conditions. Moreover, the dynamics of single cells migrating in this confinement is well understood~\cite{Brueckner2019}, providing a benchmark for the dynamics inferred for interacting cells. We compare the homotypic interaction behavior of the non-malignant MCF10A and the metastatic MDA-MB-231 mammary epithelial cell lines. While phenomenological bottom-up models have been developed to describe cell-cell interactions~\cite{Alert2020,Segerer2015,Kulawiak2016,Camley2014,Lober2015,Vedel2013}, we propose an alternative, top-down approach to learn the interacting stochastic equations of motion governing cell migration from the experimentally observed trajectories. Such an effective model captures the emergent dynamics at the cellular scale which are driven by underlying mechanisms, including the intra-cellular polarity machinery. Our inferred models for interacting cells quantitatively capture the distinct behaviors of the two cell lines. This inference reveals that the dynamics can be decomposed into a one-body motility component, which qualitatively matches that observed in single cell experiments~\cite{Brueckner2019}, and a two-body interaction term. 

The interaction terms we inferred from experiments take qualitatively different forms for the two cell lines: while MCF10A cells exhibit repulsion and effective friction, MDA-MB-231 cells exhibit attraction and a novel and surprising effective 'anti-friction' interaction. At the single-cell level, MDA-MB-231 cells are known to be more invasive than MCF10A cells~\cite{Mak2011,Kraning-Rush2013}, and express lower levels of the cell-cell adhesion protein E-cadherin~\cite{Milano2016,Sommers1991}, possibly underlying the different friction-like interactions we found for these cell lines. These two cell lines also display remarkably different collective behaviors~\cite{Carey2013,Lee2020,Kang2020}: MCF10A cells in 2D epithelial sheets exhibit aligned, directed motion and form compact spheroids in 3D culture, with few invasive branches. In contrast, MDA-MB-231 cells in 2D epithelial sheets exhibit non-aligned, random motion and form invasive, non-contiguous clusters in 3D culture, with significant single-cell dispersion from the cluster. These differences in collective behavior may relate to the distinct types of interactions we have inferred from the two-body dynamics of these cell lines.

Based on the inferred equation of motion, we predict an interaction behavior space to link the interaction terms, which govern the instantaneous stochastic dynamics, to the emergent collision behaviors. The three distinct regimes emerging in our model correspond to specific behaviors observed in experiments for various cell types: predominant reversal behavior on 1D lines has been termed \textit{contact inhibition of locomotion}~\cite{Desai2013,Scarpa2013}, a common type of cell-cell interaction~\cite{Carmona-Fontaine2008,Davis2012,Stramer2017,Abercrombie1954a}. By inhibiting intracellular Rho-signalling in neural crest cells, this reversal-dominated behavior could be tuned to following-dominated behavior~\cite{Scarpa2013}. Such following behavior has also been identified as an important mechanism in collective migration~\cite{Teddy2004,Li2018,Hayakawa2020,Fujimori2019}, and was termed \textit{contact following locomotion}~\cite{Li2018}. Finally, previous work has shown that reducing the expression levels of E-cadherin enables otherwise reversing cells to mainly slide past each other~\cite{Milano2016}. For this regime of predominant sliding interactions, we propose the term \textit{contact sliding locomotion}. Based on our interaction behavior space, we find that the 'anti-friction' interactions we identified for MDA-MB-231 cells promote such sliding behavior. The interaction behavior space could thus provide a quantitative classification of distinct modes of interaction that may be achieved through molecular perturbations in experiments~\cite{Milano2016,Scarpa2013}. On the other end of the scale, the 'anti-friction' interaction type we find here could play a role in collective systems such as the fluidization of tissue in tumor invasion~\cite{Kang2020,Palamidessi2019,Han2020}. The form of the interaction terms we inferred from experiments may thus inform models for collective cell migration~\cite{Smeets2016,Alert2020,Sepulveda2013,Basan2013,Copenhagen2018,Garcia2015}. Furthermore, the inference framework we have developed for the dynamics of interacting cell pairs can be extended to infer the dynamics of more complex collective systems, such as small clusters of cells~\cite{Segerer2015,Copenhagen2018,Dalessandro2017}, epithelial sheets~\cite{Garcia2015,Angelini2010a}, or 3D organoids~\cite{Palamidessi2019,Han2020}. In summary, our model, which we rigorously derive directly from experimental data, could potentially describe the diversity of previously observed cell-cell interaction behaviors in a unifying quantitative framework. \\

\textbf{Author Contributions} $-$ D.B.B., J.R., and C.P.B. conceived the project. A.F. performed experiments and N.A. executed the tracking. D.B.B. and N.A. analyzed the experimental data. D.B.B., N.A. and C.P.B. developed the theoretical model. P.R. contributed computer code to perform Underdamped Langevin Inference. D.B.B. and C.P.B. wrote the paper with input from all authors. \\

\textbf{Acknowledgements} $-$ We thank Ricard Alert, Edouard Hannezo and Joris Messelink for inspiring discussions. Funded by the Deutsche Forschungsgemeinschaft (DFG, German Research Foundation) - Project-ID 201269156 - SFB 1032 (Projects B01 and B12). D.B.B. is supported in part by a DFG fellowship within the Graduate School of Quantitative Biosciences Munich (QBM) and by the Joachim Herz Stiftung. P.R. is supported by a Center for the Physics of Biological Function fellowship (National Science Foundation grant No. PHY-1734030).

\section{Methods}

\textbf{Sample preparation and cell culture} $-$
Fibronectin micropatterns are made by microscale plasma-initiated protein patterning as described previously~\cite{Brueckner2019}. 

MCF10A cells (ATCC) are cultured at 37$^{\circ}$C in an atmosphere containing 5\% CO$_2$. The culturing medium DMEM/F-12 including Glutamax (Gibco) is supplemented with 5\% Horse Serum (Thermo Fisher), 20 ng/ml hEGF (Sigma), 500ng/ml Hydrocortisone (Sigma), 100ng/ml Cholera Toxin (Sigma) and 10 $\si{\micro}$g/ml Insulin (Sigma). When passaging cells, the supernatant is aspirated and centrifuged at 300rcf for 8 mins.  The adherent cells are washed once with PBS before being detached by a 12-min incubation with Accutase at 37$^{\circ}$C. Then the cell solution is re-suspended with culture medium and subsequently centrifuged at 500rcf for 6 mins. Both cell pellets are re-suspended in medium and a fraction is used for further cultivation. For experiments, a drop containing 10,000 cells is added to an ibidi $\si{\micro}$-dish (ibidi GmbH) and left to adhere for at least 4h. After that, the medium is exchanged to culture medium without phenol red. 15 nM Hoechst 33342 are added for nuclear staining. Cells are kept in a 5\% CO$_2$-atmosphere and at 37$^{\circ}$C during experiments.

MDA-MB-231 cells (DSMZ) are cultured in Minimum Essential Medium (MEM, c.c. pro), containing 10\% FBS (Gibco) and 2mM L-Glutamine (c.c. pro). Cells are grown in a 5\% CO$_2$ atmosphere at 37$^{\circ}$C. For passaging and experiments, cells are washed once with PBS and trypsinised for 3 min. This cell solution is centrifuged at 1000 rcf for 3 min. The cell pellet is re-suspended in MEM and 10,000 cells are added per $\si{\micro}$-dish and left to adhere in the incubator for 4h. The medium is then exchanged to L-15 medium containing L-glutamine (Gibco, supplemented with 10\% FCS) and 25 nM Hoechst 33342 (Invitrogen) for staining cell nuclei. Experiments are performed at 37$^{\circ}$C without CO$_2$. \\

\textbf{Microscopy and cell tracking} $-$
All measurements are performed in time-lapse mode for up to 50 h on an IMIC digital microscope (TILL Photonics) or on a Nikon Eclipse Ti microscope using a 10x objective. The samples are kept in a heated chamber (ibidi GmbH or Okolab) at 37$^{\circ}$C throughout the measurements.  Images (brightfield and DAPI) are acquired every 10 mins. Trajectories of cell pairs are obtained by selecting cells that undergo division during the experiment. Following division and subsequent re-attachment to the micropattern, we track the trajectories of the cell nuclei. A band pass filter is applied to the images of the nuclei, then images are binarised. The cell trajectories are determined by tracking the binarised images using a Matlab tracking algorithm~\cite{Blair}. For further details, see Supplementary Section S2. \\

\textbf{Contact acceleration maps} $-$
To gain insight in the general structure of the accelerations due to cell-cell interaction, we introduce contact acceleration maps. We estimate single-cell component of the dynamics from the accelerations observed at time-points where the cells are far apart $F(x,v) = \left\langle \dot{v}_i \big| x_i,v_i; | \Delta x_{ij}|>\ell \right\rangle$, where we take the threshold distance $\ell = 25 \si{\micro\meter}$. To obtain the accelerations due to cell-cell contacts, we take the time points where cells are close together and calculate the average acceleration as a function of relative position $\Delta x_{ij}=x_i-x_j$ and velocity $\Delta v_{ij}=v_i-v_j$ of cell $i$ and cell $j$: $G(\Delta x, \Delta v) \approx \left\langle \dot{v}_i - F(x_i,v_i) \big| \Delta x_{ij},\Delta v_{ij}; |\Delta x_{ij}|<\ell \right\rangle$. We show that for simple simulated examples, this approach accurately recovers the structure of the interaction terms. For more details, see Supplementary Section S3. \\

\textbf{Underdamped Langevin Inference} $-$
From the short time-scale dynamics of the measured cell trajectories $x(t)$, we infer the second order stochastic differential equation that governs the motion~\cite{Pedersen2016,Brueckner2019,Brueckner2020a,Ferretti2020}. Specifically, to infer the terms of our model (Eq.~\eqref{eq1}), we employ Underdamped Langevin Inference~\cite{Brueckner2020a}, a method which is robust with respect to the effects of the discrete sampling of $x(t)$ and the presence of measurement errors. Briefly, we fit the experimentally measured accelerations using a linear combination of basis functions $ \{b(x_i,v_i),u(|\Delta x_{ij}|)\Delta x_{ij},u(|\Delta x_{ij}|)\Delta x_{ij}\} $ using rigorous stochastic estimators~\cite{Brueckner2020a}. For the single cell terms $b(x_i,v_i)$, we use a combination of polynomials and Fourier modes, while for the interaction kernels $u(|\Delta x_{ij}|)$ we use exponential functions. The inference results do not sensitively depend on the choice of basis functions. For more details, see Supplementary Section S4. 

\bibliography{../library}

\begin{thebibliography}{60}%
\makeatletter
\providecommand \@ifxundefined [1]{%
 \@ifx{#1\undefined}
}%
\providecommand \@ifnum [1]{%
 \ifnum #1\expandafter \@firstoftwo
 \else \expandafter \@secondoftwo
 \fi
}%
\providecommand \@ifx [1]{%
 \ifx #1\expandafter \@firstoftwo
 \else \expandafter \@secondoftwo
 \fi
}%
\providecommand \natexlab [1]{#1}%
\providecommand \enquote  [1]{``#1''}%
\providecommand \bibnamefont  [1]{#1}%
\providecommand \bibfnamefont [1]{#1}%
\providecommand \citenamefont [1]{#1}%
\providecommand \href@noop [0]{\@secondoftwo}%
\providecommand \href [0]{\begingroup \@sanitize@url \@href}%
\providecommand \@href[1]{\@@startlink{#1}\@@href}%
\providecommand \@@href[1]{\endgroup#1\@@endlink}%
\providecommand \@sanitize@url [0]{\catcode `\\12\catcode `\$12\catcode
  `\&12\catcode `\#12\catcode `\^12\catcode `\_12\catcode `\%12\relax}%
\providecommand \@@startlink[1]{}%
\providecommand \@@endlink[0]{}%
\providecommand \url  [0]{\begingroup\@sanitize@url \@url }%
\providecommand \@url [1]{\endgroup\@href {#1}{\urlprefix }}%
\providecommand \urlprefix  [0]{URL }%
\providecommand \Eprint [0]{\href }%
\providecommand \doibase [0]{http://dx.doi.org/}%
\providecommand \selectlanguage [0]{\@gobble}%
\providecommand \bibinfo  [0]{\@secondoftwo}%
\providecommand \bibfield  [0]{\@secondoftwo}%
\providecommand \translation [1]{[#1]}%
\providecommand \BibitemOpen [0]{}%
\providecommand \bibitemStop [0]{}%
\providecommand \bibitemNoStop [0]{.\EOS\space}%
\providecommand \EOS [0]{\spacefactor3000\relax}%
\providecommand \BibitemShut  [1]{\csname bibitem#1\endcsname}%
\let\auto@bib@innerbib\@empty
\bibitem [{\citenamefont {Poujade}\ \emph {et~al.}(2007)\citenamefont
  {Poujade}, \citenamefont {Grasland-Mongrain}, \citenamefont {Hertzog},
  \citenamefont {Jouanneau}, \citenamefont {Chavrier}, \citenamefont {Ladoux},
  \citenamefont {Buguin},\ and\ \citenamefont {Silberzan}}]{Poujade2007a}%
  \BibitemOpen
  \bibfield  {author} {\bibinfo {author} {\bibfnamefont {M.}~\bibnamefont
  {Poujade}}, \bibinfo {author} {\bibfnamefont {E.}~\bibnamefont
  {Grasland-Mongrain}}, \bibinfo {author} {\bibfnamefont {A.}~\bibnamefont
  {Hertzog}}, \bibinfo {author} {\bibfnamefont {J.}~\bibnamefont {Jouanneau}},
  \bibinfo {author} {\bibfnamefont {P.}~\bibnamefont {Chavrier}}, \bibinfo
  {author} {\bibfnamefont {B.}~\bibnamefont {Ladoux}}, \bibinfo {author}
  {\bibfnamefont {A.}~\bibnamefont {Buguin}}, \ and\ \bibinfo {author}
  {\bibfnamefont {P.}~\bibnamefont {Silberzan}},\ }\href {\doibase
  10.1073/pnas.0705062104} {\bibfield  {journal} {\bibinfo  {journal}
  {Proceedings of the National Academy of Sciences of the United States of
  America}\ }\textbf {\bibinfo {volume} {104}},\ \bibinfo {pages} {15988}
  (\bibinfo {year} {2007})}\BibitemShut {NoStop}%
\bibitem [{\citenamefont {Stramer}\ \emph {et~al.}(2005)\citenamefont
  {Stramer}, \citenamefont {Wood}, \citenamefont {Galko}, \citenamefont {Redd},
  \citenamefont {Jacinto}, \citenamefont {Parkhurst},\ and\ \citenamefont
  {Martin}}]{Stramer2005}%
  \BibitemOpen
  \bibfield  {author} {\bibinfo {author} {\bibfnamefont {B.}~\bibnamefont
  {Stramer}}, \bibinfo {author} {\bibfnamefont {W.}~\bibnamefont {Wood}},
  \bibinfo {author} {\bibfnamefont {M.~J.}\ \bibnamefont {Galko}}, \bibinfo
  {author} {\bibfnamefont {M.~J.}\ \bibnamefont {Redd}}, \bibinfo {author}
  {\bibfnamefont {A.}~\bibnamefont {Jacinto}}, \bibinfo {author} {\bibfnamefont
  {S.~M.}\ \bibnamefont {Parkhurst}}, \ and\ \bibinfo {author} {\bibfnamefont
  {P.}~\bibnamefont {Martin}},\ }\href {\doibase 10.1083/jcb.200405120}
  {\bibfield  {journal} {\bibinfo  {journal} {Journal of Cell Biology}\
  }\textbf {\bibinfo {volume} {168}},\ \bibinfo {pages} {567} (\bibinfo {year}
  {2005})}\BibitemShut {NoStop}%
\bibitem [{\citenamefont {Weavers}\ \emph {et~al.}(2016)\citenamefont
  {Weavers}, \citenamefont {Liepe}, \citenamefont {Sim}, \citenamefont {Wood},
  \citenamefont {Martin},\ and\ \citenamefont {Stumpf}}]{Weavers2016}%
  \BibitemOpen
  \bibfield  {author} {\bibinfo {author} {\bibfnamefont {H.}~\bibnamefont
  {Weavers}}, \bibinfo {author} {\bibfnamefont {J.}~\bibnamefont {Liepe}},
  \bibinfo {author} {\bibfnamefont {A.}~\bibnamefont {Sim}}, \bibinfo {author}
  {\bibfnamefont {W.}~\bibnamefont {Wood}}, \bibinfo {author} {\bibfnamefont
  {P.}~\bibnamefont {Martin}}, \ and\ \bibinfo {author} {\bibfnamefont {M.~P.}\
  \bibnamefont {Stumpf}},\ }\href {\doibase 10.1016/j.cub.2016.06.012}
  {\bibfield  {journal} {\bibinfo  {journal} {Current Biology}\ }\textbf
  {\bibinfo {volume} {26}},\ \bibinfo {pages} {1975} (\bibinfo {year}
  {2016})}\BibitemShut {NoStop}%
\bibitem [{\citenamefont {Friedl}\ and\ \citenamefont
  {Wolf}(2003)}]{Friedl2003}%
  \BibitemOpen
  \bibfield  {author} {\bibinfo {author} {\bibfnamefont {P.}~\bibnamefont
  {Friedl}}\ and\ \bibinfo {author} {\bibfnamefont {K.}~\bibnamefont {Wolf}},\
  }\href {\doibase 10.1038/nrc1075} {\bibfield  {journal} {\bibinfo  {journal}
  {Nature reviews. Cancer}\ }\textbf {\bibinfo {volume} {3}},\ \bibinfo {pages}
  {362} (\bibinfo {year} {2003})}\BibitemShut {NoStop}%
\bibitem [{\citenamefont {Carmona-Fontaine}\ \emph {et~al.}(2008)\citenamefont
  {Carmona-Fontaine}, \citenamefont {Matthews}, \citenamefont {Kuriyama},
  \citenamefont {Moreno}, \citenamefont {Dunn}, \citenamefont {Parsons},
  \citenamefont {Stern},\ and\ \citenamefont {Mayor}}]{Carmona-Fontaine2008}%
  \BibitemOpen
  \bibfield  {author} {\bibinfo {author} {\bibfnamefont {C.}~\bibnamefont
  {Carmona-Fontaine}}, \bibinfo {author} {\bibfnamefont {H.~K.}\ \bibnamefont
  {Matthews}}, \bibinfo {author} {\bibfnamefont {S.}~\bibnamefont {Kuriyama}},
  \bibinfo {author} {\bibfnamefont {M.}~\bibnamefont {Moreno}}, \bibinfo
  {author} {\bibfnamefont {G.~A.}\ \bibnamefont {Dunn}}, \bibinfo {author}
  {\bibfnamefont {M.}~\bibnamefont {Parsons}}, \bibinfo {author} {\bibfnamefont
  {C.~D.}\ \bibnamefont {Stern}}, \ and\ \bibinfo {author} {\bibfnamefont
  {R.}~\bibnamefont {Mayor}},\ }\href {\doibase 10.1038/nature07441} {\bibfield
   {journal} {\bibinfo  {journal} {Nature}\ }\textbf {\bibinfo {volume}
  {456}},\ \bibinfo {pages} {957} (\bibinfo {year} {2008})}\BibitemShut
  {NoStop}%
\bibitem [{\citenamefont {Villar-Cervi{\~{n}}o}\ \emph
  {et~al.}(2013)\citenamefont {Villar-Cervi{\~{n}}o}, \citenamefont
  {Molano-Maz{\'{o}}n}, \citenamefont {Catchpole}, \citenamefont
  {Valdeolmillos}, \citenamefont {Henkemeyer}, \citenamefont {Mart{\'{i}}nez},
  \citenamefont {Borrell},\ and\ \citenamefont
  {Mar{\'{i}}n}}]{Villar-Cervino2013}%
  \BibitemOpen
  \bibfield  {author} {\bibinfo {author} {\bibfnamefont {V.}~\bibnamefont
  {Villar-Cervi{\~{n}}o}}, \bibinfo {author} {\bibfnamefont {M.}~\bibnamefont
  {Molano-Maz{\'{o}}n}}, \bibinfo {author} {\bibfnamefont {T.}~\bibnamefont
  {Catchpole}}, \bibinfo {author} {\bibfnamefont {M.}~\bibnamefont
  {Valdeolmillos}}, \bibinfo {author} {\bibfnamefont {M.}~\bibnamefont
  {Henkemeyer}}, \bibinfo {author} {\bibfnamefont {L.~M.}\ \bibnamefont
  {Mart{\'{i}}nez}}, \bibinfo {author} {\bibfnamefont {V.}~\bibnamefont
  {Borrell}}, \ and\ \bibinfo {author} {\bibfnamefont {O.}~\bibnamefont
  {Mar{\'{i}}n}},\ }\href {\doibase 10.1016/j.neuron.2012.11.023} {\bibfield
  {journal} {\bibinfo  {journal} {Neuron}\ }\textbf {\bibinfo {volume} {77}},\
  \bibinfo {pages} {457} (\bibinfo {year} {2013})}\BibitemShut {NoStop}%
\bibitem [{\citenamefont {Theveneau}\ \emph {et~al.}(2010)\citenamefont
  {Theveneau}, \citenamefont {Marchant}, \citenamefont {Kuriyama},
  \citenamefont {Gull}, \citenamefont {Moepps}, \citenamefont {Parsons},\ and\
  \citenamefont {Mayor}}]{Theveneau2010}%
  \BibitemOpen
  \bibfield  {author} {\bibinfo {author} {\bibfnamefont {E.}~\bibnamefont
  {Theveneau}}, \bibinfo {author} {\bibfnamefont {L.}~\bibnamefont {Marchant}},
  \bibinfo {author} {\bibfnamefont {S.}~\bibnamefont {Kuriyama}}, \bibinfo
  {author} {\bibfnamefont {M.}~\bibnamefont {Gull}}, \bibinfo {author}
  {\bibfnamefont {B.}~\bibnamefont {Moepps}}, \bibinfo {author} {\bibfnamefont
  {M.}~\bibnamefont {Parsons}}, \ and\ \bibinfo {author} {\bibfnamefont
  {R.}~\bibnamefont {Mayor}},\ }\href {\doibase 10.1016/j.devcel.2010.06.012}
  {\bibfield  {journal} {\bibinfo  {journal} {Developmental Cell}\ }\textbf
  {\bibinfo {volume} {19}},\ \bibinfo {pages} {39} (\bibinfo {year}
  {2010})}\BibitemShut {NoStop}%
\bibitem [{\citenamefont {Davis}\ \emph {et~al.}(2012)\citenamefont {Davis},
  \citenamefont {Huang}, \citenamefont {Zanet}, \citenamefont {Harrison},
  \citenamefont {Rosten}, \citenamefont {Cox}, \citenamefont {Soong},
  \citenamefont {Dunn},\ and\ \citenamefont {Stramer}}]{Davis2012}%
  \BibitemOpen
  \bibfield  {author} {\bibinfo {author} {\bibfnamefont {J.~R.}\ \bibnamefont
  {Davis}}, \bibinfo {author} {\bibfnamefont {C.~Y.}\ \bibnamefont {Huang}},
  \bibinfo {author} {\bibfnamefont {J.}~\bibnamefont {Zanet}}, \bibinfo
  {author} {\bibfnamefont {S.}~\bibnamefont {Harrison}}, \bibinfo {author}
  {\bibfnamefont {E.}~\bibnamefont {Rosten}}, \bibinfo {author} {\bibfnamefont
  {S.}~\bibnamefont {Cox}}, \bibinfo {author} {\bibfnamefont {D.~Y.}\
  \bibnamefont {Soong}}, \bibinfo {author} {\bibfnamefont {G.~A.}\ \bibnamefont
  {Dunn}}, \ and\ \bibinfo {author} {\bibfnamefont {B.~M.}\ \bibnamefont
  {Stramer}},\ }\href {\doibase 10.1242/dev.082248} {\bibfield  {journal}
  {\bibinfo  {journal} {Development}\ }\textbf {\bibinfo {volume} {139}},\
  \bibinfo {pages} {4555} (\bibinfo {year} {2012})}\BibitemShut {NoStop}%
\bibitem [{\citenamefont {Smeets}\ \emph {et~al.}(2016)\citenamefont {Smeets},
  \citenamefont {Alert}, \citenamefont {Pe{\v{s}}ek}, \citenamefont
  {Pagonabarraga}, \citenamefont {Ramon},\ and\ \citenamefont
  {Vincent}}]{Smeets2016}%
  \BibitemOpen
  \bibfield  {author} {\bibinfo {author} {\bibfnamefont {B.}~\bibnamefont
  {Smeets}}, \bibinfo {author} {\bibfnamefont {R.}~\bibnamefont {Alert}},
  \bibinfo {author} {\bibfnamefont {J.}~\bibnamefont {Pe{\v{s}}ek}}, \bibinfo
  {author} {\bibfnamefont {I.}~\bibnamefont {Pagonabarraga}}, \bibinfo {author}
  {\bibfnamefont {H.}~\bibnamefont {Ramon}}, \ and\ \bibinfo {author}
  {\bibfnamefont {R.}~\bibnamefont {Vincent}},\ }\href {\doibase
  10.1073/pnas.1521151113} {\bibfield  {journal} {\bibinfo  {journal}
  {Proceedings of the National Academy of Sciences of the United States of
  America}\ }\textbf {\bibinfo {volume} {113}},\ \bibinfo {pages} {14621}
  (\bibinfo {year} {2016})}\BibitemShut {NoStop}%
\bibitem [{\citenamefont {Stramer}\ and\ \citenamefont
  {Mayor}(2017)}]{Stramer2017}%
  \BibitemOpen
  \bibfield  {author} {\bibinfo {author} {\bibfnamefont {B.}~\bibnamefont
  {Stramer}}\ and\ \bibinfo {author} {\bibfnamefont {R.}~\bibnamefont
  {Mayor}},\ }\href {\doibase 10.1038/nrm.2016.118} {\bibfield  {journal}
  {\bibinfo  {journal} {Nature reviews. Molecular cell biology}\ }\textbf
  {\bibinfo {volume} {18}},\ \bibinfo {pages} {43} (\bibinfo {year}
  {2017})}\BibitemShut {NoStop}%
\bibitem [{\citenamefont {Astin}\ \emph {et~al.}(2010)\citenamefont {Astin},
  \citenamefont {Batson}, \citenamefont {Kadir}, \citenamefont {Charlet},
  \citenamefont {Persad}, \citenamefont {Gillatt}, \citenamefont {Oxley},\ and\
  \citenamefont {Nobes}}]{Astin2010}%
  \BibitemOpen
  \bibfield  {author} {\bibinfo {author} {\bibfnamefont {J.~W.}\ \bibnamefont
  {Astin}}, \bibinfo {author} {\bibfnamefont {J.}~\bibnamefont {Batson}},
  \bibinfo {author} {\bibfnamefont {S.}~\bibnamefont {Kadir}}, \bibinfo
  {author} {\bibfnamefont {J.}~\bibnamefont {Charlet}}, \bibinfo {author}
  {\bibfnamefont {R.~A.}\ \bibnamefont {Persad}}, \bibinfo {author}
  {\bibfnamefont {D.}~\bibnamefont {Gillatt}}, \bibinfo {author} {\bibfnamefont
  {J.~D.}\ \bibnamefont {Oxley}}, \ and\ \bibinfo {author} {\bibfnamefont
  {C.~D.}\ \bibnamefont {Nobes}},\ }\href {\doibase 10.1038/ncb2122} {\bibfield
   {journal} {\bibinfo  {journal} {Nature Cell Biology}\ }\textbf {\bibinfo
  {volume} {12}},\ \bibinfo {pages} {1194} (\bibinfo {year}
  {2010})}\BibitemShut {NoStop}%
\bibitem [{\citenamefont {Teddy}\ and\ \citenamefont
  {Kulesa}(2004)}]{Teddy2004}%
  \BibitemOpen
  \bibfield  {author} {\bibinfo {author} {\bibfnamefont {J.~M.}\ \bibnamefont
  {Teddy}}\ and\ \bibinfo {author} {\bibfnamefont {P.~M.}\ \bibnamefont
  {Kulesa}},\ }\href {\doibase 10.1242/dev.01534} {\bibfield  {journal}
  {\bibinfo  {journal} {Development}\ }\textbf {\bibinfo {volume} {131}},\
  \bibinfo {pages} {6141} (\bibinfo {year} {2004})}\BibitemShut {NoStop}%
\bibitem [{\citenamefont {Abercrombie}\ and\ \citenamefont
  {Heaysman}(1954{\natexlab{a}})}]{Abercrombie1954a}%
  \BibitemOpen
  \bibfield  {author} {\bibinfo {author} {\bibfnamefont {M.}~\bibnamefont
  {Abercrombie}}\ and\ \bibinfo {author} {\bibfnamefont {J.~E.}\ \bibnamefont
  {Heaysman}},\ }\href {\doibase 10.1016/0014-4827(54)90176-7} {\bibfield
  {journal} {\bibinfo  {journal} {Experimental Cell Research}\ }\textbf
  {\bibinfo {volume} {6}},\ \bibinfo {pages} {293} (\bibinfo {year}
  {1954}{\natexlab{a}})}\BibitemShut {NoStop}%
\bibitem [{\citenamefont {Davis}\ \emph {et~al.}(2015)\citenamefont {Davis},
  \citenamefont {Luchici}, \citenamefont {Mosis}, \citenamefont {Thackery},
  \citenamefont {Salazar}, \citenamefont {Mao}, \citenamefont {Dunn},
  \citenamefont {Betz}, \citenamefont {Miodownik},\ and\ \citenamefont
  {Stramer}}]{Davis2015}%
  \BibitemOpen
  \bibfield  {author} {\bibinfo {author} {\bibfnamefont {J.~R.}\ \bibnamefont
  {Davis}}, \bibinfo {author} {\bibfnamefont {A.}~\bibnamefont {Luchici}},
  \bibinfo {author} {\bibfnamefont {F.}~\bibnamefont {Mosis}}, \bibinfo
  {author} {\bibfnamefont {J.}~\bibnamefont {Thackery}}, \bibinfo {author}
  {\bibfnamefont {J.~A.}\ \bibnamefont {Salazar}}, \bibinfo {author}
  {\bibfnamefont {Y.}~\bibnamefont {Mao}}, \bibinfo {author} {\bibfnamefont
  {G.~A.}\ \bibnamefont {Dunn}}, \bibinfo {author} {\bibfnamefont
  {T.}~\bibnamefont {Betz}}, \bibinfo {author} {\bibfnamefont {M.}~\bibnamefont
  {Miodownik}}, \ and\ \bibinfo {author} {\bibfnamefont {B.~M.}\ \bibnamefont
  {Stramer}},\ }\href {\doibase 10.1016/j.cell.2015.02.015} {\bibfield
  {journal} {\bibinfo  {journal} {Cell}\ }\textbf {\bibinfo {volume} {161}},\
  \bibinfo {pages} {361} (\bibinfo {year} {2015})}\BibitemShut {NoStop}%
\bibitem [{\citenamefont {Moore}\ \emph {et~al.}(2013)\citenamefont {Moore},
  \citenamefont {Theveneau}, \citenamefont {Pozzi}, \citenamefont {Alexandre},
  \citenamefont {Richardson}, \citenamefont {Merks}, \citenamefont {Parsons},
  \citenamefont {Kashef}, \citenamefont {Linker},\ and\ \citenamefont
  {Mayor}}]{Moore2013}%
  \BibitemOpen
  \bibfield  {author} {\bibinfo {author} {\bibfnamefont {R.}~\bibnamefont
  {Moore}}, \bibinfo {author} {\bibfnamefont {E.}~\bibnamefont {Theveneau}},
  \bibinfo {author} {\bibfnamefont {S.}~\bibnamefont {Pozzi}}, \bibinfo
  {author} {\bibfnamefont {P.}~\bibnamefont {Alexandre}}, \bibinfo {author}
  {\bibfnamefont {J.}~\bibnamefont {Richardson}}, \bibinfo {author}
  {\bibfnamefont {A.}~\bibnamefont {Merks}}, \bibinfo {author} {\bibfnamefont
  {M.}~\bibnamefont {Parsons}}, \bibinfo {author} {\bibfnamefont
  {J.}~\bibnamefont {Kashef}}, \bibinfo {author} {\bibfnamefont
  {C.}~\bibnamefont {Linker}}, \ and\ \bibinfo {author} {\bibfnamefont
  {R.}~\bibnamefont {Mayor}},\ }\href {\doibase 10.1242/dev.098509} {\bibfield
  {journal} {\bibinfo  {journal} {Development (Cambridge)}\ }\textbf {\bibinfo
  {volume} {140}},\ \bibinfo {pages} {4763} (\bibinfo {year}
  {2013})}\BibitemShut {NoStop}%
\bibitem [{\citenamefont {Matthews}\ \emph {et~al.}(2008)\citenamefont
  {Matthews}, \citenamefont {Marchant}, \citenamefont {Carmona-Fontaine},
  \citenamefont {Kuriyama}, \citenamefont {Larra{\'{i}}n}, \citenamefont
  {Holt}, \citenamefont {Parsons},\ and\ \citenamefont {Mayor}}]{Matthews2008}%
  \BibitemOpen
  \bibfield  {author} {\bibinfo {author} {\bibfnamefont {H.~K.}\ \bibnamefont
  {Matthews}}, \bibinfo {author} {\bibfnamefont {L.}~\bibnamefont {Marchant}},
  \bibinfo {author} {\bibfnamefont {C.}~\bibnamefont {Carmona-Fontaine}},
  \bibinfo {author} {\bibfnamefont {S.}~\bibnamefont {Kuriyama}}, \bibinfo
  {author} {\bibfnamefont {J.}~\bibnamefont {Larra{\'{i}}n}}, \bibinfo {author}
  {\bibfnamefont {M.~R.}\ \bibnamefont {Holt}}, \bibinfo {author}
  {\bibfnamefont {M.}~\bibnamefont {Parsons}}, \ and\ \bibinfo {author}
  {\bibfnamefont {R.}~\bibnamefont {Mayor}},\ }\href {\doibase
  10.1242/dev.017350} {\bibfield  {journal} {\bibinfo  {journal} {Development}\
  }\textbf {\bibinfo {volume} {135}},\ \bibinfo {pages} {1771} (\bibinfo {year}
  {2008})}\BibitemShut {NoStop}%
\bibitem [{\citenamefont {Kadir}\ \emph {et~al.}(2011)\citenamefont {Kadir},
  \citenamefont {Astin}, \citenamefont {Tahtamouni}, \citenamefont {Martin},\
  and\ \citenamefont {Nobes}}]{Kadir2011}%
  \BibitemOpen
  \bibfield  {author} {\bibinfo {author} {\bibfnamefont {S.}~\bibnamefont
  {Kadir}}, \bibinfo {author} {\bibfnamefont {J.~W.}\ \bibnamefont {Astin}},
  \bibinfo {author} {\bibfnamefont {L.}~\bibnamefont {Tahtamouni}}, \bibinfo
  {author} {\bibfnamefont {P.}~\bibnamefont {Martin}}, \ and\ \bibinfo {author}
  {\bibfnamefont {C.~D.}\ \bibnamefont {Nobes}},\ }\href {\doibase
  10.1242/jcs.087965} {\bibfield  {journal} {\bibinfo  {journal} {Journal of
  Cell Science}\ }\textbf {\bibinfo {volume} {124}},\ \bibinfo {pages} {2642}
  (\bibinfo {year} {2011})}\BibitemShut {NoStop}%
\bibitem [{\citenamefont {Abercrombie}(1979)}]{Abercrombie1979}%
  \BibitemOpen
  \bibfield  {author} {\bibinfo {author} {\bibfnamefont {M.}~\bibnamefont
  {Abercrombie}},\ }\href {\doibase 10.1038/281259a0} {\bibfield  {journal}
  {\bibinfo  {journal} {Nature}\ }\textbf {\bibinfo {volume} {281}},\ \bibinfo
  {pages} {259} (\bibinfo {year} {1979})}\BibitemShut {NoStop}%
\bibitem [{\citenamefont {Milano}\ \emph {et~al.}(2016)\citenamefont {Milano},
  \citenamefont {Ngai}, \citenamefont {Muthuswamy},\ and\ \citenamefont
  {Asthagiri}}]{Milano2016}%
  \BibitemOpen
  \bibfield  {author} {\bibinfo {author} {\bibfnamefont {D.~F.}\ \bibnamefont
  {Milano}}, \bibinfo {author} {\bibfnamefont {N.~A.}\ \bibnamefont {Ngai}},
  \bibinfo {author} {\bibfnamefont {S.~K.}\ \bibnamefont {Muthuswamy}}, \ and\
  \bibinfo {author} {\bibfnamefont {A.~R.}\ \bibnamefont {Asthagiri}},\ }\href
  {\doibase 10.1016/j.bpj.2016.02.040} {\bibfield  {journal} {\bibinfo
  {journal} {Biophysical Journal}\ }\textbf {\bibinfo {volume} {110}},\
  \bibinfo {pages} {1886} (\bibinfo {year} {2016})}\BibitemShut {NoStop}%
\bibitem [{\citenamefont {Li}\ and\ \citenamefont {Wang}(2018)}]{Li2018}%
  \BibitemOpen
  \bibfield  {author} {\bibinfo {author} {\bibfnamefont {D.}~\bibnamefont
  {Li}}\ and\ \bibinfo {author} {\bibfnamefont {Y.~L.}\ \bibnamefont {Wang}},\
  }\href {\doibase 10.1073/pnas.1807543115} {\bibfield  {journal} {\bibinfo
  {journal} {Proceedings of the National Academy of Sciences of the United
  States of America}\ }\textbf {\bibinfo {volume} {115}},\ \bibinfo {pages}
  {10678} (\bibinfo {year} {2018})}\BibitemShut {NoStop}%
\bibitem [{\citenamefont {Hayakawa}\ \emph {et~al.}(2020)\citenamefont
  {Hayakawa}, \citenamefont {Hiraiwa}, \citenamefont {Wada}, \citenamefont
  {Kuwayama},\ and\ \citenamefont {Shibata}}]{Hayakawa2020}%
  \BibitemOpen
  \bibfield  {author} {\bibinfo {author} {\bibfnamefont {M.}~\bibnamefont
  {Hayakawa}}, \bibinfo {author} {\bibfnamefont {T.}~\bibnamefont {Hiraiwa}},
  \bibinfo {author} {\bibfnamefont {Y.}~\bibnamefont {Wada}}, \bibinfo {author}
  {\bibfnamefont {H.}~\bibnamefont {Kuwayama}}, \ and\ \bibinfo {author}
  {\bibfnamefont {T.}~\bibnamefont {Shibata}},\ }\href {\doibase
  10.7554/eLife.53609} {\bibfield  {journal} {\bibinfo  {journal} {eLife}\
  }\textbf {\bibinfo {volume} {9}},\ \bibinfo {pages} {e53609} (\bibinfo {year}
  {2020})}\BibitemShut {NoStop}%
\bibitem [{\citenamefont {Maiuri}\ \emph {et~al.}(2015)\citenamefont {Maiuri},
  \citenamefont {Rupprecht}, \citenamefont {Wieser}, \citenamefont {Ruprecht},
  \citenamefont {B{\'{e}}nichou}, \citenamefont {Carpi}, \citenamefont
  {Coppey}, \citenamefont {{De Beco}}, \citenamefont {Gov}, \citenamefont
  {Heisenberg}, \citenamefont {{Lage Crespo}}, \citenamefont {Lautenschlaeger},
  \citenamefont {{Le Berre}}, \citenamefont {Lennon-Dumenil}, \citenamefont
  {Raab}, \citenamefont {Thiam}, \citenamefont {Piel}, \citenamefont {Sixt},\
  and\ \citenamefont {Voituriez}}]{Maiuri2015}%
  \BibitemOpen
  \bibfield  {author} {\bibinfo {author} {\bibfnamefont {P.}~\bibnamefont
  {Maiuri}}, \bibinfo {author} {\bibfnamefont {J.~F.}\ \bibnamefont
  {Rupprecht}}, \bibinfo {author} {\bibfnamefont {S.}~\bibnamefont {Wieser}},
  \bibinfo {author} {\bibfnamefont {V.}~\bibnamefont {Ruprecht}}, \bibinfo
  {author} {\bibfnamefont {O.}~\bibnamefont {B{\'{e}}nichou}}, \bibinfo
  {author} {\bibfnamefont {N.}~\bibnamefont {Carpi}}, \bibinfo {author}
  {\bibfnamefont {M.}~\bibnamefont {Coppey}}, \bibinfo {author} {\bibfnamefont
  {S.}~\bibnamefont {{De Beco}}}, \bibinfo {author} {\bibfnamefont
  {N.}~\bibnamefont {Gov}}, \bibinfo {author} {\bibfnamefont {C.~P.}\
  \bibnamefont {Heisenberg}}, \bibinfo {author} {\bibfnamefont
  {C.}~\bibnamefont {{Lage Crespo}}}, \bibinfo {author} {\bibfnamefont
  {F.}~\bibnamefont {Lautenschlaeger}}, \bibinfo {author} {\bibfnamefont
  {M.}~\bibnamefont {{Le Berre}}}, \bibinfo {author} {\bibfnamefont {A.~M.}\
  \bibnamefont {Lennon-Dumenil}}, \bibinfo {author} {\bibfnamefont
  {M.}~\bibnamefont {Raab}}, \bibinfo {author} {\bibfnamefont {H.~R.}\
  \bibnamefont {Thiam}}, \bibinfo {author} {\bibfnamefont {M.}~\bibnamefont
  {Piel}}, \bibinfo {author} {\bibfnamefont {M.}~\bibnamefont {Sixt}}, \ and\
  \bibinfo {author} {\bibfnamefont {R.}~\bibnamefont {Voituriez}},\ }\href
  {\doibase 10.1016/j.cell.2015.01.056} {\bibfield  {journal} {\bibinfo
  {journal} {Cell}\ }\textbf {\bibinfo {volume} {161}},\ \bibinfo {pages} {374}
  (\bibinfo {year} {2015})}\BibitemShut {NoStop}%
\bibitem [{\citenamefont {Lavi}\ \emph {et~al.}(2016)\citenamefont {Lavi},
  \citenamefont {Piel}, \citenamefont {Lennon-Dum{\'{e}}nil}, \citenamefont
  {Voituriez},\ and\ \citenamefont {Gov}}]{Lavi2016}%
  \BibitemOpen
  \bibfield  {author} {\bibinfo {author} {\bibfnamefont {I.}~\bibnamefont
  {Lavi}}, \bibinfo {author} {\bibfnamefont {M.}~\bibnamefont {Piel}}, \bibinfo
  {author} {\bibfnamefont {A.-M.}\ \bibnamefont {Lennon-Dum{\'{e}}nil}},
  \bibinfo {author} {\bibfnamefont {R.}~\bibnamefont {Voituriez}}, \ and\
  \bibinfo {author} {\bibfnamefont {N.~S.}\ \bibnamefont {Gov}},\ }\href
  {\doibase 10.1038/nphys3836} {\bibfield  {journal} {\bibinfo  {journal}
  {Nature Physics}\ }\textbf {\bibinfo {volume} {12}},\ \bibinfo {pages} {1146}
  (\bibinfo {year} {2016})}\BibitemShut {NoStop}%
\bibitem [{\citenamefont {Selmeczi}\ \emph {et~al.}(2005)\citenamefont
  {Selmeczi}, \citenamefont {Mosler}, \citenamefont {Hagedorn}, \citenamefont
  {Larsen},\ and\ \citenamefont {Flyvbjerg}}]{Selmeczi2005}%
  \BibitemOpen
  \bibfield  {author} {\bibinfo {author} {\bibfnamefont {D.}~\bibnamefont
  {Selmeczi}}, \bibinfo {author} {\bibfnamefont {S.}~\bibnamefont {Mosler}},
  \bibinfo {author} {\bibfnamefont {P.~H.}\ \bibnamefont {Hagedorn}}, \bibinfo
  {author} {\bibfnamefont {N.~B.}\ \bibnamefont {Larsen}}, \ and\ \bibinfo
  {author} {\bibfnamefont {H.}~\bibnamefont {Flyvbjerg}},\ }\href {\doibase
  10.1529/biophysj.105.061150} {\bibfield  {journal} {\bibinfo  {journal}
  {Biophysical journal}\ }\textbf {\bibinfo {volume} {89}},\ \bibinfo {pages}
  {912} (\bibinfo {year} {2005})}\BibitemShut {NoStop}%
\bibitem [{\citenamefont {Li}\ \emph {et~al.}(2011)\citenamefont {Li},
  \citenamefont {Cox},\ and\ \citenamefont {Flyvbjerg}}]{Li2011}%
  \BibitemOpen
  \bibfield  {author} {\bibinfo {author} {\bibfnamefont {L.}~\bibnamefont
  {Li}}, \bibinfo {author} {\bibfnamefont {E.~C.}\ \bibnamefont {Cox}}, \ and\
  \bibinfo {author} {\bibfnamefont {H.}~\bibnamefont {Flyvbjerg}},\ }\href
  {\doibase 10.1088/1478-3975/8/4/046006} {\bibfield  {journal} {\bibinfo
  {journal} {Physical Biology}\ }\textbf {\bibinfo {volume} {8}},\ \bibinfo
  {pages} {046006} (\bibinfo {year} {2011})}\BibitemShut {NoStop}%
\bibitem [{\citenamefont {Pedersen}\ \emph {et~al.}(2016)\citenamefont
  {Pedersen}, \citenamefont {Li}, \citenamefont {Gradinaru}, \citenamefont
  {Austin}, \citenamefont {Cox},\ and\ \citenamefont
  {Flyvbjerg}}]{Pedersen2016}%
  \BibitemOpen
  \bibfield  {author} {\bibinfo {author} {\bibfnamefont {J.~N.}\ \bibnamefont
  {Pedersen}}, \bibinfo {author} {\bibfnamefont {L.}~\bibnamefont {Li}},
  \bibinfo {author} {\bibfnamefont {C.}~\bibnamefont {Gradinaru}}, \bibinfo
  {author} {\bibfnamefont {R.~H.}\ \bibnamefont {Austin}}, \bibinfo {author}
  {\bibfnamefont {E.~C.}\ \bibnamefont {Cox}}, \ and\ \bibinfo {author}
  {\bibfnamefont {H.}~\bibnamefont {Flyvbjerg}},\ }\href {\doibase
  10.1103/PhysRevE.94.062401} {\bibfield  {journal} {\bibinfo  {journal}
  {Physical Review E}\ }\textbf {\bibinfo {volume} {94}},\ \bibinfo {pages}
  {062401} (\bibinfo {year} {2016})}\BibitemShut {NoStop}%
\bibitem [{\citenamefont {Br{\"{u}}ckner}\ \emph {et~al.}(2019)\citenamefont
  {Br{\"{u}}ckner}, \citenamefont {Fink}, \citenamefont {Schreiber},
  \citenamefont {R{\"{o}}ttgermann}, \citenamefont {R{\"{a}}dler},\ and\
  \citenamefont {Broedersz}}]{Brueckner2019}%
  \BibitemOpen
  \bibfield  {author} {\bibinfo {author} {\bibfnamefont {D.~B.}\ \bibnamefont
  {Br{\"{u}}ckner}}, \bibinfo {author} {\bibfnamefont {A.}~\bibnamefont
  {Fink}}, \bibinfo {author} {\bibfnamefont {C.}~\bibnamefont {Schreiber}},
  \bibinfo {author} {\bibfnamefont {P.~J.~F.}\ \bibnamefont
  {R{\"{o}}ttgermann}}, \bibinfo {author} {\bibfnamefont {J.~O.}\ \bibnamefont
  {R{\"{a}}dler}}, \ and\ \bibinfo {author} {\bibfnamefont {C.~P.}\
  \bibnamefont {Broedersz}},\ }\href {\doibase 10.1038/s41567-019-0445-4}
  {\bibfield  {journal} {\bibinfo  {journal} {Nature Physics}\ }\textbf
  {\bibinfo {volume} {15}},\ \bibinfo {pages} {595} (\bibinfo {year}
  {2019})}\BibitemShut {NoStop}%
\bibitem [{\citenamefont {Br{\"{u}}ckner}\ \emph
  {et~al.}(2020{\natexlab{a}})\citenamefont {Br{\"{u}}ckner}, \citenamefont
  {Fink}, \citenamefont {R{\"{a}}dler},\ and\ \citenamefont
  {Broedersz}}]{Brueckner2020}%
  \BibitemOpen
  \bibfield  {author} {\bibinfo {author} {\bibfnamefont {D.~B.}\ \bibnamefont
  {Br{\"{u}}ckner}}, \bibinfo {author} {\bibfnamefont {A.}~\bibnamefont
  {Fink}}, \bibinfo {author} {\bibfnamefont {J.~O.}\ \bibnamefont
  {R{\"{a}}dler}}, \ and\ \bibinfo {author} {\bibfnamefont {C.~P.}\
  \bibnamefont {Broedersz}},\ }\href {\doibase
  http://dx.doi.org/10.1098/rsif.2019.0689} {\bibfield  {journal} {\bibinfo
  {journal} {J. R. Soc. Interface}\ }\textbf {\bibinfo {volume} {17}},\
  \bibinfo {pages} {20190689} (\bibinfo {year}
  {2020}{\natexlab{a}})}\BibitemShut {NoStop}%
\bibitem [{\citenamefont {Fink}\ \emph {et~al.}(2020)\citenamefont {Fink},
  \citenamefont {Br{\"{u}}ckner}, \citenamefont {Schreiber}, \citenamefont
  {R{\"{o}}ttgermann}, \citenamefont {Broedersz},\ and\ \citenamefont
  {R{\"{a}}dler}}]{Fink2019}%
  \BibitemOpen
  \bibfield  {author} {\bibinfo {author} {\bibfnamefont {A.}~\bibnamefont
  {Fink}}, \bibinfo {author} {\bibfnamefont {D.~B.}\ \bibnamefont
  {Br{\"{u}}ckner}}, \bibinfo {author} {\bibfnamefont {C.}~\bibnamefont
  {Schreiber}}, \bibinfo {author} {\bibfnamefont {P.~J.}\ \bibnamefont
  {R{\"{o}}ttgermann}}, \bibinfo {author} {\bibfnamefont {C.~P.}\ \bibnamefont
  {Broedersz}}, \ and\ \bibinfo {author} {\bibfnamefont {J.~O.}\ \bibnamefont
  {R{\"{a}}dler}},\ }\href {\doibase 10.1016/j.bpj.2019.11.3389} {\bibfield
  {journal} {\bibinfo  {journal} {Biophysical Journal}\ }\textbf {\bibinfo
  {volume} {118}},\ \bibinfo {pages} {552} (\bibinfo {year}
  {2020})}\BibitemShut {NoStop}%
\bibitem [{\citenamefont {Alert}\ and\ \citenamefont
  {Trepat}(2020)}]{Alert2020}%
  \BibitemOpen
  \bibfield  {author} {\bibinfo {author} {\bibfnamefont {R.}~\bibnamefont
  {Alert}}\ and\ \bibinfo {author} {\bibfnamefont {X.}~\bibnamefont {Trepat}},\
  }\href {\doibase 10.1146/annurev-conmatphys-031218-013516} {\bibfield
  {journal} {\bibinfo  {journal} {Annual Review of Condensed Matter Physics}\
  }\textbf {\bibinfo {volume} {11}},\ \bibinfo {pages} {77} (\bibinfo {year}
  {2020})}\BibitemShut {NoStop}%
\bibitem [{\citenamefont {Huang}\ \emph {et~al.}(2005)\citenamefont {Huang},
  \citenamefont {Brangwynne}, \citenamefont {Parker},\ and\ \citenamefont
  {Ingber}}]{Huang2005}%
  \BibitemOpen
  \bibfield  {author} {\bibinfo {author} {\bibfnamefont {S.}~\bibnamefont
  {Huang}}, \bibinfo {author} {\bibfnamefont {C.~P.}\ \bibnamefont
  {Brangwynne}}, \bibinfo {author} {\bibfnamefont {K.~K.}\ \bibnamefont
  {Parker}}, \ and\ \bibinfo {author} {\bibfnamefont {D.~E.}\ \bibnamefont
  {Ingber}},\ }\href {\doibase 10.1002/cm.20077} {\bibfield  {journal}
  {\bibinfo  {journal} {Cell Motility and the Cytoskeleton}\ }\textbf {\bibinfo
  {volume} {61}},\ \bibinfo {pages} {201} (\bibinfo {year} {2005})}\BibitemShut
  {NoStop}%
\bibitem [{\citenamefont {Segerer}\ \emph {et~al.}(2015)\citenamefont
  {Segerer}, \citenamefont {Th{\"{u}}roff}, \citenamefont {{Piera Alberola}},
  \citenamefont {Frey},\ and\ \citenamefont {R{\"{a}}dler}}]{Segerer2015}%
  \BibitemOpen
  \bibfield  {author} {\bibinfo {author} {\bibfnamefont {F.~J.}\ \bibnamefont
  {Segerer}}, \bibinfo {author} {\bibfnamefont {F.}~\bibnamefont
  {Th{\"{u}}roff}}, \bibinfo {author} {\bibfnamefont {A.}~\bibnamefont {{Piera
  Alberola}}}, \bibinfo {author} {\bibfnamefont {E.}~\bibnamefont {Frey}}, \
  and\ \bibinfo {author} {\bibfnamefont {J.~O.}\ \bibnamefont {R{\"{a}}dler}},\
  }\href {\doibase 10.1103/PhysRevLett.114.228102} {\bibfield  {journal}
  {\bibinfo  {journal} {Physical Review Letters}\ }\textbf {\bibinfo {volume}
  {114}},\ \bibinfo {pages} {228102} (\bibinfo {year} {2015})}\BibitemShut
  {NoStop}%
\bibitem [{\citenamefont {Desai}\ \emph {et~al.}(2013)\citenamefont {Desai},
  \citenamefont {Gopal}, \citenamefont {Chen},\ and\ \citenamefont
  {Chen}}]{Desai2013}%
  \BibitemOpen
  \bibfield  {author} {\bibinfo {author} {\bibfnamefont {R.~A.}\ \bibnamefont
  {Desai}}, \bibinfo {author} {\bibfnamefont {S.~B.}\ \bibnamefont {Gopal}},
  \bibinfo {author} {\bibfnamefont {S.}~\bibnamefont {Chen}}, \ and\ \bibinfo
  {author} {\bibfnamefont {C.~S.}\ \bibnamefont {Chen}},\ }\href {\doibase
  10.1098/rsif.2013.0717} {\bibfield  {journal} {\bibinfo  {journal} {Journal
  of the Royal Society Interface}\ }\textbf {\bibinfo {volume} {10}} (\bibinfo
  {year} {2013}),\ 10.1098/rsif.2013.0717}\BibitemShut {NoStop}%
\bibitem [{\citenamefont {Scarpa}\ \emph {et~al.}(2013)\citenamefont {Scarpa},
  \citenamefont {Roycroft}, \citenamefont {Theveneau}, \citenamefont {Terriac},
  \citenamefont {Piel}, \citenamefont {Mayor}, \citenamefont {Scarpa},
  \citenamefont {Roycroft}, \citenamefont {Theveneau}, \citenamefont {Terriac},
  \citenamefont {Piel},\ and\ \citenamefont {Mayor}}]{Scarpa2013}%
  \BibitemOpen
  \bibfield  {author} {\bibinfo {author} {\bibfnamefont {E.}~\bibnamefont
  {Scarpa}}, \bibinfo {author} {\bibfnamefont {A.}~\bibnamefont {Roycroft}},
  \bibinfo {author} {\bibfnamefont {E.}~\bibnamefont {Theveneau}}, \bibinfo
  {author} {\bibfnamefont {E.}~\bibnamefont {Terriac}}, \bibinfo {author}
  {\bibfnamefont {M.}~\bibnamefont {Piel}}, \bibinfo {author} {\bibfnamefont
  {R.}~\bibnamefont {Mayor}}, \bibinfo {author} {\bibfnamefont
  {E.}~\bibnamefont {Scarpa}}, \bibinfo {author} {\bibfnamefont
  {A.}~\bibnamefont {Roycroft}}, \bibinfo {author} {\bibfnamefont
  {E.}~\bibnamefont {Theveneau}}, \bibinfo {author} {\bibfnamefont
  {E.}~\bibnamefont {Terriac}}, \bibinfo {author} {\bibfnamefont
  {M.}~\bibnamefont {Piel}}, \ and\ \bibinfo {author} {\bibfnamefont
  {R.}~\bibnamefont {Mayor}},\ }\href {\doibase 10.1242/bio.020917} {\bibfield
  {journal} {\bibinfo  {journal} {Biology Open}\ }\textbf {\bibinfo {volume}
  {2}},\ \bibinfo {pages} {901} (\bibinfo {year} {2013})}\BibitemShut {NoStop}%
\bibitem [{\citenamefont {Lin}\ \emph {et~al.}(2015)\citenamefont {Lin},
  \citenamefont {Yin}, \citenamefont {Wu}, \citenamefont {Inoue},\ and\
  \citenamefont {Levchenko}}]{Lin2015}%
  \BibitemOpen
  \bibfield  {author} {\bibinfo {author} {\bibfnamefont {B.}~\bibnamefont
  {Lin}}, \bibinfo {author} {\bibfnamefont {T.}~\bibnamefont {Yin}}, \bibinfo
  {author} {\bibfnamefont {Y.~I.}\ \bibnamefont {Wu}}, \bibinfo {author}
  {\bibfnamefont {T.}~\bibnamefont {Inoue}}, \ and\ \bibinfo {author}
  {\bibfnamefont {A.}~\bibnamefont {Levchenko}},\ }\href {\doibase
  10.1038/ncomms7619} {\bibfield  {journal} {\bibinfo  {journal} {Nature
  Communications}\ }\textbf {\bibinfo {volume} {6}} (\bibinfo {year} {2015}),\
  10.1038/ncomms7619}\BibitemShut {NoStop}%
\bibitem [{\citenamefont {Singh}\ \emph {et~al.}(2020)\citenamefont {Singh},
  \citenamefont {Camley},\ and\ \citenamefont {Nain}}]{Singh2020}%
  \BibitemOpen
  \bibfield  {author} {\bibinfo {author} {\bibfnamefont {J.}~\bibnamefont
  {Singh}}, \bibinfo {author} {\bibfnamefont {B.~A.}\ \bibnamefont {Camley}}, \
  and\ \bibinfo {author} {\bibfnamefont {A.~S.}\ \bibnamefont {Nain}},\ }\href
  {https://www.biorxiv.org/content/10.1101/2020.05.28.122218v2} {\bibfield
  {journal} {\bibinfo  {journal} {bioRxiv}\ } (\bibinfo {year}
  {2020})}\BibitemShut {NoStop}%
\bibitem [{\citenamefont {Mayor}\ and\ \citenamefont
  {Carmona-Fontaine}(2010)}]{Mayor2010}%
  \BibitemOpen
  \bibfield  {author} {\bibinfo {author} {\bibfnamefont {R.}~\bibnamefont
  {Mayor}}\ and\ \bibinfo {author} {\bibfnamefont {C.}~\bibnamefont
  {Carmona-Fontaine}},\ }\href {\doibase 10.1016/j.tcb.2010.03.005} {\bibfield
  {journal} {\bibinfo  {journal} {Trends in Cell Biology}\ }\textbf {\bibinfo
  {volume} {20}},\ \bibinfo {pages} {319} (\bibinfo {year} {2010})}\BibitemShut
  {NoStop}%
\bibitem [{\citenamefont {Abercrombie}\ and\ \citenamefont
  {Heaysman}(1954{\natexlab{b}})}]{Abercrombie1954}%
  \BibitemOpen
  \bibfield  {author} {\bibinfo {author} {\bibfnamefont {M.}~\bibnamefont
  {Abercrombie}}\ and\ \bibinfo {author} {\bibfnamefont {J.~E.}\ \bibnamefont
  {Heaysman}},\ }\href {\doibase 10.1038/174697a0} {\bibfield  {journal}
  {\bibinfo  {journal} {Nature}\ }\textbf {\bibinfo {volume} {174}},\ \bibinfo
  {pages} {697} (\bibinfo {year} {1954}{\natexlab{b}})}\BibitemShut {NoStop}%
\bibitem [{\citenamefont {Sep{\'{u}}lveda}\ \emph {et~al.}(2013)\citenamefont
  {Sep{\'{u}}lveda}, \citenamefont {Petitjean}, \citenamefont {Cochet},
  \citenamefont {Grasland-Mongrain}, \citenamefont {Silberzan},\ and\
  \citenamefont {Hakim}}]{Sepulveda2013}%
  \BibitemOpen
  \bibfield  {author} {\bibinfo {author} {\bibfnamefont {N.}~\bibnamefont
  {Sep{\'{u}}lveda}}, \bibinfo {author} {\bibfnamefont {L.}~\bibnamefont
  {Petitjean}}, \bibinfo {author} {\bibfnamefont {O.}~\bibnamefont {Cochet}},
  \bibinfo {author} {\bibfnamefont {E.}~\bibnamefont {Grasland-Mongrain}},
  \bibinfo {author} {\bibfnamefont {P.}~\bibnamefont {Silberzan}}, \ and\
  \bibinfo {author} {\bibfnamefont {V.}~\bibnamefont {Hakim}},\ }\href
  {\doibase 10.1371/journal.pcbi.1002944} {\bibfield  {journal} {\bibinfo
  {journal} {PLoS Computational Biology}\ }\textbf {\bibinfo {volume} {9}}
  (\bibinfo {year} {2013}),\ 10.1371/journal.pcbi.1002944}\BibitemShut
  {NoStop}%
\bibitem [{\citenamefont {Basan}\ \emph {et~al.}(2013)\citenamefont {Basan},
  \citenamefont {Elgeti}, \citenamefont {Hannezo}, \citenamefont {Rappel},\
  and\ \citenamefont {Levine}}]{Basan2013}%
  \BibitemOpen
  \bibfield  {author} {\bibinfo {author} {\bibfnamefont {M.}~\bibnamefont
  {Basan}}, \bibinfo {author} {\bibfnamefont {J.}~\bibnamefont {Elgeti}},
  \bibinfo {author} {\bibfnamefont {E.}~\bibnamefont {Hannezo}}, \bibinfo
  {author} {\bibfnamefont {W.~J.}\ \bibnamefont {Rappel}}, \ and\ \bibinfo
  {author} {\bibfnamefont {H.}~\bibnamefont {Levine}},\ }\href {\doibase
  10.1073/pnas.1219937110} {\bibfield  {journal} {\bibinfo  {journal}
  {Proceedings of the National Academy of Sciences of the United States of
  America}\ }\textbf {\bibinfo {volume} {110}},\ \bibinfo {pages} {2452}
  (\bibinfo {year} {2013})}\BibitemShut {NoStop}%
\bibitem [{\citenamefont {Copenhagen}\ \emph {et~al.}(2018)\citenamefont
  {Copenhagen}, \citenamefont {Malet-engra}, \citenamefont {Yu}, \citenamefont
  {Scita}, \citenamefont {Gov},\ and\ \citenamefont
  {Gopinathan}}]{Copenhagen2018}%
  \BibitemOpen
  \bibfield  {author} {\bibinfo {author} {\bibfnamefont {K.}~\bibnamefont
  {Copenhagen}}, \bibinfo {author} {\bibfnamefont {G.}~\bibnamefont
  {Malet-engra}}, \bibinfo {author} {\bibfnamefont {W.}~\bibnamefont {Yu}},
  \bibinfo {author} {\bibfnamefont {G.}~\bibnamefont {Scita}}, \bibinfo
  {author} {\bibfnamefont {N.}~\bibnamefont {Gov}}, \ and\ \bibinfo {author}
  {\bibfnamefont {A.}~\bibnamefont {Gopinathan}},\ }\href
  {https://advances.sciencemag.org/content/4/9/eaar8483?rss=1} {\bibfield
  {journal} {\bibinfo  {journal} {Science Advances}\ }\textbf {\bibinfo
  {volume} {4}},\ \bibinfo {pages} {eaar8483} (\bibinfo {year}
  {2018})}\BibitemShut {NoStop}%
\bibitem [{\citenamefont {Garcia}\ \emph {et~al.}(2015)\citenamefont {Garcia},
  \citenamefont {Hannezo}, \citenamefont {Elgeti}, \citenamefont {Joanny},
  \citenamefont {Silberzan},\ and\ \citenamefont {Gov}}]{Garcia2015}%
  \BibitemOpen
  \bibfield  {author} {\bibinfo {author} {\bibfnamefont {S.}~\bibnamefont
  {Garcia}}, \bibinfo {author} {\bibfnamefont {E.}~\bibnamefont {Hannezo}},
  \bibinfo {author} {\bibfnamefont {J.}~\bibnamefont {Elgeti}}, \bibinfo
  {author} {\bibfnamefont {J.-F.}\ \bibnamefont {Joanny}}, \bibinfo {author}
  {\bibfnamefont {P.}~\bibnamefont {Silberzan}}, \ and\ \bibinfo {author}
  {\bibfnamefont {N.~S.}\ \bibnamefont {Gov}},\ }\href {\doibase
  10.1073/pnas.1510973112} {\bibfield  {journal} {\bibinfo  {journal}
  {Proceedings of the National Academy of Sciences}\ }\textbf {\bibinfo
  {volume} {112}},\ \bibinfo {pages} {15314} (\bibinfo {year}
  {2015})}\BibitemShut {NoStop}%
\bibitem [{\citenamefont {Kulawiak}\ \emph {et~al.}(2016)\citenamefont
  {Kulawiak}, \citenamefont {Camley},\ and\ \citenamefont
  {Rappel}}]{Kulawiak2016}%
  \BibitemOpen
  \bibfield  {author} {\bibinfo {author} {\bibfnamefont {D.~A.}\ \bibnamefont
  {Kulawiak}}, \bibinfo {author} {\bibfnamefont {B.~A.}\ \bibnamefont
  {Camley}}, \ and\ \bibinfo {author} {\bibfnamefont {W.~J.}\ \bibnamefont
  {Rappel}},\ }\href {\doibase 10.1371/journal.pcbi.1005239} {\bibfield
  {journal} {\bibinfo  {journal} {PLoS Computational Biology}\ }\textbf
  {\bibinfo {volume} {12}} (\bibinfo {year} {2016}),\
  10.1371/journal.pcbi.1005239}\BibitemShut {NoStop}%
\bibitem [{\citenamefont {Br{\"{u}}ckner}\ \emph
  {et~al.}(2020{\natexlab{b}})\citenamefont {Br{\"{u}}ckner}, \citenamefont
  {Ronceray},\ and\ \citenamefont {Broedersz}}]{Brueckner2020a}%
  \BibitemOpen
  \bibfield  {author} {\bibinfo {author} {\bibfnamefont {D.~B.}\ \bibnamefont
  {Br{\"{u}}ckner}}, \bibinfo {author} {\bibfnamefont {P.}~\bibnamefont
  {Ronceray}}, \ and\ \bibinfo {author} {\bibfnamefont {C.~P.}\ \bibnamefont
  {Broedersz}},\ }\href {\doibase 10.1103/PhysRevLett.125.058103} {\bibfield
  {journal} {\bibinfo  {journal} {Physical Review Letters}\ }\textbf {\bibinfo
  {volume} {125}},\ \bibinfo {pages} {58103} (\bibinfo {year}
  {2020}{\natexlab{b}})}\BibitemShut {NoStop}%
\bibitem [{\citenamefont {Camley}\ and\ \citenamefont
  {Rappel}(2014)}]{Camley2014}%
  \BibitemOpen
  \bibfield  {author} {\bibinfo {author} {\bibfnamefont {B.~A.}\ \bibnamefont
  {Camley}}\ and\ \bibinfo {author} {\bibfnamefont {W.~J.}\ \bibnamefont
  {Rappel}},\ }\href {\doibase 10.1103/PhysRevE.89.062705} {\bibfield
  {journal} {\bibinfo  {journal} {Physical Review E - Statistical, Nonlinear,
  and Soft Matter Physics}\ }\textbf {\bibinfo {volume} {89}},\ \bibinfo
  {pages} {062705} (\bibinfo {year} {2014})}\BibitemShut {NoStop}%
\bibitem [{\citenamefont {L{\"{o}}ber}\ \emph {et~al.}(2015)\citenamefont
  {L{\"{o}}ber}, \citenamefont {Ziebert},\ and\ \citenamefont
  {Aranson}}]{Lober2015}%
  \BibitemOpen
  \bibfield  {author} {\bibinfo {author} {\bibfnamefont {J.}~\bibnamefont
  {L{\"{o}}ber}}, \bibinfo {author} {\bibfnamefont {F.}~\bibnamefont
  {Ziebert}}, \ and\ \bibinfo {author} {\bibfnamefont {I.~S.}\ \bibnamefont
  {Aranson}},\ }\href {\doibase 10.1038/srep09172} {\bibfield  {journal}
  {\bibinfo  {journal} {Scientific Reports}\ }\textbf {\bibinfo {volume} {5}},\
  \bibinfo {pages} {1} (\bibinfo {year} {2015})}\BibitemShut {NoStop}%
\bibitem [{\citenamefont {Vedel}\ \emph {et~al.}(2013)\citenamefont {Vedel},
  \citenamefont {Tay}, \citenamefont {Johnston}, \citenamefont {Bruus},\ and\
  \citenamefont {Quake}}]{Vedel2013}%
  \BibitemOpen
  \bibfield  {author} {\bibinfo {author} {\bibfnamefont {S.}~\bibnamefont
  {Vedel}}, \bibinfo {author} {\bibfnamefont {S.}~\bibnamefont {Tay}}, \bibinfo
  {author} {\bibfnamefont {D.~M.}\ \bibnamefont {Johnston}}, \bibinfo {author}
  {\bibfnamefont {H.}~\bibnamefont {Bruus}}, \ and\ \bibinfo {author}
  {\bibfnamefont {S.~R.}\ \bibnamefont {Quake}},\ }\href {\doibase
  10.1073/pnas.1204291110} {\bibfield  {journal} {\bibinfo  {journal}
  {Proceedings of the National Academy of Sciences}\ }\textbf {\bibinfo
  {volume} {110}},\ \bibinfo {pages} {129} (\bibinfo {year}
  {2013})}\BibitemShut {NoStop}%
\bibitem [{\citenamefont {Mak}\ \emph {et~al.}(2011)\citenamefont {Mak},
  \citenamefont {Reinhart-King},\ and\ \citenamefont {Erickson}}]{Mak2011}%
  \BibitemOpen
  \bibfield  {author} {\bibinfo {author} {\bibfnamefont {M.}~\bibnamefont
  {Mak}}, \bibinfo {author} {\bibfnamefont {C.~A.}\ \bibnamefont
  {Reinhart-King}}, \ and\ \bibinfo {author} {\bibfnamefont {D.}~\bibnamefont
  {Erickson}},\ }\href {\doibase 10.1371/journal.pone.0020825} {\bibfield
  {journal} {\bibinfo  {journal} {PLoS ONE}\ }\textbf {\bibinfo {volume} {6}},\
  \bibinfo {pages} {e20825} (\bibinfo {year} {2011})}\BibitemShut {NoStop}%
\bibitem [{\citenamefont {Kraning-Rush}\ \emph {et~al.}(2013)\citenamefont
  {Kraning-Rush}, \citenamefont {Carey}, \citenamefont {Lampi},\ and\
  \citenamefont {Reinhart-King}}]{Kraning-Rush2013}%
  \BibitemOpen
  \bibfield  {author} {\bibinfo {author} {\bibfnamefont {C.~M.}\ \bibnamefont
  {Kraning-Rush}}, \bibinfo {author} {\bibfnamefont {S.~P.}\ \bibnamefont
  {Carey}}, \bibinfo {author} {\bibfnamefont {M.~C.}\ \bibnamefont {Lampi}}, \
  and\ \bibinfo {author} {\bibfnamefont {C.~A.}\ \bibnamefont
  {Reinhart-King}},\ }\href {\doibase 10.1039/c3ib20196a} {\bibfield  {journal}
  {\bibinfo  {journal} {Integrative Biology}\ }\textbf {\bibinfo {volume}
  {5}},\ \bibinfo {pages} {606} (\bibinfo {year} {2013})}\BibitemShut {NoStop}%
\bibitem [{\citenamefont {Sommers}\ \emph {et~al.}(1991)\citenamefont
  {Sommers}, \citenamefont {Thompson}, \citenamefont {Torri}, \citenamefont
  {Kemler}, \citenamefont {Gelmann},\ and\ \citenamefont
  {Byers}}]{Sommers1991}%
  \BibitemOpen
  \bibfield  {author} {\bibinfo {author} {\bibfnamefont {C.~L.}\ \bibnamefont
  {Sommers}}, \bibinfo {author} {\bibfnamefont {E.~W.}\ \bibnamefont
  {Thompson}}, \bibinfo {author} {\bibfnamefont {J.~A.}\ \bibnamefont {Torri}},
  \bibinfo {author} {\bibfnamefont {R.}~\bibnamefont {Kemler}}, \bibinfo
  {author} {\bibfnamefont {E.~P.}\ \bibnamefont {Gelmann}}, \ and\ \bibinfo
  {author} {\bibfnamefont {S.~W.}\ \bibnamefont {Byers}},\ }\href
  {http://cgd.aacrjournals.org/cgi/reprint/2/8/365} {\bibfield  {journal}
  {\bibinfo  {journal} {Cell growth {\&} differentiation}\ }\textbf {\bibinfo
  {volume} {2}},\ \bibinfo {pages} {365} (\bibinfo {year} {1991})}\BibitemShut
  {NoStop}%
\bibitem [{\citenamefont {Carey}\ \emph {et~al.}(2013)\citenamefont {Carey},
  \citenamefont {Starchenko}, \citenamefont {McGregor},\ and\ \citenamefont
  {Reinhart-King}}]{Carey2013}%
  \BibitemOpen
  \bibfield  {author} {\bibinfo {author} {\bibfnamefont {S.~P.}\ \bibnamefont
  {Carey}}, \bibinfo {author} {\bibfnamefont {A.}~\bibnamefont {Starchenko}},
  \bibinfo {author} {\bibfnamefont {A.~L.}\ \bibnamefont {McGregor}}, \ and\
  \bibinfo {author} {\bibfnamefont {C.~A.}\ \bibnamefont {Reinhart-King}},\
  }\href {\doibase 10.1007/s10585-013-9565-x} {\bibfield  {journal} {\bibinfo
  {journal} {Clinical and Experimental Metastasis}\ }\textbf {\bibinfo {volume}
  {30}},\ \bibinfo {pages} {615} (\bibinfo {year} {2013})}\BibitemShut
  {NoStop}%
\bibitem [{\citenamefont {Lee}\ \emph {et~al.}(2020)\citenamefont {Lee},
  \citenamefont {Vitolo}, \citenamefont {Losert},\ and\ \citenamefont
  {Martin}}]{Lee2020}%
  \BibitemOpen
  \bibfield  {author} {\bibinfo {author} {\bibfnamefont {R.~M.}\ \bibnamefont
  {Lee}}, \bibinfo {author} {\bibfnamefont {M.~I.}\ \bibnamefont {Vitolo}},
  \bibinfo {author} {\bibfnamefont {W.}~\bibnamefont {Losert}}, \ and\ \bibinfo
  {author} {\bibfnamefont {S.~S.}\ \bibnamefont {Martin}},\ }\href {\doibase
  10.1101/2020.06.04.135178} {\bibfield  {journal} {\bibinfo  {journal}
  {bioRxiv}\ } (\bibinfo {year} {2020}),\
  10.1101/2020.06.04.135178}\BibitemShut {NoStop}%
\bibitem [{\citenamefont {Kang}\ \emph {et~al.}(2020)\citenamefont {Kang},
  \citenamefont {Ferruzzi}, \citenamefont {Spatarelu}, \citenamefont {Han},
  \citenamefont {Sharma}, \citenamefont {Koehler}, \citenamefont {Butler},
  \citenamefont {Roblyer}, \citenamefont {Zaman}, \citenamefont {Guo},
  \citenamefont {Chen}, \citenamefont {Pegoraro},\ and\ \citenamefont
  {Fredberg}}]{Kang2020}%
  \BibitemOpen
  \bibfield  {author} {\bibinfo {author} {\bibfnamefont {W.}~\bibnamefont
  {Kang}}, \bibinfo {author} {\bibfnamefont {J.}~\bibnamefont {Ferruzzi}},
  \bibinfo {author} {\bibfnamefont {C.-P.}\ \bibnamefont {Spatarelu}}, \bibinfo
  {author} {\bibfnamefont {Y.~L.}\ \bibnamefont {Han}}, \bibinfo {author}
  {\bibfnamefont {Y.}~\bibnamefont {Sharma}}, \bibinfo {author} {\bibfnamefont
  {S.~A.}\ \bibnamefont {Koehler}}, \bibinfo {author} {\bibfnamefont {J.~P.}\
  \bibnamefont {Butler}}, \bibinfo {author} {\bibfnamefont {D.}~\bibnamefont
  {Roblyer}}, \bibinfo {author} {\bibfnamefont {M.~H.}\ \bibnamefont {Zaman}},
  \bibinfo {author} {\bibfnamefont {M.}~\bibnamefont {Guo}}, \bibinfo {author}
  {\bibfnamefont {Z.}~\bibnamefont {Chen}}, \bibinfo {author} {\bibfnamefont
  {A.~F.}\ \bibnamefont {Pegoraro}}, \ and\ \bibinfo {author} {\bibfnamefont
  {J.~J.}\ \bibnamefont {Fredberg}},\ }\href {\doibase
  10.1101/2020.04.28.066845} {\bibfield  {journal} {\bibinfo  {journal}
  {bioRxiv}\ ,\ \bibinfo {pages} {2020.04.28.066845}} (\bibinfo {year}
  {2020})}\BibitemShut {NoStop}%
\bibitem [{\citenamefont {Fujimori}\ \emph {et~al.}(2019)\citenamefont
  {Fujimori}, \citenamefont {Nakajima}, \citenamefont {Shimada},\ and\
  \citenamefont {Sawai}}]{Fujimori2019}%
  \BibitemOpen
  \bibfield  {author} {\bibinfo {author} {\bibfnamefont {T.}~\bibnamefont
  {Fujimori}}, \bibinfo {author} {\bibfnamefont {A.}~\bibnamefont {Nakajima}},
  \bibinfo {author} {\bibfnamefont {N.}~\bibnamefont {Shimada}}, \ and\
  \bibinfo {author} {\bibfnamefont {S.}~\bibnamefont {Sawai}},\ }\href
  {\doibase 10.1073/pnas.1815063116} {\bibfield  {journal} {\bibinfo  {journal}
  {Proceedings of the National Academy of Sciences of the United States of
  America}\ }\textbf {\bibinfo {volume} {116}},\ \bibinfo {pages} {4291}
  (\bibinfo {year} {2019})}\BibitemShut {NoStop}%
\bibitem [{\citenamefont {Palamidessi}\ \emph {et~al.}(2019)\citenamefont
  {Palamidessi}, \citenamefont {Malinverno}, \citenamefont {Frittoli},
  \citenamefont {Corallino}, \citenamefont {Barbieri}, \citenamefont
  {Sigismund}, \citenamefont {Beznoussenko}, \citenamefont {Martini},
  \citenamefont {Garre}, \citenamefont {Ferrara}, \citenamefont {Tripodo},
  \citenamefont {Ascione}, \citenamefont {Cavalcanti-Adam}, \citenamefont {Li},
  \citenamefont {{Di Fiore}}, \citenamefont {Parazzoli}, \citenamefont
  {Giavazzi}, \citenamefont {Cerbino},\ and\ \citenamefont
  {Scita}}]{Palamidessi2019}%
  \BibitemOpen
  \bibfield  {author} {\bibinfo {author} {\bibfnamefont {A.}~\bibnamefont
  {Palamidessi}}, \bibinfo {author} {\bibfnamefont {C.}~\bibnamefont
  {Malinverno}}, \bibinfo {author} {\bibfnamefont {E.}~\bibnamefont
  {Frittoli}}, \bibinfo {author} {\bibfnamefont {S.}~\bibnamefont {Corallino}},
  \bibinfo {author} {\bibfnamefont {E.}~\bibnamefont {Barbieri}}, \bibinfo
  {author} {\bibfnamefont {S.}~\bibnamefont {Sigismund}}, \bibinfo {author}
  {\bibfnamefont {G.~V.}\ \bibnamefont {Beznoussenko}}, \bibinfo {author}
  {\bibfnamefont {E.}~\bibnamefont {Martini}}, \bibinfo {author} {\bibfnamefont
  {M.}~\bibnamefont {Garre}}, \bibinfo {author} {\bibfnamefont
  {I.}~\bibnamefont {Ferrara}}, \bibinfo {author} {\bibfnamefont
  {C.}~\bibnamefont {Tripodo}}, \bibinfo {author} {\bibfnamefont
  {F.}~\bibnamefont {Ascione}}, \bibinfo {author} {\bibfnamefont {E.~A.}\
  \bibnamefont {Cavalcanti-Adam}}, \bibinfo {author} {\bibfnamefont
  {Q.}~\bibnamefont {Li}}, \bibinfo {author} {\bibfnamefont {P.~P.}\
  \bibnamefont {{Di Fiore}}}, \bibinfo {author} {\bibfnamefont
  {D.}~\bibnamefont {Parazzoli}}, \bibinfo {author} {\bibfnamefont
  {F.}~\bibnamefont {Giavazzi}}, \bibinfo {author} {\bibfnamefont
  {R.}~\bibnamefont {Cerbino}}, \ and\ \bibinfo {author} {\bibfnamefont
  {G.}~\bibnamefont {Scita}},\ }\href
  {http://dx.doi.org/10.1038/s41563-019-0425-1} {\bibfield  {journal} {\bibinfo
   {journal} {Nature Materials}\ }\textbf {\bibinfo {volume} {18}},\ \bibinfo
  {pages} {1252} (\bibinfo {year} {2019})}\BibitemShut {NoStop}%
\bibitem [{\citenamefont {Han}\ \emph {et~al.}(2020)\citenamefont {Han},
  \citenamefont {Pegoraro}, \citenamefont {Li}, \citenamefont {Li},
  \citenamefont {Yuan}, \citenamefont {Xu}, \citenamefont {Gu}, \citenamefont
  {Sun}, \citenamefont {Hao}, \citenamefont {Gupta}, \citenamefont {Li},
  \citenamefont {Tang}, \citenamefont {Kang}, \citenamefont {Teng},
  \citenamefont {Fredberg},\ and\ \citenamefont {Guo}}]{Han2020}%
  \BibitemOpen
  \bibfield  {author} {\bibinfo {author} {\bibfnamefont {Y.~L.}\ \bibnamefont
  {Han}}, \bibinfo {author} {\bibfnamefont {A.~F.}\ \bibnamefont {Pegoraro}},
  \bibinfo {author} {\bibfnamefont {H.}~\bibnamefont {Li}}, \bibinfo {author}
  {\bibfnamefont {K.}~\bibnamefont {Li}}, \bibinfo {author} {\bibfnamefont
  {Y.}~\bibnamefont {Yuan}}, \bibinfo {author} {\bibfnamefont {G.}~\bibnamefont
  {Xu}}, \bibinfo {author} {\bibfnamefont {Z.}~\bibnamefont {Gu}}, \bibinfo
  {author} {\bibfnamefont {J.}~\bibnamefont {Sun}}, \bibinfo {author}
  {\bibfnamefont {Y.}~\bibnamefont {Hao}}, \bibinfo {author} {\bibfnamefont
  {S.~K.}\ \bibnamefont {Gupta}}, \bibinfo {author} {\bibfnamefont
  {Y.}~\bibnamefont {Li}}, \bibinfo {author} {\bibfnamefont {W.}~\bibnamefont
  {Tang}}, \bibinfo {author} {\bibfnamefont {H.}~\bibnamefont {Kang}}, \bibinfo
  {author} {\bibfnamefont {L.}~\bibnamefont {Teng}}, \bibinfo {author}
  {\bibfnamefont {J.~J.}\ \bibnamefont {Fredberg}}, \ and\ \bibinfo {author}
  {\bibfnamefont {M.}~\bibnamefont {Guo}},\ }\href {\doibase
  10.1038/s41567-019-0680-8} {\bibfield  {journal} {\bibinfo  {journal} {Nature
  Physics}\ }\textbf {\bibinfo {volume} {16}},\ \bibinfo {pages} {101}
  (\bibinfo {year} {2020})}\BibitemShut {NoStop}%
\bibitem [{\citenamefont {D'alessandro}\ \emph {et~al.}(2017)\citenamefont
  {D'alessandro}, \citenamefont {Solon}, \citenamefont {Hayakawa},
  \citenamefont {Anjard}, \citenamefont {Detcheverry}, \citenamefont {Rieu},\
  and\ \citenamefont {Rivi{\`{e}}re}}]{Dalessandro2017}%
  \BibitemOpen
  \bibfield  {author} {\bibinfo {author} {\bibfnamefont {J.}~\bibnamefont
  {D'alessandro}}, \bibinfo {author} {\bibfnamefont {A.~P.}\ \bibnamefont
  {Solon}}, \bibinfo {author} {\bibfnamefont {Y.}~\bibnamefont {Hayakawa}},
  \bibinfo {author} {\bibfnamefont {C.}~\bibnamefont {Anjard}}, \bibinfo
  {author} {\bibfnamefont {F.}~\bibnamefont {Detcheverry}}, \bibinfo {author}
  {\bibfnamefont {J.~P.}\ \bibnamefont {Rieu}}, \ and\ \bibinfo {author}
  {\bibfnamefont {C.}~\bibnamefont {Rivi{\`{e}}re}},\ }\href {\doibase
  10.1038/nphys4180} {\bibfield  {journal} {\bibinfo  {journal} {Nature
  Physics}\ }\textbf {\bibinfo {volume} {13}},\ \bibinfo {pages} {999}
  (\bibinfo {year} {2017})}\BibitemShut {NoStop}%
\bibitem [{\citenamefont {Angelini}\ \emph {et~al.}(2010)\citenamefont
  {Angelini}, \citenamefont {Hannezo}, \citenamefont {Trepat}, \citenamefont
  {Fredberg},\ and\ \citenamefont {Weitz}}]{Angelini2010a}%
  \BibitemOpen
  \bibfield  {author} {\bibinfo {author} {\bibfnamefont {T.~E.}\ \bibnamefont
  {Angelini}}, \bibinfo {author} {\bibfnamefont {E.}~\bibnamefont {Hannezo}},
  \bibinfo {author} {\bibfnamefont {X.}~\bibnamefont {Trepat}}, \bibinfo
  {author} {\bibfnamefont {J.~J.}\ \bibnamefont {Fredberg}}, \ and\ \bibinfo
  {author} {\bibfnamefont {D.~A.}\ \bibnamefont {Weitz}},\ }\href {\doibase
  10.1103/PhysRevLett.104.168104} {\bibfield  {journal} {\bibinfo  {journal}
  {Physical Review Letters}\ }\textbf {\bibinfo {volume} {104}},\ \bibinfo
  {pages} {1} (\bibinfo {year} {2010})}\BibitemShut {NoStop}%
\bibitem [{\citenamefont {Blair}\ and\ \citenamefont {Dufresne}()}]{Blair}%
  \BibitemOpen
  \bibfield  {author} {\bibinfo {author} {\bibfnamefont {D.}~\bibnamefont
  {Blair}}\ and\ \bibinfo {author} {\bibfnamefont {E.}~\bibnamefont
  {Dufresne}},\ }\href {http://site.physics.georgetown.edu/matlab/} {\bibinfo
  {journal} {http://site.physics.georgetown.edu/matlab/}\ }\BibitemShut
  {NoStop}%
\bibitem [{\citenamefont {Ferretti}\ \emph {et~al.}(2020)\citenamefont
  {Ferretti}, \citenamefont {Chard{\`{e}}s}, \citenamefont {Mora},
  \citenamefont {Walczak},\ and\ \citenamefont {Giardina}}]{Ferretti2020}%
  \BibitemOpen
\bibfield  {journal} {  }\bibfield  {author} {\bibinfo {author} {\bibfnamefont
  {F.}~\bibnamefont {Ferretti}}, \bibinfo {author} {\bibfnamefont
  {V.}~\bibnamefont {Chard{\`{e}}s}}, \bibinfo {author} {\bibfnamefont
  {T.}~\bibnamefont {Mora}}, \bibinfo {author} {\bibfnamefont {A.~M.}\
  \bibnamefont {Walczak}}, \ and\ \bibinfo {author} {\bibfnamefont
  {I.}~\bibnamefont {Giardina}},\ }\href {\doibase 10.1103/physrevx.10.031018}
  {\bibfield  {journal} {\bibinfo  {journal} {Physical Review X}\ }\textbf
  {\bibinfo {volume} {10}},\ \bibinfo {pages} {031018} (\bibinfo {year}
  {2020})}\BibitemShut {NoStop}%
\end{thebibliography}%



\section*{Supplementary material}

# Supplementary Material:

**Learning the dynamics of cell-cell interactions in confined cell migration**

David B. Brückner, Nicolas Arlt, Alexandra Fink, Pierre Ronceray,
Joachim O. Rädler and Chase P. Broedersz

# Contents

# 1 Movie descriptions

**Supplementary Movies 1&2**
Pairs of MCF10A cells transitioning repeatedly between the square adhesion sites of a two-state micropattern. The cell nucleus is fluorescently labelled to allow automated tracking of cell positions. Scale bar: 25 µm.

**Supplementary Movies 3&4**
Pairs of MDA-MB-231 cells transitioning repeatedly between the square adhesion sites of a two-state micropattern. The cell nucleus is fluorescently labelled to allow automated tracking of cell positions. Scale bar: 25 µm.

**Supplementary Movie 5**
A pair of MDA-MB-231 cells transfected with LifeAct-GFP to visualize actin on a two-state micropattern. The outline of the micropattern is drawn in white as a reference. Actin hotspots are visible at the tip of the lamellipodia. Scale bar: 25 µm.

**Supplementary Movies 6&7**
Pairs of MDA-MB-231 cells migrating on a rectangular micropattern without constriction with the same overall dimensions as the two-state micropatterns. The cell nucleus is fluorescently labelled to allow automated tracking of cell positions. Scale bar: 25 µm.

# 2 Further experimental details

## 2.1 Micropattern design

The adhesive sites of the two-state micropatterns have square dimensions $(36.7 \pm 0.6)^2$ µm$^2$. The connecting bridge has length $(35.3 \pm 0.5)$ µm and width $(6.9 \pm 0.6)$ µm. The quoted errors correspond to the deviations in the dimensions of final protein patterns which are due to the intrinsic variance of the manual stamping process and the measurement uncertainty associated with the limited resolution of the brightfield images.

## 2.2 Cell exclusion criteria

We track the trajectories of a large number of cell pairs confined to two-state micropatterns. Following previous work [1], we apply the following inclusion criteria in our analysis:

1. Two-cell trajectories are obtained by selecting cells which undergo division during the experiment. Tracking begins after both daughter cells have re-attached to the pattern after division. Tracking is terminated when one of the two cells rounds up for division again.

2. Both cells and their protrusions are confined within the borders of the micropattern.

3. Both cells show no abnormalities such as multiple nuclei or the occurrence of cell death or detachment from the substrate.

4. At least one transition occurs during the experiment (i.e. at least one of the two cells performs a transition across the bridge of the micropattern).

## 2.3 Cell transfections

For live-cell imaging of actin, approximately 10,000 MDA-MB-231 cells are seeded in patterned $\mu$-dishes and left to adhere for 12 h. As a cell culturing medium, MEM with Glutamax (Gibco) supplemented with 10% FCS is used. 500ng LiveAct-GFP mRNA (prepared in-house) is resuspended in OptiMEM (Gibco) to a final volume of 150 μL. This solution is then added to a mix of 1.25 μL Lipofectamine 2000 (Invitrogen) and 123.75 μL OptiMEM, and left to incubate for 20 minutes at room temperature. Subsequently, cells are rinsed once with PBS and the transfection mix is added and left on the cells for at least 5 h, before being replaced by L-15 medium. Cells are imaged every 10 minutes on the Nikon Ti Eclipse microscope using a 60x oil-immersion objective.

## 2.4 Tracking procedure

Brightfield and fluorescence images of the stained nuclei are acquired every 10 min. A band pass filter is applied to the images of the nuclei, then images are binarised. The cell trajectories are determined by tracking the nuclei using a Matlab tracking algorithm [2]. After application of the tracking algorithm, each trajectory is inspected manually to verify that only two particles have been identified, and that the trajectories exhibit no gaps that consist of more than one missing frame. In the rare case of a single missing frame, we interpolate linearly between the previous and the subsequent coordinate. Furthermore, we verify that the tracking algorithm has correctly identified the identity of the two cells. In the cases where the two cells have been mixed up, this is corrected manually. After tracking, the reference boundaries of the patterns are determined manually by means of the bright-field images, on which the micropatterns are visible.

# 3 Analysis of the experimental dynamics

## 3.1 Error analysis

All errorbars throughout the paper correspond to bootstrap errors, as described in refs. [3, 4]. Briefly, from our data set of $N$ cell pair trajectories $\{\mathbf{x}_k\}$, where $k = 1...N$, we generate $N_{\text{BS}}$ bootstrap realizations by randomly sampling $N$ cell pair trajectories with replacement for each realization. To obtain the error in an observable $\theta_{\text{expt}}$ measured from the experimental data set, we estimate the value of $\theta$ for each bootstrap realization and take the standard deviation of all obtained $\theta$s as our estimate for the error in $\theta_{\text{expt}}$. To obtain the error in an observable $\theta_{\text{model}}$ predicted by our model, we perform ULI on each of the $N_{\text{BS}}$ bootstrap realizations to obtain

$N_{\text{BS}}$ bootstrapped models. Then, we simulate a large number of trajectories for each bootstrap model, and estimate $\theta$ for each set of trajectories. The standard deviation of these $\theta$s is our estimate for the error in $\theta_{\text{model}}$.

## 3.2 Stochastic switching dynamics and cross-correlation functions

**Survival probability functions** − To quantify the transition dynamics of cell pairs within the two-state micropattern, we define two configurations: one where both cells occupy the same side of the pattern, and one where they occupy opposite sides, with the two sides defined by $x < 0$ and $x > 0$ (Fig. S1a). We obtain similar results if the boundaries are instead defined by $x < -L/2$ and $x > L/2$, where $L$ is the length of the bridge. We define a switching event as a switch from a same-side to an opposite-side configuration, and vice versa. To gain insight into the dynamics of these configuration switches, we calculate the times between switches, which yields the average dwell times of the same-side configuration $\{\tau_{\text{same}}\}$ and the opposite-side configuration $\{\tau_{\text{opp}}\}$. Thus, we can calculate the survival probability distribution function of each configuration, i.e. the probability that a switch has not occurred after time $t$, given that the system is initially in state $k = \{\text{same}, \text{opp}\}$:

$$S_k(t) = 1 - \int_0^t p(\tau_k) \mathrm{d}\tau_k \tag{S1}$$

where $p(\tau_k)$ is the probability distribution of $\tau_k$. These distributions show that the same-side configuration is typically occupied for shorter times than the opposite-side configuration (Fig. S1b,d). Both survival probability functions exhibit an initially exponential decay (Fig. S1c,e), but become noisy at long times due to limited sampling.

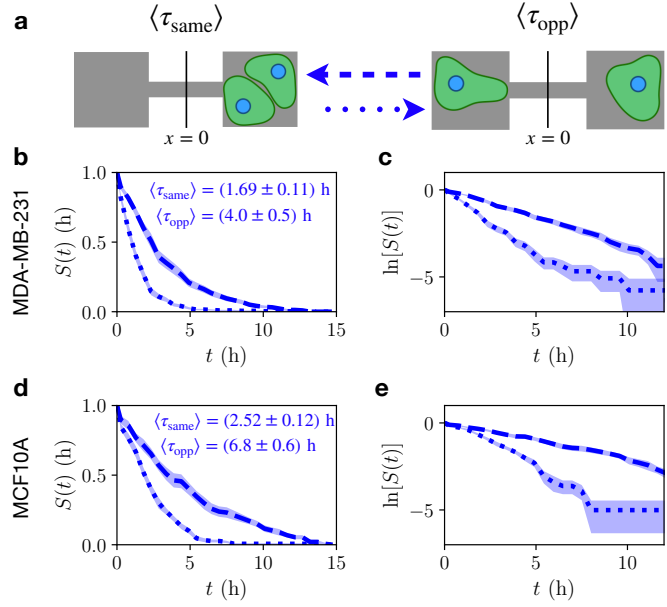

Figure S1: **Stochastic switching dynamics. a.** Sketch of the same- and opposite-side configurations. **b.** Survival probability functions of the same- (dotted line) and opposite-side (dashed line) configurations, for MDA-MB-231 cells. The average dwell times of each state are given. **c.** Log-linear plot of panel **b**. **d.** Survival probability functions of the same- (dotted line) and opposite-side (dashed line) configurations, for MCF10A cells. The average dwell times of each state are given. **e.** Log-linear plot of panel **d**. Shaded regions and error intervals for the average dwell times denote bootstrap errors.

**Position cross-correlation function** − The cross-correlation function of cell positions, defined by:

$$\langle x_1(t) x_2(t') \rangle := \frac{1}{2 \sum_{j=1}^{N_{\text{pairs}}} T_j} \sum_{j=1}^{N_{\text{pairs}}} \sum_{t'=1}^{T_j} x_1(t) x_2(t') \tag{S2}$$

where $N_{\text{pairs}}$ is the number of tracked cell pairs, and $T_j$ is the total number of time-points in the trajectory of pair $j$.

**Velocity cross-correlation function** − To investigate the correlations in migration velocities when the cells are in contact, we calculate the cross-correlation function of cell velocities $v(t) = (x(t) - x(t + \Delta t))/\Delta t$, subject to the constraint that the cells are in the same-side configuration:

$$\langle v_1(t) v_2(t') \rangle_{\text{same}} := \frac{1}{2 \sum_{j=1}^{N_{\text{pairs}}} T_j^{\text{same}}} \sum_{j=1}^{N_{\text{pairs}}} \sum_{\{\theta_{\text{same}}\}} v_1(t) v_2(t') \tag{S3}$$

where $\{\theta_{\text{same}}\}$ is the set of time-point combinations $(t, t')$, where the cells are in the same-side configuration at both time $t$ and $t'$, and $T_j^{\text{same}}$ is the total number of such time-point combinations of cell pair $j$.

## 3.3 Movement in the second dimension

The two-state micropattern is designed in such a way that most of the behavior occurs in the $x$-direction. We find that most of the interaction behavior is indeed captured by the $x$-components of the trajectories: the variance in $y$-motion is small (Fig. S2a,b), and the joint probability distribution $p(y_1, y_2)$ is peaked around $(0,0)$ and exhibits no special structure (Fig. S2d,f). Furthermore, the cross-correlation function of $y$-positions $\langle y_1(t) y_2(t') \rangle$ vanishes (Fig. S2g,h). Thus, we find that the $x$-component captures most of the behavior displayed by these interacting cells, and thus provides a simple, low-dimensional representation of the system's behavioral dynamics.

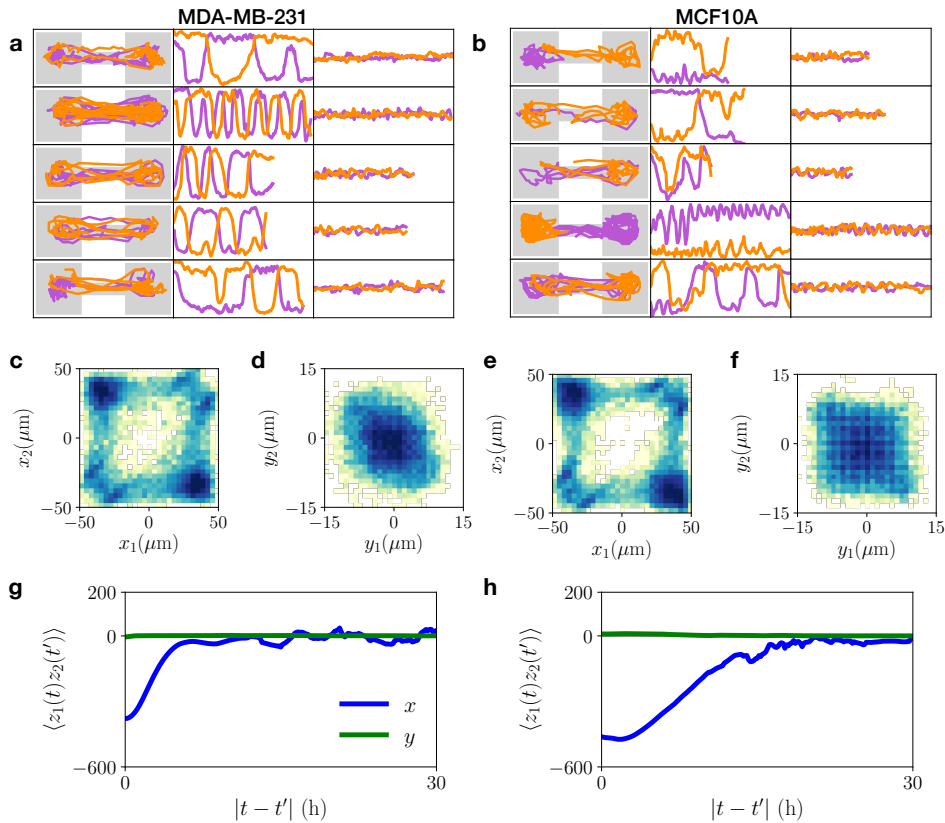

Figure S2: **Motion in 2D. a,b.** Several examples of 2D trajectories. Left: $xy$-trajectory plotted on top of the micropattern dimension (shown in grey). Axis limits are $-50$ μm $< x < 50$ μm and $-20$ μm $< y < 20$ μm; $(x = 0, y = 0)$ corresponds to the center of the bridge. Middle: $x$-trajectory as a function of time $t$. Axis limits are $-50$ μm $< x < 50$ μm and $0 < t < 30$ h. Right: $y$-trajectory as a function of time $t$. Axis limits are $-50$ μm $< y < 50$ μm and $0 < t < 30$ h, to allow direct comparison with the $x$-trajectory. **c,e.** Joint probability distributions $p(x_1, x_2)$ of the $x$-positions, plotted logarithmically on the same colour scale as in Fig. 2 in the main text. Here shown without the Gaussian interpolation employed in Fig. 2 in the main text. **d,f.** Joint probability distributions $p(y_1, y_2)$ of the $y$-positions, plotted logarithmically on the same colour scale as in Fig. 2 in the main text. **g,h.** Position cross-correlation function for $x$ (blue) and $y$-components (green). For all panels, the left hand side corresponds to MDA-MB-231 cells, while the right hand side corresponds to MCF10A cells.

6

## 3.4 Collision events

To investigate the outcomes of cellular collisions, we identified collision events and defined criteria to classify them as reversal, sliding, and following events. Similar classifications have been developed in the literature before [5, ?, 6, 7]. Here, we define a collision as an event where the two cell nuclei move to within a threshold distance $\Delta x_c$ of each other (Fig. S3a). To classify reversals and sliding events, we are interested in whether the cells have moved past each other as a response to the collision event. Thus, we identify collisions as reversals if the cells do not switch positions shortly after the collision, and as sliding events if they do switch positions. In practice, we find that we can easily identify these three events by inspecting the trajectories within a fixed time window $dT$ following the collision. Thus we identify reversals if cells do not switch positions after a time interval $dT$, while for sliding events cells switch positions at least once within $dT$ after the collision. To avoid artefacts where cells enter and exit the threshold distance $\Delta x_c$ repeatedly in a short time-period, we only identify the first collision event in a time-interval $dT$. Thus, we assume that the time-scale between subsequent transitions is $\gg dT$. Indeed, we find that subsequent traversals of the threshold distance $\Delta x_c$ usually do not correspond to new collisions, but occur while the cells still remain in contact. For following events, we are interested in identifying head-tail collisions which result in adhesion and subsequent joint migration [8]. In practice, we find that these events are robustly selected by identifying events where two cells perform the same transition across the micropatterned bridge within a time interval $dT$.

We choose the threshold distance $\Delta x_c = 20$ μm throughout, i.e. slightly larger than a typical cell radius when the cell occupies the island. For the time interval, we choose $dT = 6\Delta t = 1$ h throughout. This is a reasonable choice as this interval needs to be sufficiently larger than our time resolution $\Delta t = 10$ min, but smaller than the shortest switching time-scale, i.e. $\langle \tau_{\text{same}} \rangle \approx$ 1.7 h for the MDA-MB-231 cells. By visual inspection of a large number of trajectories, we find that these values for $\Delta x_c$ and $dT$ robustly identify reversal, sliding and following events for both cell types (Fig. S3b-f). Furthermore, we find that the measured percentages of events are robust within a wide range of values around those chosen for our analysis (Fig. S3c-f). We achieve this level of robustness due to the clear separation of time-scales between $dT$ and the transition time, and due to our criterion that no two collisions can occur within an interval $dT$.

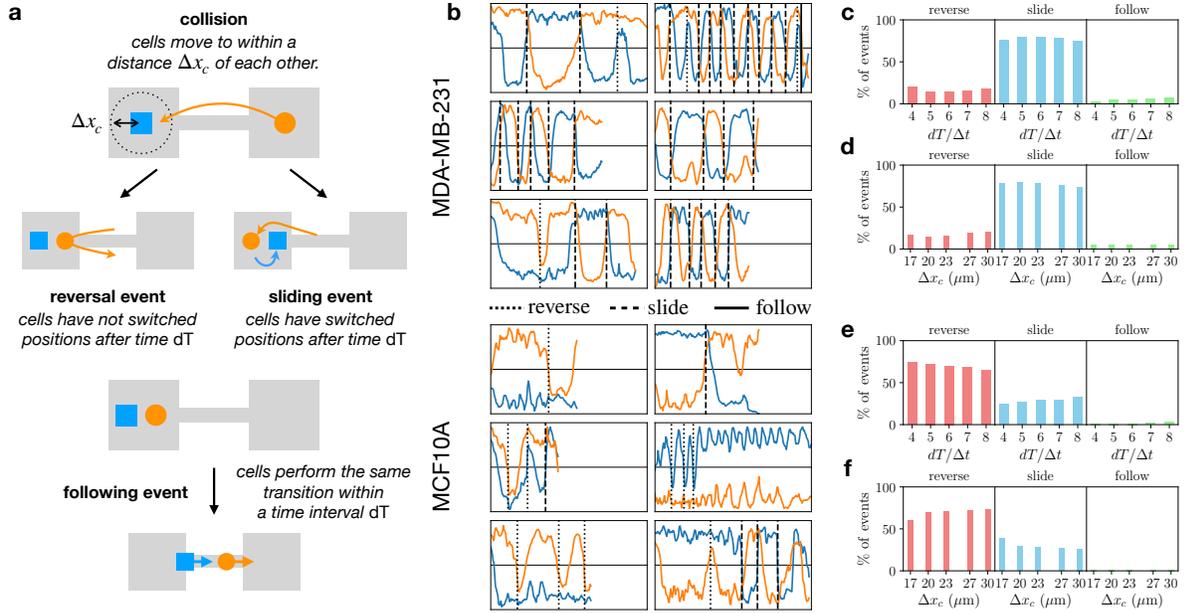

Figure S3: **Definition and identification of collision events. a.** Sketches of the definitions of reversal, sliding and following events. The blue square and orange circle show the nuclear positions of the two cells. **b.** Example trajectories with collision events shown by vertical lines (dotted: reversal, dashed: sliding, solid: following). Top 6 trajectories correspond to MDA-MB-231 cells, bottom 6 trajectories to MCF10A cells. Each trajectory plot shows the $x$-position of the cell nucleus on the $y$-axis, with axis limits $-50$ μm $< x < 50$ μm; the $x$-axis shows the time in the interval $0 < t < 33$ h. Following events occur rarely in our data set; an example can be seen in the top right trajectory. **c,e.** Percentages of reversal, sliding, and following events obtained for different values of the threshold $dT$, using the standard value $\Delta x_c = 20$ μm, for MDA-MB-231 and MCF10A cells, respectively. **d,f.** Percentages of reversal, sliding, and following events obtained for different values of the threshold $\Delta x_c$, using the standard value $dT = 6\Delta t = 1$ h, for MDA-MB-231 and MCF10A cells, respectively.

## 3.5 Contact acceleration maps

**Construction of CAMs** − To obtain insight into the detailed interaction dynamics of the cell pairs, we measure 'contact acceleration maps' (CAMs), which give the interaction component of the acceleration of the cells as a function of the separation $\Delta x_{ij} = x_i - x_j$ of cells $i$ and $j$, and their relative velocity $\Delta v_{ij} = v_i - v_j$. To measure CAMs, we assume that the dynamics can be split into separate single-cell and interaction parts, resulting in the equation of motion for cell $i$:

$$\dot{x}_i = v_i$$
$$\dot{v}_i = F(x_i, v_i) + G(\Delta x_{ij}, \Delta v_{ij}) + \sigma \eta_i(t) \quad (S4)$$

where $\langle \eta_i(t) \rangle = 0$, and $\langle \eta_i(t)\eta_j(t') \rangle = \delta_{ij}\delta(t - t')$. As show by the rigorously inferred model using Underdamped Langevin Inference (ULI), the dynamics can indeed be described by such an equation of motion.

For a system of the form given by Eq. (S4), a simple way of obtaining an estimate of the interacting component $G(\Delta x, \Delta v)$ is through a conditional averaging procedure. Specifically, for

8

interactions which decay beyond a threshold distance $\ell$, and the one-body term $F(x,v)$ can be estimated as

$$F(x,v) \approx \langle \dot{v}_i | x_i, v_i; |\Delta x_{ij}| > \ell \rangle \tag{S5}$$

Next, we estimate the interaction term by calculating

$$G(\Delta x, \Delta v) \approx \langle \dot{v}_i - F(x_i, v_i) | \Delta x_{ij}, \Delta v_{ij}; |\Delta x_{ij}| < \ell \rangle \tag{S6}$$

Note that more general binning approaches, e.g. conditioning on $(x_1, x_2)$ or $(v_1, v_2)$, are not able to recover simple dependencies on $\Delta x$ or $\Delta v$, as non-trivial correlations exist in these phase spaces. As we show in the next paragraph, this construction yields a good estimate for the interactions in the type of system we are considering here. To give this approach further credence, we show in section 4.3.2 that our model inferred from ULI, which does not rely on the assumption of short-ranged interactions, captures the experimentally measured CAMs well.

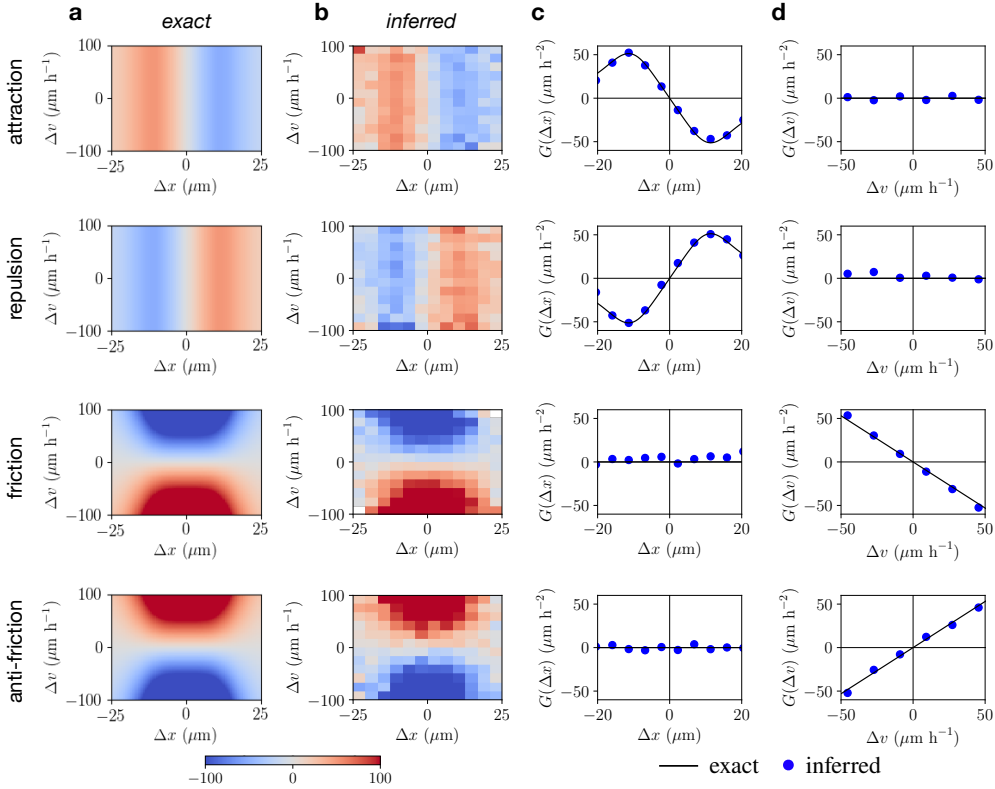

Figure S4: **Testing CAM inference.** For all cases, we used 100 trajectories with 200 time-points each, sampled at a time-interval $\Delta t = 10$ min, which is comparable to the experimental case. We use a cutoff of $\ell = 25$ μm. **a.** Exact interaction terms $G(\Delta x, \Delta v) = f_0 g(|\Delta x|)\Delta x + \gamma_0 g(|\Delta x|)\Delta v$ in units of μm h$^{-2}$. For attraction/repulsion, we take $f_0 = \pm 6$ h$^{-2}$, $\gamma_0 = 0$. For friction/anti-friction, we take $f_0 = 0$, $\gamma_0 = \pm 2$ h$^{-1}$. **b.** Inferred CAM. **c.** Inferred (blue dots) and exact (black line) $\Delta x$-dependence of the interaction. **d.** Inferred (blue dots) and exact (black line) $\Delta v$-dependence of the interaction.

**Testing on simulated data** − First, we test whether CAMs yield a good indicator for the form of interaction terms governing the type of system we are considering here. Specifically, we estimate CAMs from simulated trajectories with similar length and time resolution as in the experiment, obtained by simulating the inferred single-cell term with known, artificial interactions. To obtain a simulated data set with known interactions, which resembles the experimental data

9

set, we use the single-cell term $F_{231}(x_i, v_i)$ and noise amplitude $\sigma_{231}$ inferred from MDA-MB-231 experiments, with additional cohesion or frictional interactions. Thus, we simulate the model

$$\dot{v}_i = F_{231}(x_i, v_i) + f_0 g(|\Delta x_{ij}|)\Delta x_{ij} + \gamma_0 g(|\Delta x_{ij}|)\Delta v_{ij} + \sigma_{231}\eta_i(t) \tag{S7}$$

with $g(|\Delta x_{ij}|) = 1/[(|\Delta x_{ij}|/R_0)^4 + 1]$, $R_0 = 15$ μm, and we vary the coefficients $f_0$ and $\gamma_0$. Independently of the precise choice of parameters or functional forms, we find that the inferred CAMs recover the known interactions (Fig. S4).

**Application to experimental data** − We calculate CAMs from the experimental data using a threshold $\ell = 25$ μm. We obtain a confinement term that is qualitatively very similar to that obtained from single cell experiments (Fig. S5) [1]. Next, we obtain the CAMs as shown in Fig. S5, and in Fig. 3 of the main text. The resulting maps are robust with respect to changing the threshold $\ell$ within a reasonable range. To obtain insight into the type of interactions implied by the measured CAMs, we first compare them to CAMs plotted for simple analytical interactions (Fig. S4). By inspection, we find that the CAM for MDA-MB-231 cells looks very similar to an anti-friction interaction, while that for MCF10A cells looks similar to a repulsive interaction (compare Figs. S5b, S4a). By performing averaging conditioned on only one variable, we show that the $\Delta x$ and $\Delta v$ dependencies of the contact accelerations confirm this conclusion (Fig. 3b,e in the main text). In addition, we also find that MDA-MB-231 cells exhibit an attractive component, and MCF10A cells exhibit an additional weak friction component (Fig. 3c,f in the main text). We use these results to guide our rigorous inference approach (section 4.1).

To further test these findings without relying on conditional averaging of the CAMs, we calculate the first moments of the contact accelerations with respect to $\Delta x$ and $\Delta v$. For the cohesive component, we calculate:

$$p_{\Delta x} = \langle [\dot{v}_i - F(x_i, v_i)]\Delta x_{ij} | |\Delta x_{ij}| < \ell \rangle \tag{S8}$$

For the case of simple monotonic interactions, the sign of this quantity indicates the type of interaction observed: we expect $p_{\Delta x} < 0$ for simple monotonic attractive interactions, and $p_{\Delta x} > 0$ for repulsive interactions. Thus, this quantity summarizes the sign of the interaction. For the friction component, we calculate:

$$p_{\Delta v} = \langle [\dot{v}_i - F(x_i, v_i)]\Delta v_{ij} | |\Delta x_{ij}| < \ell \rangle \tag{S9}$$

which gives $p_{\Delta v} < 0$ for friction interactions, and $p_{\Delta v} > 0$ for anti-friction interactions. Indeed, we find that the dipole moments further support our conclusions based on the CAMs: MDA-MB-231 cells exhibit contact accelerations indicating attraction and anti-friction, while MCF10A cells exhibit repulsion and weak friction.

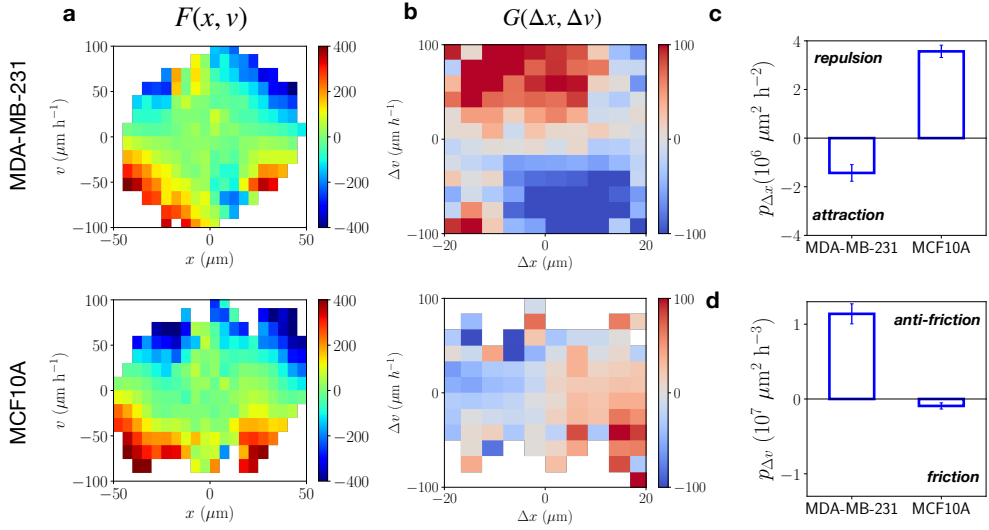

Figure S5: **Calculation of contact acceleration maps and dipole moments. a.** Estimated deterministic single-cell contribution to the dynamics $F(x,v)$. **b.** Contact acceleration map $G(\Delta x, \Delta v)$. In the main text, we show an $8 \times 8$ grid, in the range $-45 < \Delta v < 45$ for MCF10A and $-70 < \Delta v < 70$ for MDA-MB-231. For comparison, we here show $10 \times 10$ grids in the range $-100 < \Delta v < 100$, which show the same qualitative features. **c.** First moment of the contact accelerations with respect to $\Delta x$, $p_{\Delta x}$. **d.** First moment of the contact accelerations with respect to $\Delta v$, $p_{\Delta v}$. Error bars in **c**, **d** correspond to bootstrap errors. *Top row:* MDA-MB-231 cells. *Bottom row:* MCF10A cells.

## 4 Inference method and model selection

### 4.1 Application of Underdamped Langevin Inference

To infer an interacting stochastic equation of motion for confined migrating cell pairs, we employ a rigorous inference method, Underdamped Langevin Inference (ULI) [9]. In this section, we lay out the details of applying ULI to our system. For further details on the method itself, see ref. [9]. Our inference ansatz is to postulate that the system can be described by the general equation of motion for cell $i$ with position $x_i(t)$, velocity $v_i(t) = dx_i/dt$, and acceleration $\dot{v}_i(t) = dv_i/dt$:

$$\dot{x}_i = v_i$$
$$\dot{v}_i = F(x_i, v_i) + f(|\Delta x_{ij}|)\Delta x_{ij} + \gamma(|\Delta x_{ij}|)\Delta v_{ij} + \sigma \eta_i(t) \quad (S10)$$

where $\Delta x_{ij} = x_i - x_j$, $\Delta v_{ij} = v_i - v_j$, $\langle \eta_i(t) \rangle = 0$, and $\langle \eta_i(t) \eta_j(t') \rangle = \delta_{ij}\delta(t-t')$.

Using ULI, the stochastic equation of motion of such an interacting system can be reconstructed by projecting the dynamics onto a set of $n_b$ basis functions $\{b_\alpha(x,v)\}_{\alpha=1...n_b}$, which are subjected to an orthonormalization scheme $\hat{c}_\alpha(x,v) = \sum_{\beta=1}^{n_b} \hat{B}_{\alpha\beta}^{-1/2} b_\beta(x,v)$ such that $\langle \hat{c}_\alpha \hat{c}_\beta \rangle = \delta_{\alpha\beta}$. The total deterministic contribution $F^{(\text{total})} = F(x_i, v_i) + f(|\Delta x_{ij}|)\Delta x_{ij} + \gamma(|\Delta x_{ij}|)\Delta v_{ij}$ of the system can then be approximated as a linear combination of these basis functions, $F^{(\text{total})} \approx \sum_{\alpha=1}^{n_b} F_\alpha^{(\text{total})} \hat{c}_\alpha(x,v)$. Using ULI, we estimate the coefficients of this expansion of the deter-

11

ministic term $\hat{F}_\alpha^{(\text{total})}$ and the noise amplitude $\hat{\sigma}$ using the increments of the observed position trajectories $x_i(t)$.

For interacting systems, we separate single-particle and interaction contributions into separate sets of basis functions. We approximate the cohesion and friction terms $f(|\Delta x_{ij}|)$ and $\gamma(|\Delta x_{ij}|)$ using a set of interaction kernels $\{u_\alpha(|\Delta x_{ij}|)\}$ (see section 4.2). We fit the single-cell term $F(x_i, v_i)$ with a basis consisting of Fourier components in $x_i$ and polynomials in $v_i$ including terms up to third order [1]:

$$F(x_i, v_i) \approx \sum_{n=0}^{N} \sum_{m=0}^{M} [A_{nm} \cos(2\pi n x_i / w) + B_{nm} \sin(2\pi n x_i / w)] v_i^m \tag{S11}$$

where $N = M = 3$ and $w = 100$ μm. As we show in section 4.2, our inference results are not sensitive to the precise choise of basis employed.

A key assumption of our model (Eq. (S10)) is that the noise $\eta_i(t)$ is uncorrelated in time. To self-consistently test this assumption, we calculate the trajectories of the noise increments $\Delta W_i(t) = \int_t^{t+\Delta t} \eta_i(s)\,ds$. An empirical estimator for $\Delta W_i(t)$ is [10, 11, 1]:

$$\widehat{\Delta W_i}(t) \approx \frac{\Delta t}{\hat{\sigma}} \left[ \dot{v}_i(t) - \hat{F}^{(\text{total})}(x_i, v_i) \right] \tag{S12}$$

Thus, we calculate the auto-correlation function of the noise as $\hat{\phi}_{\Delta W} = \langle \widehat{\Delta W_i}(t) \widehat{\Delta W_i}(t') \rangle$. We find that for both cell lines, the noise decays to zero within a single time-step, confirming our white noise assumption. The weak negative correlation at $|t - t'| = \Delta t$ is due to the presence of measurement errors in the positions, as discussed in refs. [12, 1].

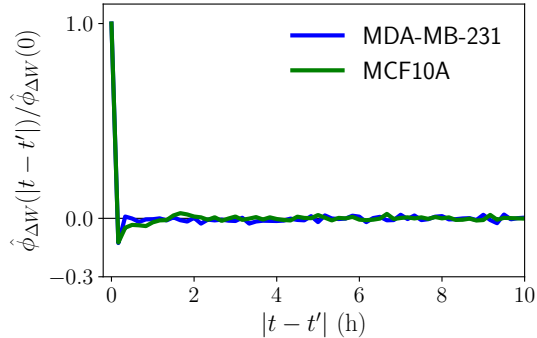

Figure S6: **Inferred noise correlation functions.** The correlation functions are normalized by their value at $|t - t'| = 0$. The blue curve corresponds to MDA-MB-231 cells, the green curve to MCF10A cells.

Three conditions for accurate inference from stochastic underdamped systems are (i) sufficiently long trajectories to constrain the $n_b$ parameters of the fitted model, (ii) a sufficiently small measurement time interval $\Delta t$ to resolve the dynamics and (iii) measurement errors on the positions that are smaller than the typical displacement in a single time-step:

**(i) Trajectory length** − Inference from a finite data set relies on having sufficient information to accurately resolve the features of the underlying dynamical terms of the equation of motion. The information contained in a set of trajectories for a system of the type of Eq. (S10) can be

|            | $N_{\text{pairs}}$ | $N_{\text{pts}}$ | $\hat{I}_b$ (nats) | $\hat{\sigma}$ ( μm h$^{3/2}$) | $\langle|\hat{v}|\rangle\Delta t$ ( μm) | $|\hat{m}|$ ( μm) |
|---|---|---|---|---|---|---|
| MDA-MB-231 | 90  | 15,979 | $11,800 \pm 150$ | 51.4 | 2.6 | 1.3 |
| MCF10A     | 155 | 19,470 | $11,900 \pm 160$ | 47.9 | 2.4 | 1.4 |

Table 1: **Statistics of the stochastic trajectory data sets for both cell lines**. From left to right, the columns denote: (i) The number of tracked cell pairs. (ii) The total number of recorded time-points. (iii) The empirical estimate of the information content of the data set, obtained by projecting the dynamics onto our standard basis choice (see section 4.2). The error in the inferred information content is estimated as $\delta \hat{I}_b \approx [2\hat{I}_b + n_b^2/4]^{1/2}$ [9]. (iv) The inferred noise amplitude. (v) The typical displacement in a single time-step. (vi) The inferred amplitude of the measurement error, which is in line with previous estimates for single cell migration in the same setup [1].

empirically estimated as $\hat{I}_b = \frac{\tau}{2\hat{\sigma}^2} \sum_{\alpha=1}^{n_b} \left(\hat{F}_\alpha^{(\text{total})}\right)^2$, where $\tau$ is the total length of the trajectory. The parameters of the expansion can be accurately inferred if $\hat{I}_b \gg n_b$, where $\hat{I}_b$ is given in natural information units (1 nat = $1/\log 2$ bits) [9]. Here, we employ a basis with $n_b = 34$ parameters (28 parameters for the single cell term and 6 parameters for the interaction kernels). As shown in table 1, our data sets contain enough information to constrain these parameters.

**(ii) Discretization** — To ensure a sufficiently accurate temporal sampling of the observed signal, we ensured that the measurement time interval $\Delta t$ should be small enough to resolve the time-scales of the collision dynamics, i.e. the switching time $\langle \tau_{\text{same}} \rangle = (1.69 \pm 0.11)$ h of MDA-MB-231 cells. Our measurement time interval is $\Delta t = 10$ min, and thus sufficiently small to resolve this time-scale. Additionally, the time interval plays an important role as velocities and accelerations are obtained as discrete derivatives from the position trajectories $x_i(t)$. Indeed, even for small $\Delta t$, inference from underdamped systems exhibits systematic discretization biases [12, 9, 13]. The leading order term of the bias is removed through the construction of the ULI estimators [9]. We show empirically that higher order biases do not strongly affect our inference results by performing a self-consistency test (see section 4.3.1).

**(iii) Measurement error** — In any tracking experiment, the observed position trajectories are subject to time-uncorrelated measurement error $m(t)$, such that the observed signal is $y(t) = x(t) + m(t)$, where $x(t)$ is the true signal. ULI yields accurate inference results in the regime $|m| < v\Delta t$, where $v\Delta t$ is the typical displacement in a single time-step. We can evaluate this condition from the data, using the average speed of the cells, and comparing it to the measurement error amplitude inferred from the trajectories [9]. As shown in table 1, this condition is fulfilled for both data sets.

## 4.2 Robustness with respect to the projection basis

To infer the interaction terms of the dynamics, we approximate the cohesion and friction terms $f(|\Delta x_{ij}|)$ and $\gamma(|\Delta x_{ij}|)$ using a set of interaction kernels $\{u_\alpha(|\Delta x_{ij}|)\}$. Physically, we expect cell-cell interactions to be spatially local. Thus, to ensure accurate inference in the region of interest, we choose kernels which decay at large distances, $u_\alpha(|\Delta x_{ij}| \to \infty) \to 0$. A simple choice for such kernels is a set of exponentials $u_n(|\Delta x_{ij}|) = \exp(-|\Delta x_{ij}|/nr_0)$ with $n = 1...N$.

This basis therefore has two hyperparameters that have to be chosen, the number for kernels $N$ and the maximum decay length $r_{\max} = Nr_0$. Alternatively, we also test a basis consisting of Gaussian functions $u_n(|\Delta x_{ij}|) = \exp(-(|\Delta x_{ij}| - nr_0)^2/2W^2)$ with $n = 1...N$. This basis therefore has three hyperparameters $N, r_{\max} = Nr_0$, and $W$. While this inference scheme could be supplemented by an additional optimization of the hyperparameters, we find this not to be necessary in this case, as the inferred interaction terms do not sensitively depend on the choice of hyperparameters or basis functions (Figs. S7,S8). Furthermore, the predictive power of the inferred model is not sensitively affected by the choice of basis (Fig. S9). Throughout the main text, we choose an exponential basis with an intermediate value of $N = 3$ functions and a maximum decay length $r_{\max} = 20$ μm (black line in Figs. Figs. S7,S8).

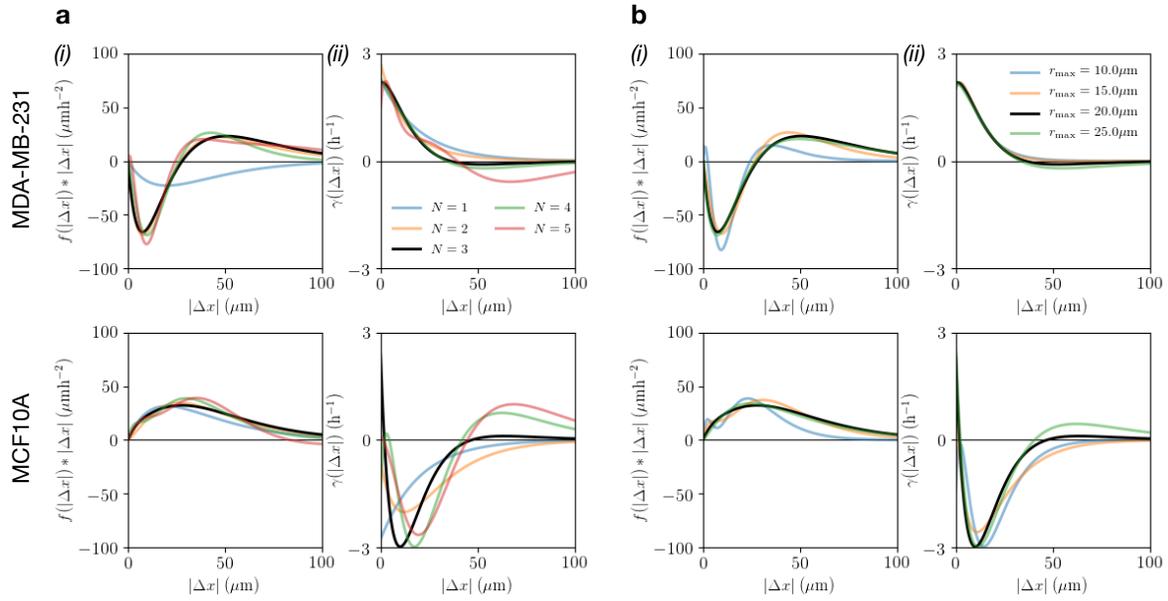

Figure S7: **Inference results for exponential interaction kernels. a**, Varying the number of kernels $N$, using $r_{\max} = 20$ μm **b**, Varying the maximum decay length $r_{\max}$, using $N = 3$. *(i)*, Cohesive component $f(|\Delta x_{ij}|)|\Delta x_{ij}|$. *(ii)*, Friction kernel $\gamma(|\Delta x_{ij}|)$. *Top row:* MDA-MB-231 cells. *Bottom row:* MCF10A cells. Black line corresponds to the curves shown in Fig. 4 of the main text, using $N = 3$ and $r_{\max} = 20$ μm.

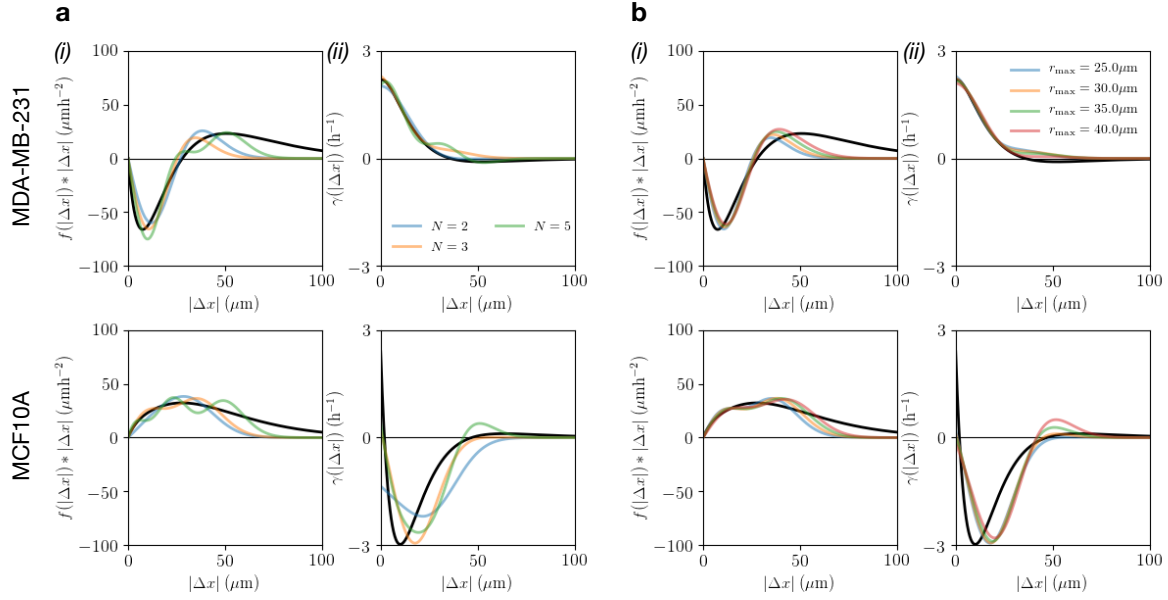

Figure S8: **Inference results for Gaussian interaction kernels. a**, Varying the number of kernels $N$, using $r_{max} = 25$ µm and $W = 4$ µm. **b**, Varying the maximum decay length $r_{max}$, using $N = 3$ and $W = 4$ µm. *(i)*, Cohesive component $f(|\Delta x_{ij}|)|\Delta x_{ij}|$. *(ii)*, Friction kernel $\gamma(|\Delta x_{ij}|)$. *Top row:* MDA-MB-231 cells. *Bottom row:* MCF10A cells. Black line corresponds to the curves shown in Fig. 4 of the main text, using an exponential basis with $N = 3$ and $r_{max} = 20$ µm.

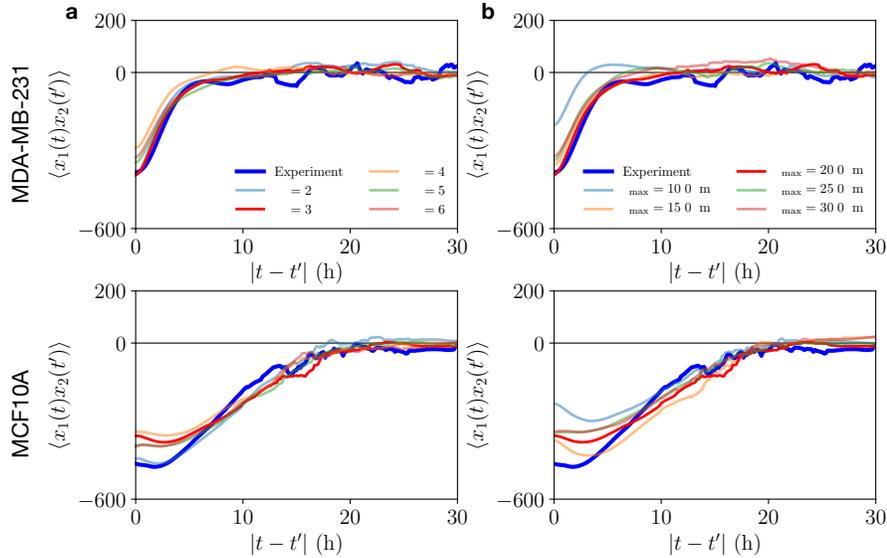

Figure S9: **Predicted position cross-correlation functions for various exponential bases. a**, Varying the number of kernels $N$. **b**, Varying the maximum decay length $r_{max}$. *Top row:* MDA-MB-231 cells. *Bottom row:* MCF10A cells.

15

## 4.3 Simulations of the inferred model

An important step in performing inference from data is to test the inferred model by simulating stochastic trajectories based on the inferred model terms, and to compare their statistical properties to those observed experimentally. We simulate the dynamics using Verlet integration with a small time step $dt$. To compare the statistics of these simulated trajectories to those observed experimentally, we sample the simulated trajectories with the same discrete time interval as in experiments, $\Delta t = 10$ min $\gg dt$.

### 4.3.1 Self-consistency test

First, we determine whether the inferred model is self-consistent: for a self-consistent inference, re-inferring a model from simulated trajectories should yield the same model. Here, we use the same number of simulated trajectories as experimentally observed trajectories, with a similar trajectory length, and the same sampling interval $\Delta t$ as in the experiment. We apply this test to the inferred models for MDA-MB-231 and MCF10A cells, and find that the re-inferred model accurately matches the original inferred model (Fig. S10), showing that our inference approach is numerically stable.

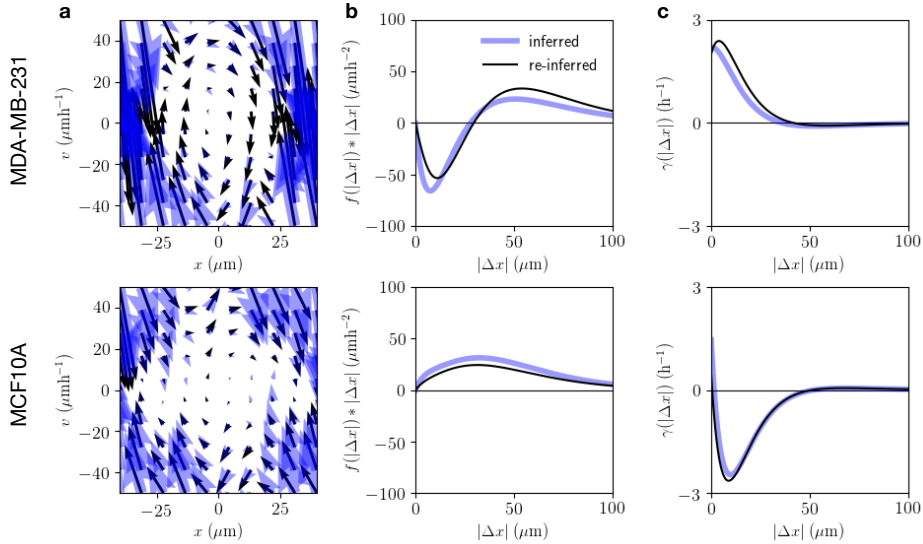

Figure S10: **Stability test of the inferred model. a**, Flow field of the confinement term $F(x_i, v_i)$. Blue arrows: inferred from experimental data, black arrows: re-inferred from simulated trajectories. **b**, Cohesive component $f(|\Delta x_{ij}|)|\Delta x_{ij}|$. **c**, Friction kernel $\gamma(|\Delta x_{ij}|)$. *Top row:* MDA-MB-231 cells. *Bottom row:* MCF10A cells.

### 4.3.2 Testing the predictive power of the model

To test the predictive power of the model, we apply the same analysis routines that were applied to the experimental data to our simulated data (results shown in Fig. 2 of the main text). The inferred model is fully constrained by the short time-scale accelerations of the dynamics.

16

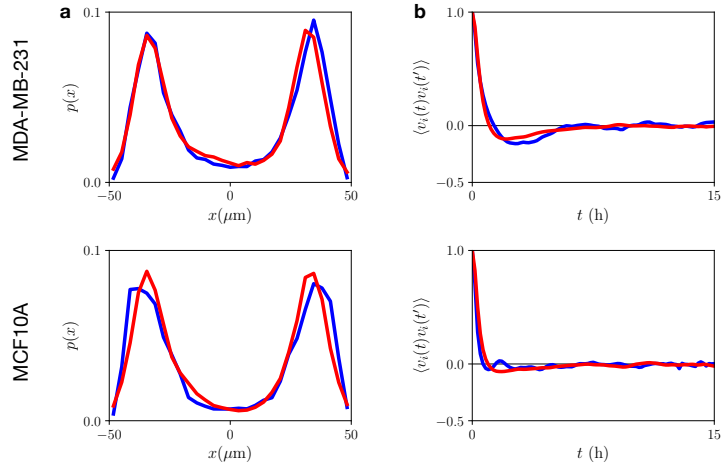

Figure S11: **Experimental and predicted dynamics of the inferred model (Eq. 1 in the main text).** **a**, Probability distribution of all cell positions $p(x)$ (experiment shown in blue, model predictions in red). **b**, Normalized velocity auto-correlation function $\langle v_i(t)v_i(t')\rangle$. *Top row:* MDA-MB-231 cells. *Bottom row:* MCF10A cells.

Thus, comparing the predicted long time-scale features such as correlation functions to the experimental data provides an independent test of the model. In addition to the statistics shown in the main text, here we show several additional statistics to test experiment-model agreement. To test how accurately the model captures the dynamics at the single-cell level, we plot the distribution of all cell positions $p(x)$, and the velocity auto-correlation function $\langle v_i(t)v_i(t')\rangle$. As shown in Fig. S11a,b, these statistics are well captured by the model.

In our model, we assume that the cell-cell interactions separate into a cohesive contribution $f(|\Delta x|)\Delta x$ and an effective linear friction $\gamma(|\Delta x|)\Delta v$. This choice is motivated by the observation that the $\Delta v$-dependent component of the contact acceleration maps is linear (Fig. 3c,f in the main text). We find that the contact acceleration maps predicted by the model are qualitatively very similar to those inferred from experiments (Fig. S12), indicating that this assumption is valid.

Next, we show side-by-side comparisons of the full joint probability distribution of positions $p(x_1, x_2)$ and velocities $p(v_1, v_2)$ (Fig. S13). The experimental distributions $p(x_1, x_2)$ exhibit several features (Fig. S13a). First, there is a clear minimum around $(0, 0)$, corresponding to both cells occupying the connecting bridge. Second, we find peaks where each cell occupies one island, and fainter peaks where both cells occupy the same island. This reflects the mutual exclusion behavior exhibited by these cells. These peaks are connected by horizontal and vertical 'paths', indicating that during transitions, typically, only one cell performs a transition at a time. Finally, we find that the peaks corresponding to both cells occupying the same island are 'split', and exhibit two distinct close-by maxima. Our model captures almost all of these features, including the relative occupation of the same- and opposite-side configuration, and the path-structure of the map (Fig. S13b). However, the model does not exhibit the same splitting of the same-side probabilities, which may be due to movement in the second dimension (the short axis of the micropattern, $y$), which is not captured by the model. Our model further captures the structure of the velocity distributions $p(v_1, v_2)$ (Fig. S13c,d).

17

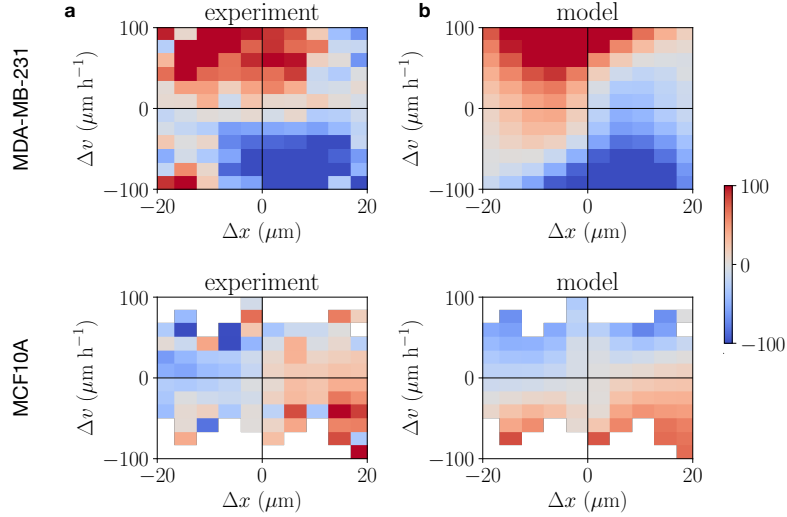

Figure S12: **Experimental and predicted contact acceleration maps. a,** Experimentally measured contact acceleration map. **b,** Contact acceleration map measured from simulation data, plotted for the same region of phase space sampled in the experiment. *Top row:* MDA-MB-231 cells. *Bottom row:* MCF10A cells.

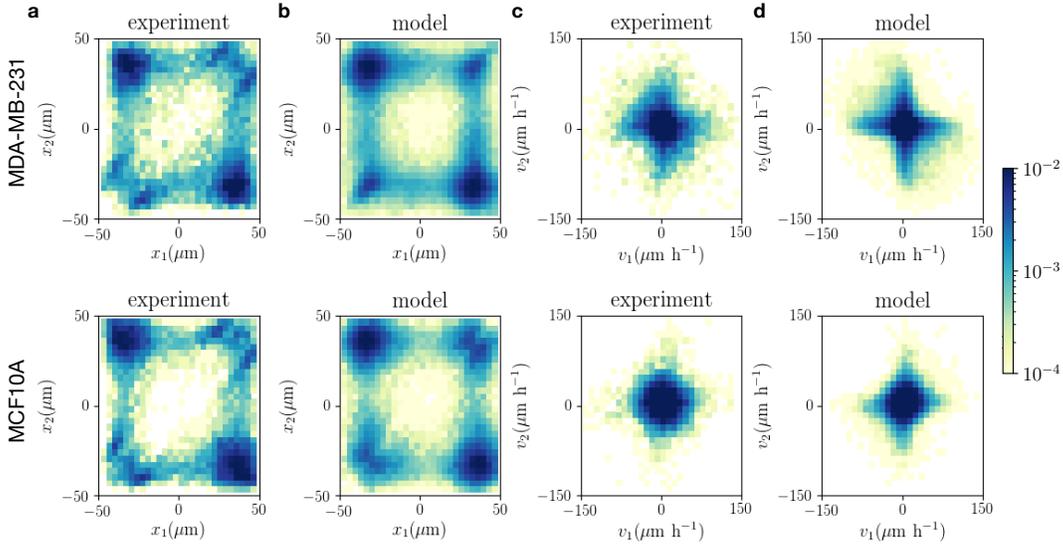

Figure S13: **Experimental and predicted joint probability distributions. a,b.** Experimental and predicted joint probability distribution of positions $p(x_1, x_2)$. **c,d.** Experimental and predicted joint probability distribution of velocities $p(v_1, v_2)$. *Top row:* MDA-MB-231 cells. *Bottom row:* MCF10A cells.

### 4.3.3 Ruling out simpler models

We arrived at our model (Eq. (S10)) by first excluding simpler alternatives. First, we consider the non-interacting case, consisting only of the single-cell term:

$$\dot{v}_i = F(x_i, v_i) + \sigma \eta_i(t) \tag{S13}$$

18

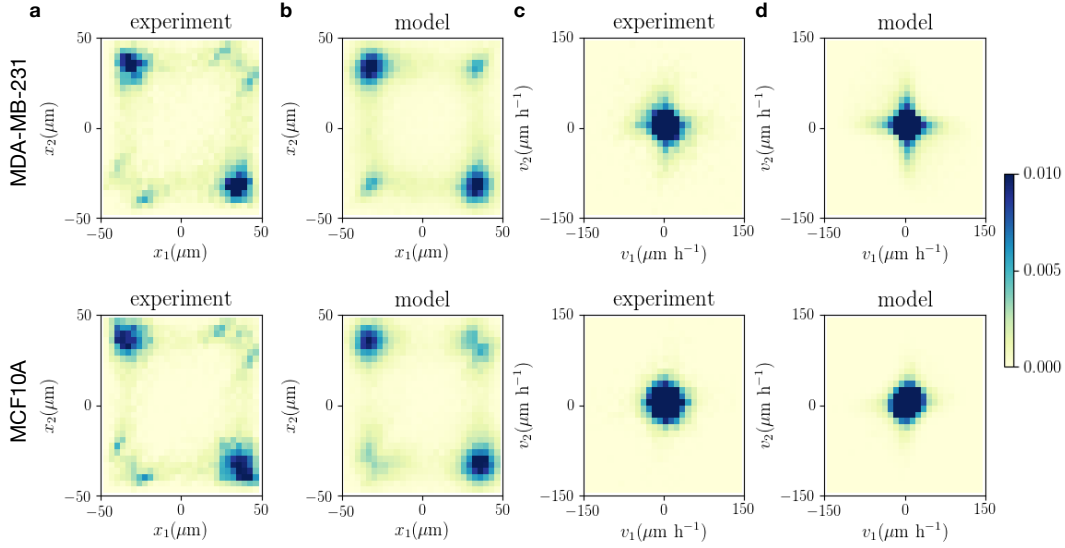

Figure S14: **Experimental and predicted joint probability distributions, plotted on a linear scale.** Same panels as shown in Fig. S13, but with a linear colour scale.

As expected, such a model is unable to capture the correlations in the system, and can therefore be ruled out (Fig. S15). This model is still able to capture the distinct minimum in the joint probability density around $(x_1, x_2) = (0, 0)$, suggesting that this feature is due to the single-cell term: due to the confinement very little occupancy is expected near the center of the connecting bridge.

Next, we consider a model including only a cohesive term:

$$\dot{v}_i = F(x_i, v_i) + f(|\Delta x_{ij}|)\Delta x_{ij} + \sigma \eta_i(t) \tag{S14}$$

While this model can approximately capture the dynamics of MCF10A cells, except for the velocity cross-correlation function, it completely fails to describe the MDA-MB-231 statistics (Fig. S16). In fact it predicts that cells are more likely to occupy the same-side configuration, in qualitative disagreement with our experimental observations, likely due to the attractive nature of the cohesive interaction in MDA-MB-231 cells.

Finally, we consider a model including frictional interactions, but no cohesion:

$$\dot{v}_i = F(x_i, v_i) + \gamma(|\Delta x_{ij}|)\Delta v_{ij} + \sigma \eta_i(t) \tag{S15}$$

This model qualitatively fails to account for the MCF10A statistics (Fig. S17): it predicts that cells are more likely to occupy the same-side configuration, likely due to the regular friction between MCF10A cells, which acts to slow cells down when they are close to each other.

In conclusion, we find that the simplest model within the class of models considered here, which can accurately capture the statistics of both MCF10A and MDA-MB-231 cell pairs, requires both cohesive and friction interactions.

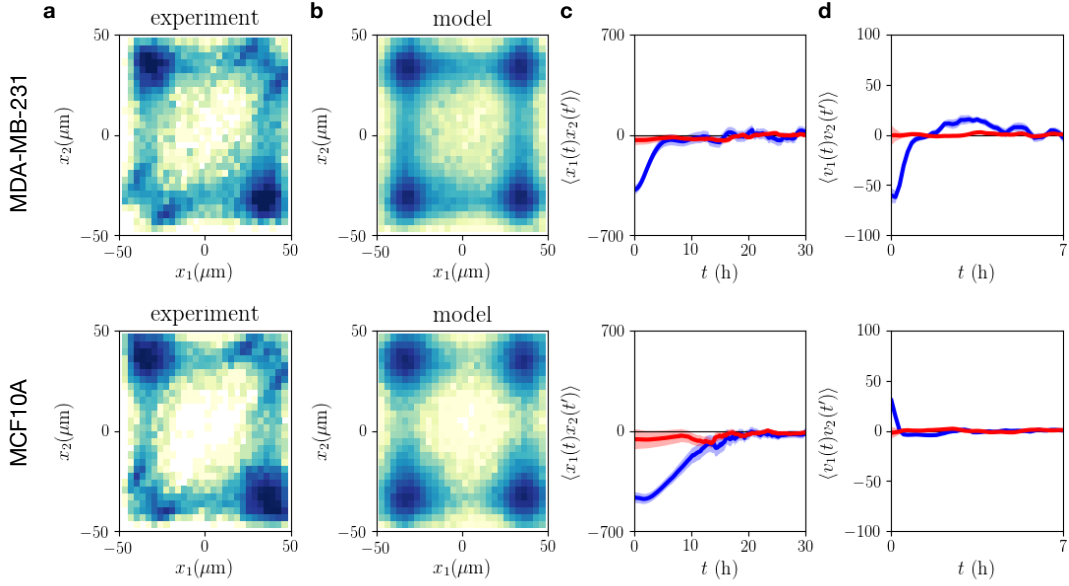

Figure S15: **Experimental and predicted dynamics for an inferred model without interactions. a.** Experimental joint probability distribution $p(x_1, x_2)$. The colour bar corresponds to that shown in Fig. S13. **b.** Model prediction of the joint probability distribution $p(x_1, x_2)$. **c.** Position cross-correlation functions for the experiment (blue) and model prediction (red). **d.** Velocity cross-correlation functions for same-side configurations. *Top row:* MDA-MB-231 cells. *Bottom row:* MCF10A cells.

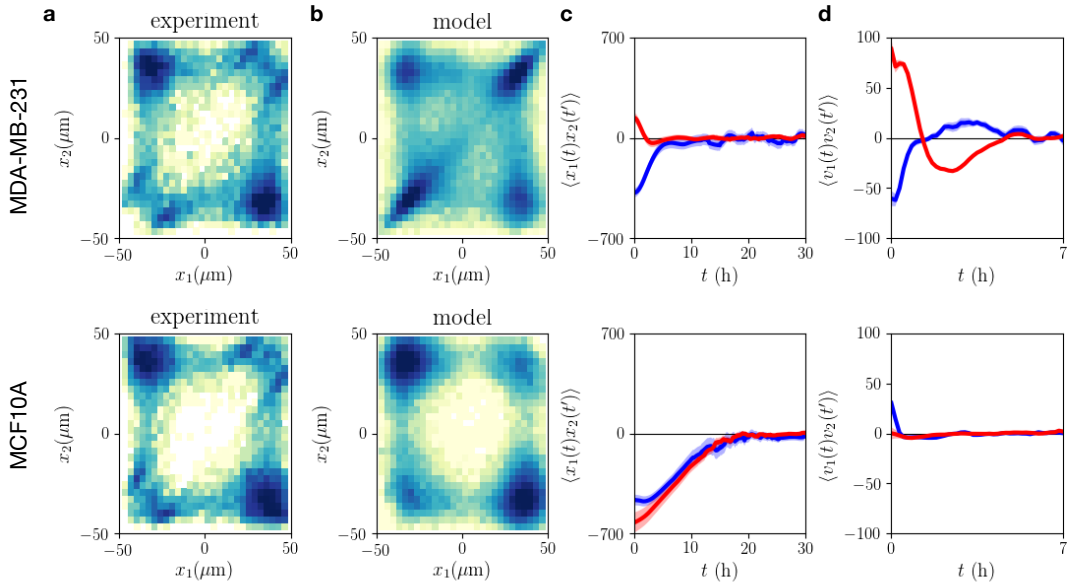

Figure S16: **Experimental and predicted dynamics for an inferred model with only cohesive, but no friction interactions.** See Fig. S15 for captions.

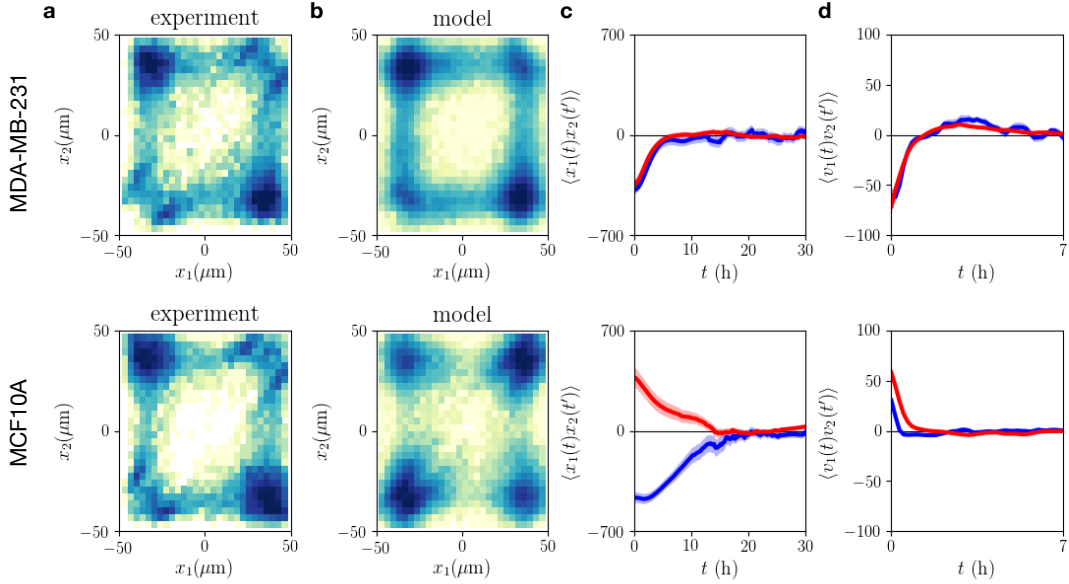

Figure S17: **Experimental and predicted dynamics for an inferred model with only friction, but no cohesive interactions.** See Fig. S15 for captions.

### 4.4 Separation of single-cell and interaction terms

Here, we directly compare the single-cell term inferred from experiments with interacting cell pairs ($F(x_i, v_i)$ in Eq. (S10)) to the deterministic term inferred from experiments in which only a single cell occupies the pattern [1], denoted $F_{\text{sc}}(x, v)$. In Fig. S18, the terms are compared side by side. Furthermore, we show the deterministic flow field $(\dot{x}, \dot{v}) = (v, F(x, v))$ superimposed for both experiments. These results indicate a remarkable similarity of the inferred terms, indicating that the contributions of single-cell dynamics (corresponding to the internal motility of the cell and its interaction with the local micro-environment placed by the micropattern) are not strongly affected by the presence of another cell.

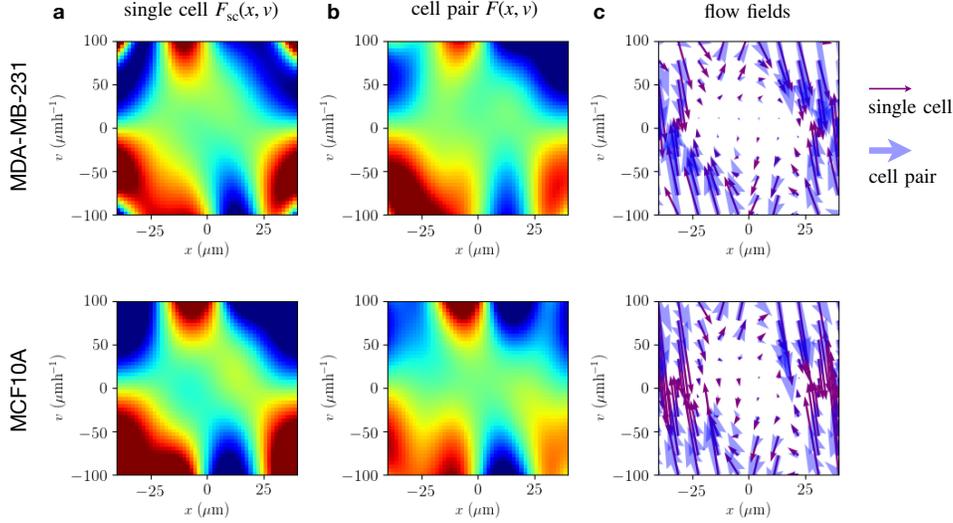

Figure S18: **Disentangling single-cell and interaction contributions.** **a**, Deterministic term $F_{sc}(x, v)$ inferred from experiments with single cells confined to two-state micro-patterns [1], obtained by applying ULI with the same basis expansion as used for cell pair experiments (Eq. (S11)), without interaction terms. Plotted with the same colour scale as in Fig. 4 in the main text. **b**, Single-cell term $F(x, v)$ inferred from cell pair experiments (as shown in Fig. 4 in the main text). **c**, Direct comparison of the flow fields of both terms. Fat blue arrows: inferred from cell pair data, thin darkviolet arrows: inferred from single-cell experiments. *Top row:* MDA-MB-231 cells. *Bottom row:* MCF10A cells.

### 4.5 Generalization of the inference approach to higher dimensions and heterotypic interactions

#### 4.5.1 Inference in 2D and 3D multi-cellular systems

The cell-cell interaction inference procedure we have developed here can be readily generalized to 2D and 3D systems, including assemblies of many cells. Specifically, our model (Eq. (S10)) can be generalized to higher dimensions as follows:

$$\dot{\mathbf{r}}_i = \mathbf{v}_i$$
$$\dot{\mathbf{v}}_i = \mathbf{F}(\mathbf{x}_i, \mathbf{v}_i) + \sum_{j \neq i} \left[ f(r_{ij}) \mathbf{r}_{ij} + \gamma(r_{ij}) \mathbf{v}_{ij} \right] + \sigma \boldsymbol{\eta}_i(t) \quad \text{(S16)}$$

where $\mathbf{r}_{ij} = \mathbf{r}_i - \mathbf{r}_j$, $r_{ij} = |\mathbf{r}_{ij}|$, and $\mathbf{v}_{ij} = \mathbf{v}_i - \mathbf{v}_j$, and the sum goes over all particles $j = 1...N$. The Gaussian white noise in this case has the property $\langle \eta_{i,\mu}(t) \eta_{j,\nu}(t') \rangle = \delta_{\mu\nu} \delta_{ij} \delta(t - t')$, where $\{i, j\}$ are particle indices and $\{\mu, \nu\} = 1...d$, where $d$ is the dimensionality of the system.

Here, the single-cell contribution $\mathbf{F}(\mathbf{x}_i, \mathbf{v}_i)$ reflects the properties of the environment in which cells migrate. In a spatially unstructured system, such as in the case of an epithelial monolayer, there will be no space-dependence, $\mathbf{F}(\mathbf{x}_i, \mathbf{v}_i) \equiv \mathbf{F}(\mathbf{v}_i)$. For example, for a simple persistent random motion model, the single-cell contribution would take the form $\mathbf{F}(\mathbf{v}_i) = -\tau_p^{-1} \mathbf{v}_i$, where $\tau_p$ is the persistence time of the cell. For the interactions terms in Eq. (S16), we assume radially symmetric interactions, i.e. that $f$ and $\gamma$ only depend on $r_{ij}$, similar to the model we have

22

inferred here. In this case, the interactions we have inferred here from confined 1D migration of cells could be directly generalized to a 2D or 3D scenario, by taking $f(r_{ij}) = f(|\Delta x_{ij}|)$ and $\gamma(r_{ij}) = \gamma(|\Delta x_{ij}|)$. The Underdamped Langevin Inference method we have used here for 1D inference can be directly generalized to 2D or 3D data by assuming that interactions are in the radial direction only. Note, however, that this method could also infer non-radial forces that depend also, for instance, on the velocity vector of each of the particles through alignment or avoidance torques. For a demonstration, see ref. [9].

### 4.5.2 Heterotypic interactions

Our model (Eq. (S10)) could furthermore be generalized to account for heterotypic interactions, where cells of different types interact with one another. In this case, we can no longer assume that the single-cell, interaction, and noise terms are identical for all cells, but they will depend on the cell type of each cell $s_i$:

$$\dot{\mathbf{r}}_i = \mathbf{v}_i$$
$$\dot{\mathbf{v}}_i = \mathbf{F}_{s_i}(\mathbf{x}_i, \mathbf{v}_i) + \sum_{j \neq i} \left[ f_{s_i s_j}(r_{ij})\mathbf{r}_{ij} + \gamma_{s_i s_j}(r_{ij})\mathbf{v}_{ij} \right] + \sigma_{s_i} \boldsymbol{\eta}_i(t) \quad \text{(S17)}$$

where for a system with $N$ cells with $n$ different cell types, $i, j = 1...N$, and $s_i \in \{1...n\}$. In this case, for each cell the single-cell and noise contribution is determined by its cell type: $\mathbf{F}_{s_i}$ and $\sigma_{s_i}$. The interaction components $f_{s_i s_j}$ and $\gamma_{s_i s_j}$ depend on the cell type of both cells in a pairwise interaction. In the example of two cell types interacting, there are thus four types of interactions: $f_{s_1 s_1}, f_{s_1 s_2}, f_{s_2 s_1}, f_{s_2 s_2}$, and similarly for $\gamma$. To infer these interactions, the same set of basis functions $\{c_\alpha(x, v)\}$ can be used for both cell types, but a different set of coefficients will be inferred for each type $s_i$, i.e. $F_{s_i}^{(\text{total})} \approx \sum_{\alpha=1}^{n_b} F_{\alpha, s_i}^{(\text{total})} \hat{c}_\alpha(x, v)$. Note that if $f_{s_2 s_1} \neq f_{s_1 s_2}$, the cells will exhibit non-reciprocal interactions.

## 5 Cell-cell interactions on a micropatterned track

To further test the generality of our results, we investigate the dynamics of MDA-MB-231 cell pairs in a different confinement geometry: a short track with the same overall dimensions as the two-state micropatterns, but without a constriction (Fig. S19a). Specifically, this micropattern has average dimensions $((103.4 \pm 0.3)$ μm$) \times ((34.8 \pm 0.2)$ μm$)$. We track a total of 84 cell pairs, resulting in a large data set of coupled cell trajectories (Fig. S19b). By applying the same inference scheme (using the same basis expansion) as for the two-state data, we determine the single cell contribution $F(x, v)$ and the cohesive and frictional interactions (Fig. S19c-e). As before, the single cell contribution is similar to the deterministic contribution inferred from experiments with only a single cell in the pattern, $F_{\text{sc}}(x, v)$ (Inset Fig. S19c; see section 4.4 and ref. [1]). Importantly, we find that the inferred model exhibits the same types of interactions as in the two-state geometry, including a short-range attractive and an anti-friction component. The model performs well on predicting the statistics of the trajectories: the joint probability distribution, the collision statistics, the position probability distribution, and the velocity auto-correlation function are well captured (Fig. S19f-h,j). The cross-correlation of positions deviates slightly, but decays on a similar time-scale (Fig. S19i). This deviation is likely due to a larger

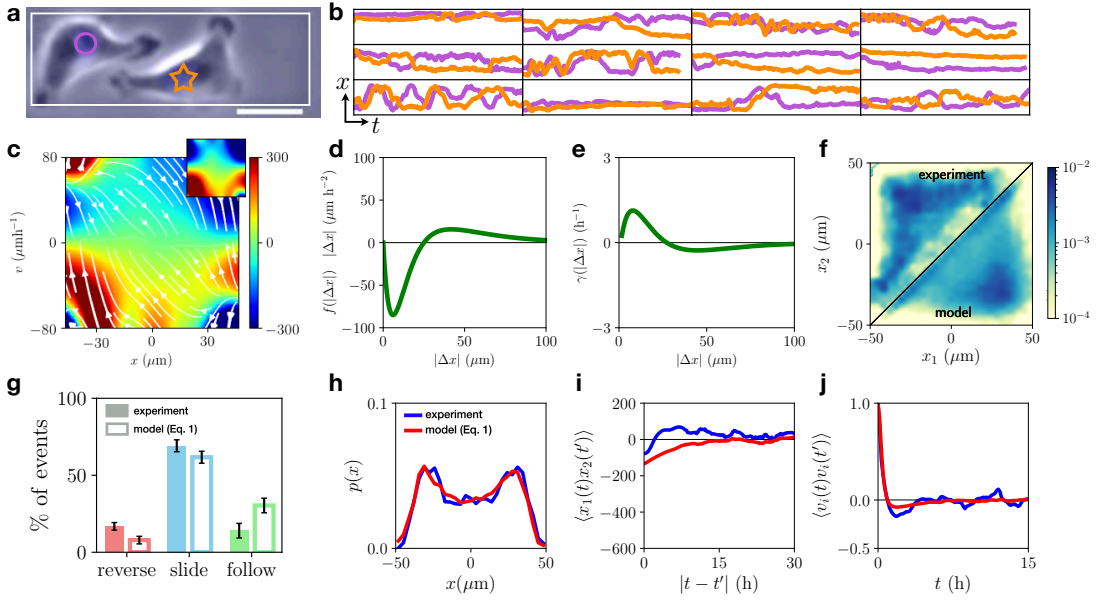

Figure S19: **Cell-cell interaction dynamics on a micropatterned track (MDA-MB-231 cells). a**, Microscopy image of an interacting cell pair confined to a micropatterned track. Micropattern outline is shown in white. Scale bar: 25 µm. **b**, Sample set of nuclear trajectories $x_{1,2}(t)$ as a function of time, shown for 12 cell pairs. Axes limits are $0 < t < 30$ h and -60 µm $< x <$ 60 µm, with $x = 0$ at the center of the pattern. A total of 84 cell pairs were tracked. **c**, Single-cell contribution $F(x, v)$ to the interacting dynamics, measured in units of µm/h². White lines indicate the flow field given by $(\dot{x}, \dot{v}) = (v, F(x,v))$. Inset: corresponding term inferred from experiments with single cells [1]. **d**, Cohesive interaction term $f(|\Delta x|)\Delta x$. Positive values indicate repulsive interactions, while negative values correspond to attraction. **e**, Effective frictional interaction term $\gamma(|\Delta x|)$. Here, positive values indicate an effective anti-friction, and negative values an effective frictional interaction. **f**, Joint probability distributions $p(x_1, x_2)$ of cell positions, plotted logarithmically. The top triangle of the symmetric distribution shows the experimental result, the bottom triangle shows the model prediction. **g**, Percentages of each of the three types of collision events observed. **h**, Probability distribution of all cell positions $p(x)$ (experiment shown in blue, model predictions in red). **i**, Cross-correlation function of cell positions $\langle x_1(t)x_2(t')\rangle$, plotted on the same scale as the plots in Fig. 2 of the main text. **j**, Normalized velocity auto-correlation function $\langle v_i(t)v_i(t')\rangle$.

freedom to explore the $y$-dimension of the pattern than in the two-state geometry, which is not accounted for by our 1D model. Taken together, these results demonstrate that our inference procedure can be generalized to other geometries, and that the results of the inferred cell-cell interactions do not sensitively depend on the precise confinement geometry.

We challenge our inference approach further by attempting to predict the collision statistics of interacting cells on a track geometry using only observations from other experiments to constrain our model. Specifically, we combine the single-cell and noise terms inferred from experiments where only a single cell migrates on the track together with the interactive terms

inferred from cell pair experiments on two-state micropatterns:

$$\begin{aligned}\dot{x}_i &= v_i \\ \dot{v}_i &= F_{\text{sc}}^{\text{track}}(x_i, v_i) + f^{\text{two-state}}(|\Delta x_{ij}|)\Delta x_{ij} + \gamma^{\text{two-state}}(|\Delta x_{ij}|)\Delta v_{ij} + \sigma_{\text{sc}}^{\text{track}}\eta_i(t)\end{aligned} \quad (S18)$$

Strikingly, this model quantitatively predicts the relative percentage of collision events observed in experiment, highlighting the potential generalizability of the inferred interactions.

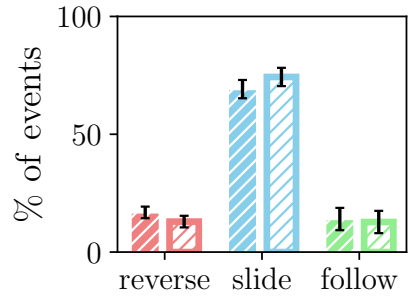

Figure S20: **Predicted collision behaviors using the inferred interactions from two-state micropatterns.** Experimental (solid) and predicted (empty) collision percentages using Eqn. S10. These correspond to the results shown in Fig. 2d of the main text.

# 6 Construction of the interaction behavior space (Fig. 5)

By inferring a model directly from experimental data, we have shown that the migration of cells in two-state micropatterns is well captured by an equation of motion of the form Eq. (S10). In order to predict behaviors beyond those sampled in our experiments, we extrapolate this description by scanning the interaction space spanned by the prefactors of cohesion and friction. Thus, we simulate trajectories using the equation of motion

$$\dot{v}_i = F(x_i, v_i) + f_0 g_c(|\Delta x_{ij}|)\Delta x_{ij} + \gamma_0 g_f(|\Delta x_{ij}|)\Delta v_{ij} + \sigma \eta_i(t) \tag{S19}$$

Here, we employ the confinement term $F(x_i, v_i)$ and the noise strength $\sigma$ inferred from MDA-MB-231 experiments. For the inference on our experimental data (section 4), we employed a functional expansion of the interactions using $N = 3$ exponentials, which were required to capture the quantitative details of the correlation functions (Fig. 2 in the main text). The collision behavior is already well captured with a simple expansion consisting a single exponential ($N = 1$). Thus, for simplicity, in the construction of the IBS we use single exponentials. For the result in the main text (Fig. 5), we use $g_c = g_f = e^{-|\Delta x_{ij}|/R_0}$ with $R_0 = 30$ µm.

To construct the IBS, we scan the $(f_0, \gamma_0)$-plane to obtain a set of models, each of which can be used to simulate a large set of stochastic trajectories. We apply the same analysis routines to these trajectories as to the experimental data, and identify for each value of $(f_0, \gamma_0)$ the relative percentage of reversal, sliding and following events (Fig. S21). We find that the interaction space exhibits well-defined behavioral regimes, in which one of the three behaviors dominates. For example, for a combination of repulsive and frictional interactions, we find that reversal events dominate over all other events (bottom right in Fig. S21a). Thus, we can identify for each value of $(f_0, \gamma_0)$ the collision type that dominates the observed behavior, plotted in Fig. S21d. Using this construction, we can therefore connect the interaction parameters governing the instantaneous short-time dynamics to the emergent long-time behavior. We therefore term this construction *interaction behavior space* (IBS).

Importantly, if we perform the same simulations with the MCF10A single-cell term, we obtain a qualitatively similar IBS (Fig. S22), indicating that our predictions are not sensitive to the details of the single-cell term, but are determined by the interactions. To determine where to approximately place the MCF10A and MDA-MB-231 cells within the IBS, we perform an inference with only a single ($N = 1$) exponential basis function. In accord with our more general inference (using $N = 3$), we find that MCF10A exhibit repulsive and frictional interactions, while MDA-MB-231 cells exhibit attractive and anti-friction interactions (stars in Fig. S22). Furthermore, the qualitative structure of the IBS is not sensitive to the choice of interaction decay length $R_0$ within a reasonable range, or to using non-exponential kernels (Fig. S23).

These results demonstrate two things: First, a large variety of behaviors can be captured by our model (Eq. (S19)). Second, with new experiments, similar models can be inferred, and the inferred interactions can be placed within the IBS, assuming no additional parameters are required to describe their behavior. Thus, the IBS could provide a way to connect different experimental observations within a single framework.

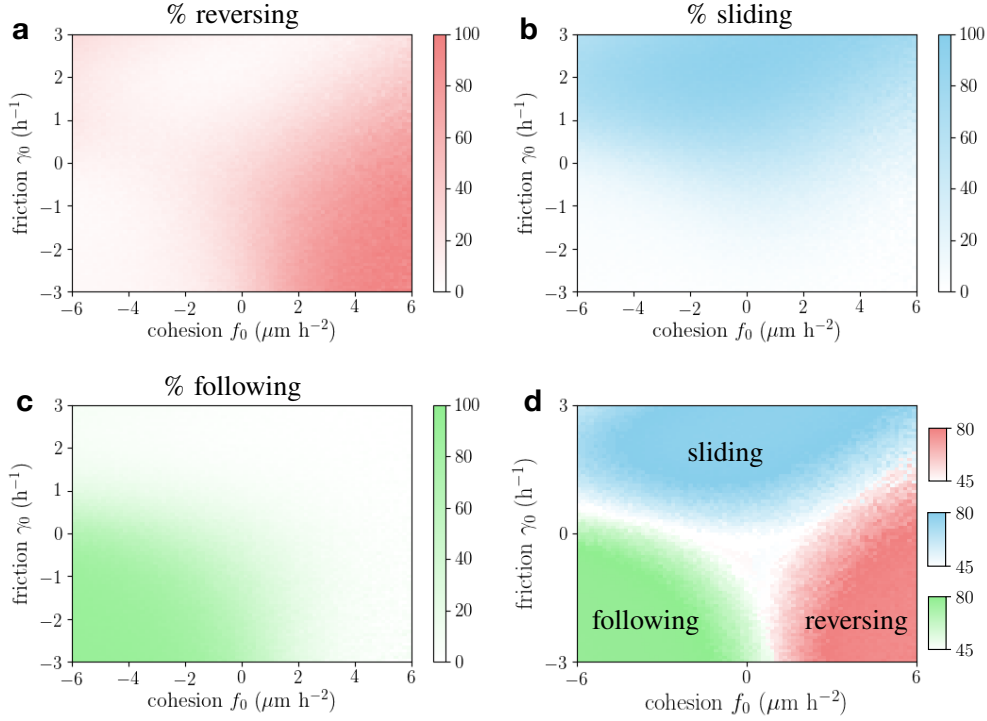

Figure S21: **Identification of behavioral regimes in the interaction behavior space. a**, Predicted percentage of reversal events as a function of the cohesion and friction prefactors. Here, the same interaction kernel is used as in Fig. 5 of the main text: $g_{c,f} = e^{-|\Delta x_{ij}|/R_0}$ with $R_0 = 30$ µm. **b**, Predicted percentage of sliding events. **c**, Predicted percentage of following events **d**, To construct the behavioral regimes, we identify the behavior with the maximal percentage in each bin, and plot its percentage in the respective colour scheme. The colour scheme is constructed such that percentages around 50%, where no single behavior contributes the majority of events, are plotted in white.

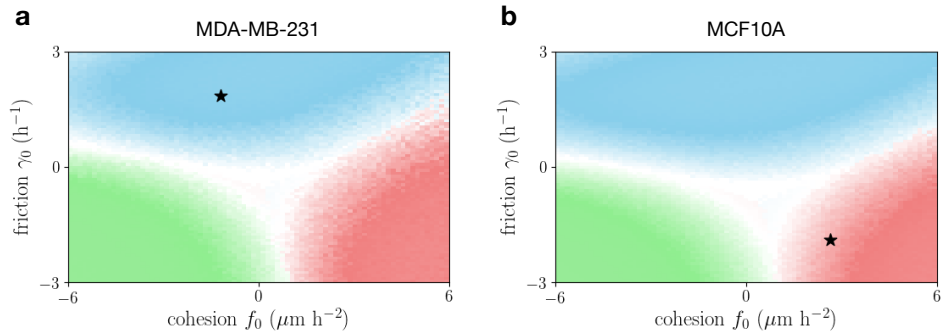

Figure S22: **Robustness of the IBS with respect to the single-cell term $F(x_i, v_i)$. a**, Predicted behavioral regimes using the MDA-MB-231 single-cell term. **b**, Predicted behavioral regimes using the MCF10A single-cell term. In both cases, we use $g_{c,f} = e^{-|\Delta x_{ij}|/R_0}$ with $R_0 = 30$ µm. Black stars indicate the inferred values for $(f_0, \gamma_0)$ for each cell type using an interaction basis consisting only of a single exponential $u_{\text{single}}(|\Delta x_{ij}|) = e^{-|\Delta x_{ij}|/R_0}$ with $R_0 = 30$ µm.

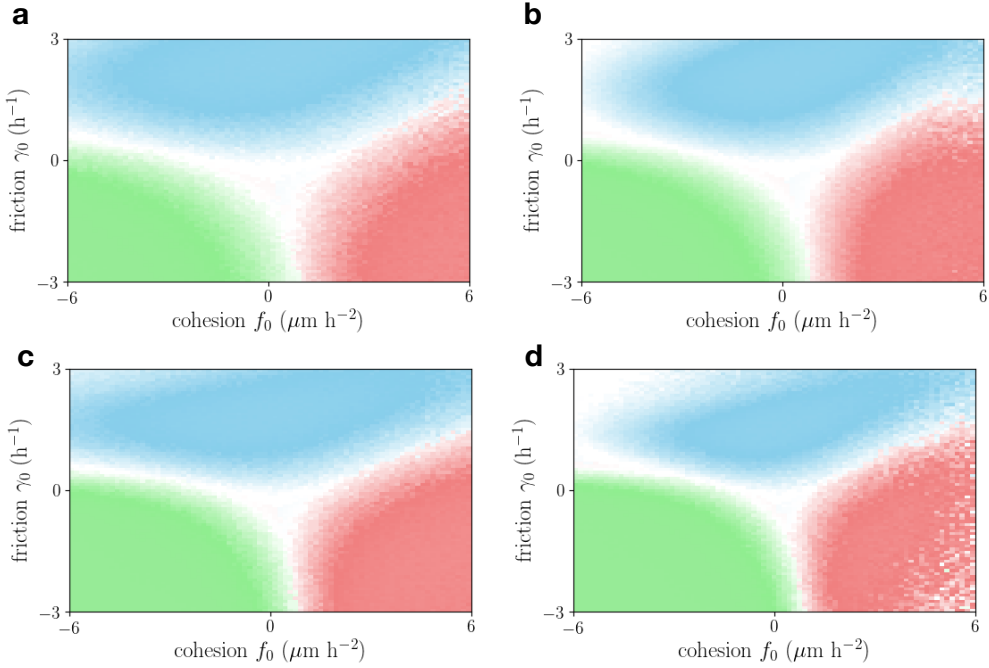

Figure S23: **Robustness of the IBS with respect to the functional form $g_{c,f}(|\Delta x_{ij}|)$ of the interaction kernels.** **a**, Predicted behavioral regimes using $g_{c,f} = e^{-|\Delta x_{ij}|/R_0}$ with $R_0 = 30$ µm (shown in main text). **b**, Predicted behavioral regimes using $g_{c,f} = e^{-|\Delta x_{ij}|/R_0}$ with $R_0 = 40$ µm. **c**, Predicted behavioral regimes using $g_{c,f} = 1/[(|\Delta x_{ij}|/R_0)^4 + 1]$ with $R_0 = 30$ µm. **d**, Predicted behavioral regimes using $g_{c,f} = 1/[(|\Delta x_{ij}|/R_0)^4 + 1]$ with $R_0 = 40$ µm.

# References

[1] D. B. Brückner, A. Fink, C. Schreiber, P. J. F. Röttgermann, J. O. Rädler, and C. P. Broedersz, "Stochastic nonlinear dynamics of confined cell migration in two-state systems," *Nature Physics*, vol. 15, no. 6, pp. 595–601, 2019.

[2] D. Blair and E. Dufresne, "The Matlab Particle Tracking Code Repository," *http://site.physics.georgetown.edu/matlab/*, 2008.

[3] B. Efron, "Bootstrap Methods: Another Look at the Jackknife," *The Annals of Statistics*, vol. 7, pp. 1–26, jul 1979.

[4] B. Efron and R. Tibshirani, *An Introduction to the Bootstrap*. CRC Press, 1994.

[5] D. F. Milano, N. A. Ngai, S. K. Muthuswamy, and A. R. Asthagiri, "Regulators of Metastasis Modulate the Migratory Response to Cell Contact under Spatial Confinement," *Biophysical Journal*, vol. 110, no. 8, pp. 1886–1895, 2016.

[6] R. A. Desai, S. B. Gopal, S. Chen, and C. S. Chen, "Contact inhibition of locomotion probabilities drive solitary versus collective cell migration," *Journal of the Royal Society Interface*, vol. 10, no. 88, 2013.

[7] J. Singh, B. A. Camley, and A. S. Nain, "Rules of Contact Inhibition of Locomotion for Cells on Suspended Nanofibers," *bioRxiv*, 2020.

[8] D. Li and Y. L. Wang, "Coordination of cell migration mediated by sitedependent cell-cell contact," *Proceedings of the National Academy of Sciences of the United States of America*, vol. 115, no. 42, pp. 10678–10683, 2018.

[9] D. B. Brückner, P. Ronceray, and C. P. Broedersz, "Inferring the dynamics of underdamped stochastic systems," *Physical Review Letters*, vol. 125, no. 5, p. 58103, 2020.

[10] D. Selmeczi, S. Mosler, P. H. Hagedorn, N. B. Larsen, and H. Flyvbjerg, "Cell motility as persistent random motion: theories from experiments.," *Biophysical journal*, vol. 89, no. 2, pp. 912–931, 2005.

[11] G. J. Stephens, B. Johnson-Kerner, W. Bialek, and W. S. Ryu, "Dimensionality and Dynamics in the Behavior of C. elegans," *PLoS Comput Biol*, vol. 4, no. 4, p. e1000028, 2008.

[12] J. N. Pedersen, L. Li, C. Gradinaru, R. H. Austin, E. C. Cox, and H. Flyvbjerg, "How to connect time-lapse recorded trajectories of motile microorganisms with dynamical models in continuous time," *Physical Review E*, vol. 94, no. 6, p. 062401, 2016.

[13] F. Ferretti, V. Chardès, T. Mora, A. M. Walczak, and I. Giardina, "Building General Langevin Models from Discrete Datasets," *Physical Review X*, vol. 10, no. 3, p. 031018, 2020.



\end{document}